\documentclass[12pt,aps,prd,preprint,tightenlines,superscriptaddress,
amsmath,amssymb,nofootinbib]{revtex4-1} 

\RequirePackage[colorlinks=true
,urlcolor=blue
,anchorcolor=blue
,citecolor=blue
,filecolor=blue
,linkcolor=blue
,menucolor=blue
,pagecolor=blue
,linktocpage=true
,pdfproducer=medialab
,pdfa=true
]{hyperref}

\allowdisplaybreaks

\usepackage{amsmath,amssymb,amsthm,amsfonts}
\usepackage{graphicx}
\usepackage{subfigure}
\usepackage{dcolumn}	
\usepackage{hyperref}      	
\usepackage{bm}		
\usepackage{epsfig}
\usepackage{epstopdf}
\usepackage{setspace}
\usepackage[usenames, dvipsnames]{color}
\usepackage{slashed}
\usepackage{comment}
\usepackage{enumitem}
\usepackage{siunitx}
\usepackage{multirow}
\usepackage{datetime}
\usepackage{lineno}
\usepackage{float}
\usepackage{xspace}

\newcommand{\PRE}[1]{{#1}} 

\newcommand{\be}{\begin{equation}\begin{aligned}}
\newcommand{\ee}{\end{aligned}\end{equation}}
\newcommand{\beq}{\begin{equation}}
\newcommand{\eeq}{\end{equation}}
\newcommand{\beqa}{\begin{eqnarray}}
\newcommand{\eeqa}{\end{eqnarray}}

\newcommand{\ifb}{\text{fb}^{-1}}

\newcommand{\gev}{\text{GeV}}
\newcommand{\tev}{\text{TeV}}
\newcommand{\fb}{\text{fb}}

\newcommand{\mm}{\text{mm}}
\newcommand{\cm}{\text{cm}}
\newcommand{\m}{\text{m}}

\newcommand{\fs}{\text{fs}\xspace}
\newcommand{\ps}{\text{ps}\xspace}
\newcommand{\ns}{\text{ns}\xspace}
\newcommand{\mus}{\ensuremath{\mu}\text{s}\xspace}
\newcommand{\s}{\text{s}}

\newcommand{\kHz}{\text{kHz}\xspace}
\newcommand{\MHz}{\text{MHz}\xspace}

\renewcommand{\eqref}[1]{Eq.~(\ref{#1})}

\newcommand{\secref}[1]{Sec.~\ref{sec:#1}}

\newcommand{\Secref}[1]{Section~\ref{sec:#1}}
\newcommand{\figref}[1]{Fig.~\ref{fig:#1}}
\newcommand{\figsref}[2]{Figs.~\ref{fig:#1} and \ref{fig:#2}}
\newcommand{\Figref}[1]{Figure~\ref{fig:#1}}
\newcommand{\tableref}[1]{Table~\ref{table:#1}}

\begin{document}
 
\preprint{Submitted to the LHCC, 7 November 2018 \hspace*{1.4in} CERN-LHCC-2018-036, LHCC-P-013}
\preprint{UCI-TR-2018-22, KYUSHU-RCAPP-2018-07}

\title{\PRE{\vspace*{0.2in}}
TECHNICAL PROPOSAL \\
\PRE{\vspace*{0.4in}}
{\Large FASER} \\ \vspace*{0.2in}
{\Large FORWARD SEARCH EXPERIMENT AT THE LHC}
\PRE{\vspace*{0.3in}}}

\author{Akitaka Ariga}
\affiliation{Universit\"at Bern, Sidlerstrasse 5, CH-3012 Bern, Switzerland}

\author{Tomoko Ariga}
\affiliation{Universit\"at Bern, Sidlerstrasse 5, CH-3012 Bern, Switzerland}
\affiliation{Kyushu University, Nishi-ku, 819-0395 Fukuoka, Japan}

\author{Jamie Boyd}
\email[Contact email: ]{Jamie.Boyd@cern.ch}
\affiliation{CERN, CH-1211 Geneva 23, Switzerland}

\author{Franck Cadoux}
\affiliation{D\'epartement de Physique Nucl\'eaire et Corpusculaire, 
University of Geneva, CH-1211 Geneva 4, Switzerland}

\author{David~W.~Casper}
\affiliation{Department of Physics and Astronomy, 
University of California, Irvine, CA 92697-4575, USA}

\author{Francesco Cerutti}
\affiliation{CERN, CH-1211 Geneva 23, Switzerland}

\author{Salvatore Danzeca}
\affiliation{CERN, CH-1211 Geneva 23, Switzerland}

\author{Liam Dougherty}
\affiliation{CERN, CH-1211 Geneva 23, Switzerland} 

\author{Yannick Favre}
\affiliation{D\'epartement de Physique Nucl\'eaire et Corpusculaire, 
University of Geneva, CH-1211 Geneva 4, Switzerland}

\author{Jonathan~L.~Feng}
\email[Contact email: ]{jlf@uci.edu}
\affiliation{Department of Physics and Astronomy, 
University of California, Irvine, CA 92697-4575, USA}

\author{Didier Ferrere}
\affiliation{D\'epartement de Physique Nucl\'eaire et Corpusculaire, 
University of Geneva, CH-1211 Geneva 4, Switzerland}

\author{Jonathan Gall}
\affiliation{CERN, CH-1211 Geneva 23, Switzerland}

\author{Iftah Galon}
\affiliation{New High Energy Theory Center, 
Rutgers, The State University of New Jersey, 
Piscataway, New Jersey 08854-8019, USA}

\author{Sergio Gonzalez-Sevilla}
\affiliation{D\'epartement de Physique Nucl\'eaire et Corpusculaire, 
University of Geneva, CH-1211 Geneva 4, Switzerland}

\author{Shih-Chieh Hsu}
\affiliation{Department of Physics, University of Washington, PO Box 351560, Seattle, WA 98195-1560, USA}

\author{Giuseppe Iacobucci}
\affiliation{D\'epartement de Physique Nucl\'eaire et Corpusculaire, 
University of Geneva, CH-1211 Geneva 4, Switzerland}

\author{Enrique Kajomovitz}
\affiliation{
Technion---Israel Institute of Technology, Haifa 32000, Israel}

\author{Felix Kling}
\affiliation{Department of Physics and Astronomy, 
University of California, Irvine, CA 92697-4575, USA}

\author{Susanne Kuehn}
\affiliation{CERN, CH-1211 Geneva 23, Switzerland}

\author{Mike Lamont}
\affiliation{CERN, CH-1211 Geneva 23, Switzerland}

\author{Lorne Levinson}
\affiliation{
Weizmann Institute of Science, Rehovot 761001, Israel}

\author{Hidetoshi Otono}
\affiliation{Kyushu University, Nishi-ku, 819-0395 Fukuoka, Japan}

\author{John Osborne}
\affiliation{CERN, CH-1211 Geneva 23, Switzerland}

\author{Brian Petersen}
\affiliation{CERN, CH-1211 Geneva 23, Switzerland}

\author{Osamu Sato}
\affiliation{Nagoya University, Furo-cho, Chikusa-ku, Nagoya 464-8602, Japan}

\author{Marta Sabat\'e-Gilarte}
\affiliation{CERN, CH-1211 Geneva 23, Switzerland}
\affiliation{University of Seville, Seville, Spain}

\author{Matthias Schott}
\affiliation{Institut für Physik, Universität Mainz, Mainz, Germany}

\author{Anna Sfyrla}
\affiliation{D\'epartement de Physique Nucl\'eaire et Corpusculaire, 
University of Geneva, CH-1211 Geneva 4, Switzerland}

\author{Jordan Smolinsky}
\affiliation{Department of Physics and Astronomy, 
University of California, Irvine, CA 92697-4575, USA}

\author{Aaron~M.~Soffa}
\affiliation{Department of Physics and Astronomy, 
University of California, Irvine, CA 92697-4575, USA}

\author{Yosuke Takubo}
\affiliation{Institute of Particle and Nuclear Study, 
KEK, Oho 1-1, Tsukuba, Ibaraki 305-0801, Japan}

\author{Pierre Thonet}
\affiliation{CERN, CH-1211 Geneva 23, Switzerland}

\author{Eric Torrence}
\affiliation{University of Oregon, Eugene, OR 97403, USA}

\author{Sebastian Trojanowski}
\affiliation{\mbox{National Centre for Nuclear Research, Ho{\. z}a 69, 00-681 Warsaw, Poland}}
\affiliation{Consortium for Fundamental Physics, School of Mathematics and  Statistics, University of Sheffield, Hounsfield Road, Sheffield, S3 7RH, UK}

\author{Gang Zhang\PRE{\vspace*{.2in}}}
\affiliation{
Tsinghua University, Beijing, China
}


\begin{abstract}
\PRE{\vspace*{0.2in}}
FASER is a proposed small and inexpensive experiment designed to search for light, weakly-interacting particles during Run 3 of the LHC from 2021--23.  Such particles may be produced in large numbers along the beam collision axis, travel for hundreds of meters without interacting, and then decay to standard model particles. To search for such events, FASER will be located 480 m downstream of the ATLAS IP in the unused service tunnel TI12 and be sensitive to particles that decay in a cylindrical volume with radius $R= 10~\cm$ and length $L = 1.5~\m$.  FASER will complement the LHC's existing physics program, extending its discovery potential to a host of new, light particles, with potentially far-reaching implications for particle physics and cosmology.

This document describes the technical details of the FASER detector components: the magnets, the tracker, the scintillator system, and the calorimeter, as well as the trigger and readout system. The preparatory work that is needed to install and operate the detector, including civil engineering, transport, and integration with various services is also presented. The information presented includes preliminary cost estimates for the detector components and the infrastructure work, as well as a timeline for the design, construction, and installation of the experiment.
\end{abstract}


\maketitle

\renewcommand{\baselinestretch}{0.95}\normalsize
\tableofcontents
\renewcommand{\baselinestretch}{1.0}\normalsize

\clearpage
\section{Introduction}
\label{sec:introduction}

FASER is a proposed small and inexpensive experiment designed to search for light, weakly-interacting particles at the LHC.  Such particles are dominantly produced along the beam collision axis and may be long-lived particles (LLPs), traveling hundreds of meters before decaying.  To exploit both of these properties, FASER is to be located along the beam collision axis, 480 m downstream from the ATLAS interaction point, in the unused service tunnel TI12.  FASER will be sensitive to particles that decay in a small cylindrical volume with radius $R= 10~\cm$ and length $L = 1.5~\m$. Despite its small size, FASER will significantly extend the LHC's discovery potential to a host of new, light particles.  We propose that FASER be installed in TI12 in Long Shutdown 2 in time to collect $150~\ifb$ of data from 2021-23 during Run 3 of the 14 TeV LHC. 

The basic physics of the FASER experiment and the concept of the detector have been described in FASER's Letter Of Intent (LOI)~\cite{Ariga:2018zuc}. This document gives much more detail about the technical aspects of the experiment. In \Secref{overview} we give a brief overview of the physics and experiment.  We then describe the detector environment in \Secref{environment}.  In Sections~\ref{sec:magnet}--\ref{sec:tdaq}, we describe the components of the detector in turn: the magnets, the tracker, the scintillators, the calorimeter, the detector support structure and the readout and trigger.  In Sections~\ref{sec:civil}--\ref{sec:safety} we give details about the civil engineering required to install the detector, installation and integration, commissioning, and safety aspects.  We discuss off-line software and computing in \Secref{software} and summarize the overall costing and schedule in \Secref{costing}.

\clearpage
\section{Overview of FASER}
\label{sec:overview}

\subsection{Physics Goals}

For decades, the leading examples of new physics targets at particle colliders were particles with TeV-scale masses and ${\cal O}(1)$ couplings to the standard model (SM).  More recently, however, there is a growing and complementary interest in new particles that are much lighter and more weakly coupled~\cite{Battaglieri:2017aum}.  Among their many motivations, such particles may yield dark matter with the correct thermal relic density and resolve outstanding discrepancies between theory and low-energy experiments~\cite{Bennett:2006fi, Pohl:2010zza, Krasznahorkay:2015iga}.  Perhaps most importantly, new particles that are light and weakly coupled can be discovered by relatively inexpensive, small, and fast experiments.

If new light and weakly-interacting particles exist, they are typically produced parallel to the beam line and may travel hundreds of meters without interacting before decaying to visible particles, such as electrons and positrons.  The existing detectors at the LHC, such as ATLAS and CMS, are therefore not well-matched to these particles, as they have holes along the beam line to let the proton beams in, and new light, weakly-interacting particles would escape through these holes undetected.  

The goal of FASER is to target this ``blind spot'' by being located along the beam collision axis, far downstream from the ATLAS interaction point (IP).  At this point, the proton beams are bent by magnets and so pass by FASER unhindered. However, new light, weakly-interacting particles will travel in straight lines and can decay to visible particles in FASER.  Moreover, such particles are highly collimated.  For example, new particles that are produced in pion decays are typically produced within angles of $\theta \sim \Lambda_{\text{QCD}} / E$ of the beam collision axis, where $E$ is the energy of the particle.  For $E \sim \text{TeV}$, this implies that even $\sim 480~\m$ downstream, such particles have only spread out $\sim 10~\cm$ in the transverse plane.  A small and inexpensive detector placed in the very forward region may therefore be capable of highly sensitive searches. 

FASER's physics reach has now been investigated for a host of light, weakly-interacting particles.  In these studies, the most typical signals are those of LLPs that are produced at or close to the IP, travel along the beam collision axis, and decay visibly in FASER:
\begin{equation}
  p p  \to \text{LLP} +X, \quad  \text{LLP travels} \ \sim 480~\text{m}, \quad \text{LLP} \to \text{charged tracks} + X \text{ (or $\gamma \gamma + X$)} \ .
\end{equation}
These signals are striking: two oppositely charged tracks (or two photons) with very high energy ($\sim \tev$) that emanate from a common vertex inside the detector and which have a combined momentum that points back through 10 m of concrete and 90 m of rock to the IP. Studies of dark photons, dark Higgs bosons, heavy neutral leptons, light $B-L$ gauge bosons, axion-like particles, and others~\cite{Feng:2017uoz, Feng:2017vli, Batell:2017kty, Kling:2018wct, Helo:2018qej, Bauer:2018onh, Feng:2018noy, Berlin:2018jbm, Dercks:2018eua} have demonstrated that FASER and a possible follow-up experiment, FASER 2, have a full physics program, with significant discovery potential in a variety of models. A systematic study of the physics potential is presented in~\cite{Ariga:2018uku}.

In addition to searches for new physics, FASER may also provide interesting probes of standard model physics.  As an example, in Run 3, the number of muon neutrinos passing through FASER with energies above 100 GeV is $\sim 10^{13}$, with roughly 600 interacting in the 10 cm-thick block of lead that is near the front of FASER (see \secref{exptoverview}).  Furthermore, a few tau neutrinos with $\sim \tev$ energies are expected to interact in FASER.  Although more study is required, these event rates imply that FASER may also provide interesting information about SM particles by detecting the first neutrinos at the LHC and, for example, constraining neutrino interaction rates in the energy range $E_{\nu} \sim 400~\gev - 4~\tev$, where they are currently unconstrained. (See, for example, Appendix 1 of Ref.~\cite{Ariga:2018zuc}.)

\subsection{The Experiment}
\label{sec:exptoverview}

The proposed location of FASER and the LHC infrastructure between the IP and the detector are shown in \figref{Infrastructure}.  A sketch of the proposed detector to be installed in the TI12 tunnel\footnote{In the LOI we were expecting the TI18 tunnel to be the best place for FASER, but additional measurements made by the CERN survey team during LHC Technical Stop 2 show that the TI12 tunnel on the other side of the LHC interaction point, IP1, will allow for a longer detector on the beam collision axis. TI12 is the same distance from IP1 as TI18, but is on the side towards the LHCb experiment, rather than towards ALICE. } is shown in \figref{DetectorLayout}, and a brief overview is given below.

\begin{figure}[tbp]
\centering
\includegraphics[width=0.99\textwidth]{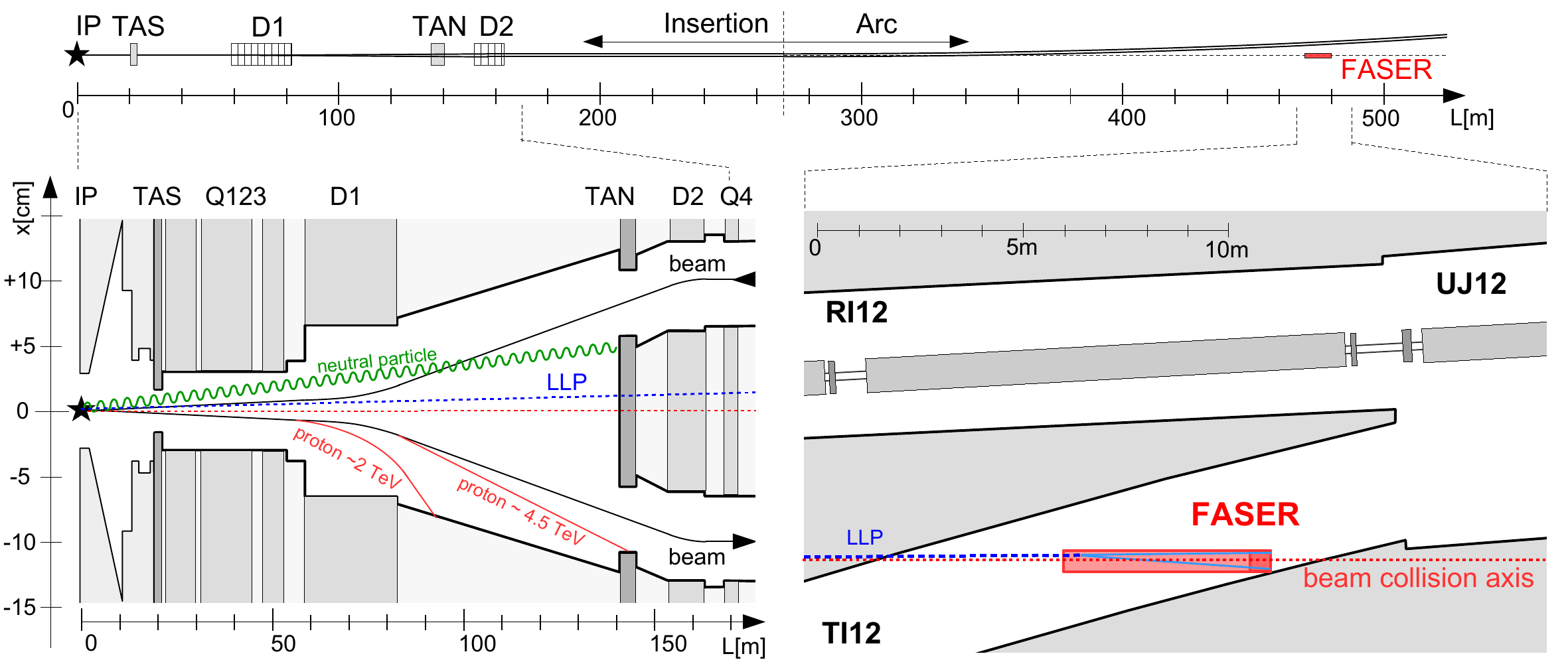} 
\caption{Schematic view of the far-forward region downstream of ATLAS and various particle trajectories. {\bf Upper panel}: FASER is located $480~\m$ downstream of ATLAS along the beam collision axis (dotted line) after the main LHC tunnel curves away.  {\bf Lower left panel}: High-energy particles produced at the IP in the far-forward direction.  Charged particles are deflected by LHC magnets, and neutral hadrons are absorbed by either the TAS or TAN, but LLPs pass through the LHC infrastructure without interacting. Note the extreme difference in horizontal and vertical scales.  {\bf Lower right panel}: LLPs may then travel $\sim 480~\m$ further downstream, passing through 10 m of concrete and 90 m of rock, and decay within FASER in TI12. 
} 
\label{fig:Infrastructure}
\end{figure}

\begin{figure}[tbp]
\centering
\includegraphics[width=0.98\textwidth]{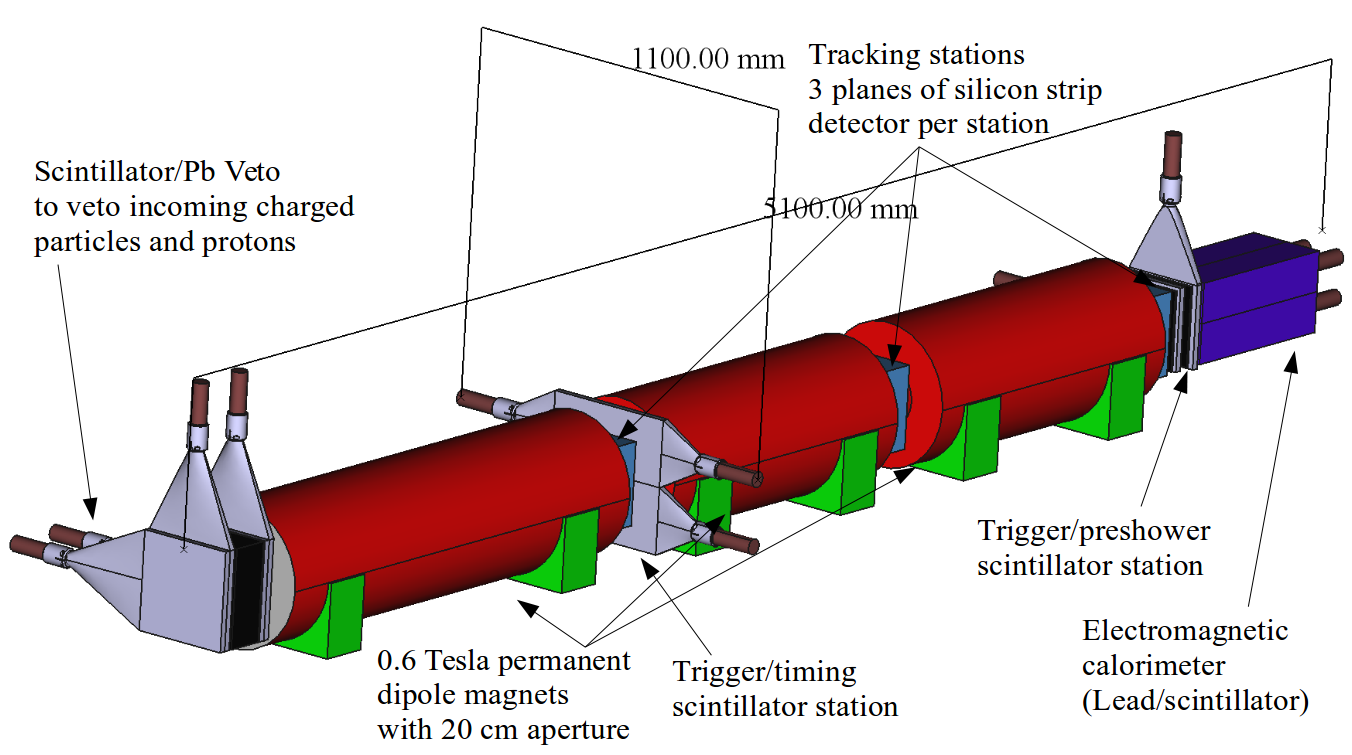} 
\caption{
Layout of the proposed FASER detector. LLPs enter from the left.  The detector components include scintillators (gray), dipole magnets (red), tracking stations (blue), and a calorimeter (dark purple). 
}
\label{fig:DetectorLayout}
\end{figure}

At the entrance to the detector, two scintillator stations are used to veto charged particles coming through the cavern wall from the IP; these are primarily high-energy muons. Each station consists of two layers of scintillators. Between the stations is a 20-radiation-lengths-thick layer of lead that converts photons produced in the wall into electromagnetic showers that can be efficiently vetoed by the scintillators.

The veto stations are followed by a $1.5~\m$ long, 0.6 T permanent dipole magnet with a $10~\cm$ aperture radius. This is the decay volume for LLPs decaying into a pair of charged particles, where the magnet providing a horizontal kick to separate the decay products to a detectable distance. The decay volume is not foreseen to be under vacuum.

After the decay volume is a spectrometer consisting of two $1~\m$ long, 0.6 T dipole magnets with three tracking stations, which are located at either end and in between the magnets. Each tracking station is composed of layers of precision silicon strip detectors. The three magnets have their fields aligned to give the maximum separation for charged particles in the bending plane. Scintillator stations for triggering and precision time measurements are located at the entrance and exit of the spectrometer. The primary purpose of the spectrometer is to observe the characteristic signal of two oppositely-charged particles pointing back towards the IP, measure their momenta, and sweep out low-momentum charged particles before they reach the final layer of the spectrometer. 

The final component is the electromagnetic calorimeter. This will identify high-energy electrons and photons and measure the total electromagnetic energy. The primary signals are two close-by electrons or photons with too-small separation for the calorimeter to resolve individually.

\clearpage
\section{Detector Environment}
\label{sec:environment}

\subsection{Detector Length Constraints}

Detailed measurements from the CERN survey team have mapped out the beam collision axis or line of sight (LOS) in both the TI18 and TI12 tunnels. This LOS assumes no crossing angle between the beams at IP1.  In reality, the LHC runs with a crossing angle in the range of about 150~$\mu$rad\footnote{This value is the half crossing angle, which is typically what is quoted and what is used for the rest of this document.}, the effect of which is discussed below. The TI18 and TI12 tunnels connect the LHC to the much shallower SPS, and they therefore slope steeply upwards as they leave the LHC tunnel.  Because of this geometry, the LOS is below the tunnel floor as it enters the tunnel, and then emerges from the floor. Given this, to maximize the detector length that can be centered on the LOS, it will be desirable to lower the floor.  Measurements from the CERN survey team show that, with the allowed digging that can be done in LS2 (limited to 460~mm, as discussed in \secref{civil}), a significantly longer detector can be installed in TI12 than in TI18.  In particular, based on our design, we could fit a roughly 5 m-long detector in TI12, compared to a roughly 3 m-long detector in TI18. 

A more detailed modeling of the tunnel and the detector shows that the length of the detector becomes limited by the back part hitting the wall of the tunnel, as shown in \figref{det-position}.

\begin{figure}[bp]
\centering
\includegraphics[width=0.98\textwidth]{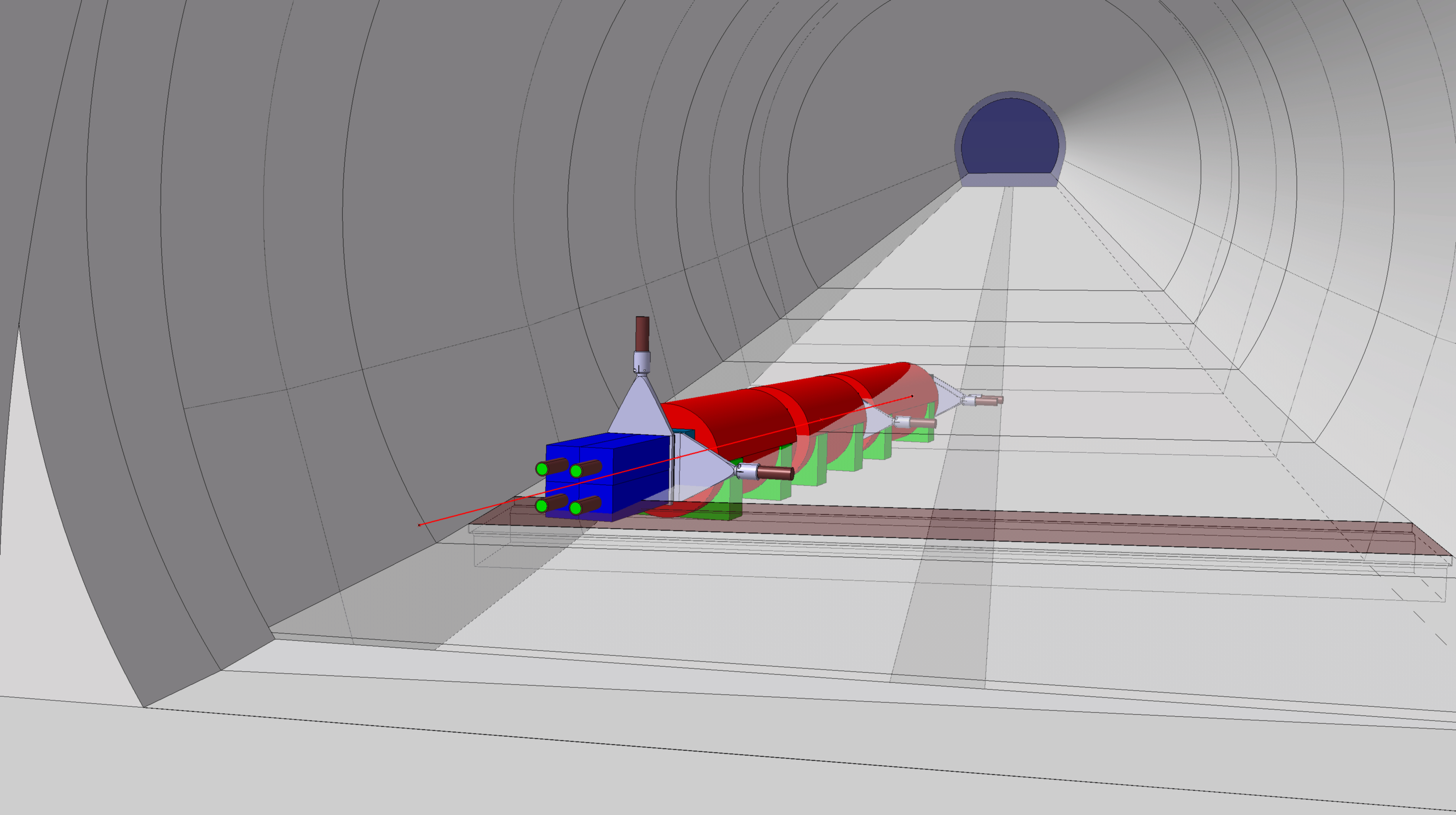}
\caption{A model of the FASER detector situated at the proposed location (centered on the nominal LOS) in the TI12 tunnel.}
\label{fig:det-position}
\end{figure}

\subsection{Beam Configuration Effects}

\subsubsection{Effect of IP1 beam crossing angle}

To avoid long range beam-beam effects and parasitic collisions inside the common beam pipe, the LHC runs with a half crossing angle at IP1 of about 150~$\mu$rad. In the LHC design, and for all running to date, the crossing angle is in the vertical plane. For Run 2, it was decided to flip the crossing angle direction (sign) periodically (e.g., once per year) to distribute the collision debris more evenly and prolong the lifetime of the focusing magnets, which are a potential limitation on the total integrated luminosity that can be delivered before LS3. In addition, since the start of 2017, the crossing angle was reduced during the physics fills from an initial value of 160~$\mu$rad to 120~$\mu$rad at the end of the fill. At the FASER location (480~m from IP1) a half crossing angle of 160~$\mu$rad corresponds to a shift of the collision axis (compared to the nominal LOS, assuming no crossing angle) of 7.7~cm, and 120~$\mu$rad will shift the axis by about 2~cm less.

For Run 3 the optics that will be used in the LHC has not been finalized. Two options are under consideration: (i) {\em round beams}, in which the $\beta^\ast$ is the same in the horizontal and vertical planes, and (ii) {\em flat beams}, in which the $\beta^\ast$ is larger in the crossing plane and which can give higher luminosities. In the case of {\em round beams}, the crossing angle will be similar to that used in Run 2, being in the vertical plane and with the sign possibly changed each year of running. With {\em flat beams}, the crossing angle will be changed to be in the horizontal plane, and the sign will not be changed (this will be fixed to the sign for which the LOS points towards the outside of the LHC ring). In all cases, the crossing angle values will be similar to those used in Run 2 and will be reduced in a similar way during the physics fills.

For FASER the considerations above mean that we need to be ready for the LOS to be displaced from the nominal LOS by up to 8~cm in either the horizontal plane or in the vertical plane (and switching between the displacement being up or down every year). The baseline strategy to deal with this would be to keep the detector centered on the nominal LOS and sacrifice some signal rate from the fact that the actual LOS is shifted with respect to the nominal LOS by the crossing angle. Simulation studies show that this leads to a minor loss in physics reach; for example, for dark photons, it leads to a loss in signal acceptance of roughly 25\%, corresponding to an almost imperceptible change in sensitivity reach in the $(\text{mass}, \text{coupling})$ plane. It may be possible to align the detector to be more centered on the LOS taking into account the crossing angle (especially in the case of flat beams when this will not be changed between years), but for now we consider that we will center FASER on the nominal LOS.

\subsubsection{Effect of beam divergence}

The beam divergence is a measure of the intrinsic transverse momentum spread of the collision system due to the machine optics. It is given by $D = \sqrt{ \epsilon / \beta^* }$, where $\epsilon$ is the transverse emittance of the beam, and $\beta^*$ is the value of the $\beta$ function at the IP. Whereas the crossing angle shifts the LOS, the effect of the divergence is to spread the per-collision collision products out around the nominal LOS. For the typical values of parameters expected to be used in Run 3 ($\beta* \approx 30$~cm and $\epsilon \approx 3 \times 10^{-10}$~m), we expect $D\approx 30~\mu$rad. In the case of flat optics with $\beta^*_x\approx 60$~cm and $\beta^*_y\approx 15$~cm, this will give $D_x\approx 20~\mu$rad and $D_y\approx 40~\mu$rad. Given the LHC machine parameters, then, the value of the divergence is usually of the order of 10\% of the crossing angle, and so we expect the effect of the divergence to be very small for FASER. Of course, the effect of the divergence can be taken into account in the signal simulations.

\subsubsection{Effect of filling scheme}

The effect of the LHC filling scheme on FASER is expected to be minimal. The background particle flux coming from the IP will be related to the instantaneous luminosity, but these rates are much lower than the bunch crossing rate, and so the effect of pileup is completely negligible. The signal rate is also proportional to the instantaneous luminosity, but again, for the same luminosity, it will not be dependent on the filling scheme.

\subsection{Particle Flux}
\label{sec:particleFlux}

FLUKA simulation~\cite{Ferrari:2005zk,Bohlen:2014buj} studies have been carried out to estimate the expected particle fluxes entering the FASER detector~\cite{FLUKAstudy}.  These studies include particles arising from collisions in IP1 and from beam-related backgrounds in the LHC. In addition, detectors have been installed in the TI18 and TI12 tunnels during LHC Technical Stops in 2018 to measure the particle flux and validate the simulations. These fluxes are important as they will determine the trigger rate and the number of particles that can be used for detector alignment and calibration. The current simulations and most of the {\em in situ} measurements are for TI18.  The preferred FASER location is now the TI12 tunnel, but the expectation is that the particle flux will be the same in TI12 and TI18, as the collision system and LHC infrastructure is symmetric around IP1. First {\em in situ} measurements from TI12 demonstrate that this is the case, and future simulations will be carried out for TI12.

\subsubsection{FLUKA simulation}

FLUKA simulations have been carried out by the EN-STI CERN group as part of the Physics Beyond Colliders effort. 
The simulations were made for the TI18 tunnel, normalized to Run 3 conditions (we assume a constant luminosity of $2 \times 10^{34}~\cm^{-2}~\s^{-1}$). The construction of the TI12 tunnel simulation model is envisaged, although its symmetry a priori suggests no major impact on the results.

The simulations show, as expected, that muons and neutrinos are the only particles that traverse the FASER detector locations with an appreciable rate. The muons and neutrinos are either produced directly from proton-proton interactions at the IP or in upstream showers from particles produced at the IP hitting machine components (for example, high energy neutrons hitting the TAN 140~m from the IP). The FLUKA simulation contains a realistic geometry of the LHC tunnel and the LHC optics, and is therefore expected to model such processes well. Similar simulations using this setup have been used for understanding particle fluences and radiation levels in different areas of the LHC complex, and in many cases have been validated with measurements.

The simulated particle fluences as functions of energy for muons and neutrinos are shown in \figref{particleFlux} for an instantaneous luminosity of $2 \times 10^{34}~\cm^{-2}~\s^{-1}$ and collision energy of 13 TeV. This shows there is a large difference in the spectra between positive and negative charged muons, due to the bending of the LHC magnets.  FLUKA studies show that only some percent of the muons at FASER come from the IP, but rather originate in showers from particles produced in the IP1 collisions and later interacting with the LHC machine elements (for example the TAN 140~m from the IP). Whereas the neutrinos that go through FASER are mostly originating directly from the IP1 collisions (typically from pion decay, where the pions are produced in the IP1 collisions).  The rate for all muons with energy above different thresholds is presented in Table~\ref{tab:FLUKAflux}. For particles with energy above 10~GeV, this gives a charged particle rate through the 10~cm-radius detector volume of around 100~Hz. 

\begin{figure}[tbp]
\centering
\includegraphics[width=0.87\textwidth]{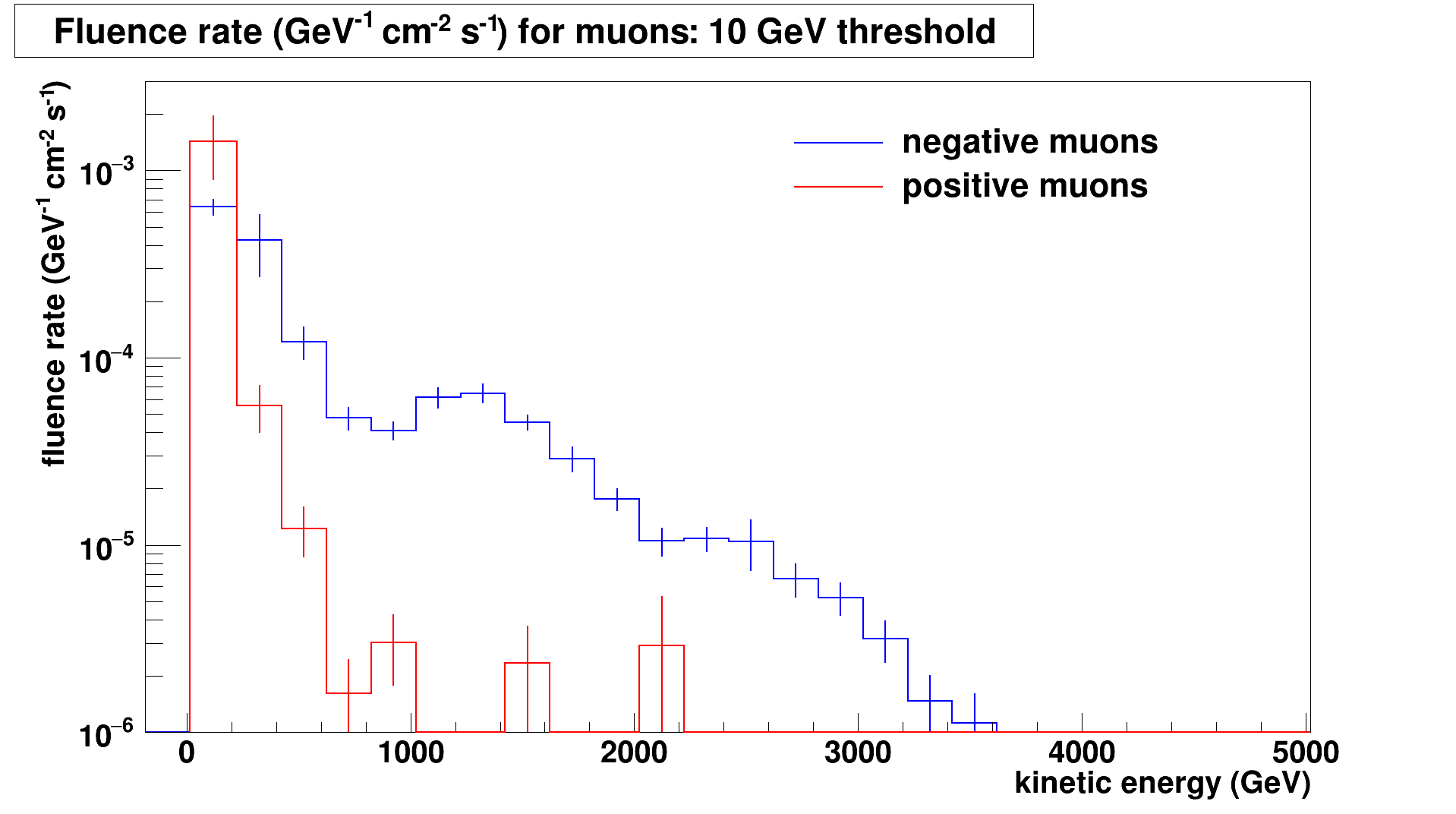}
\newline
\includegraphics[width=0.87\textwidth]{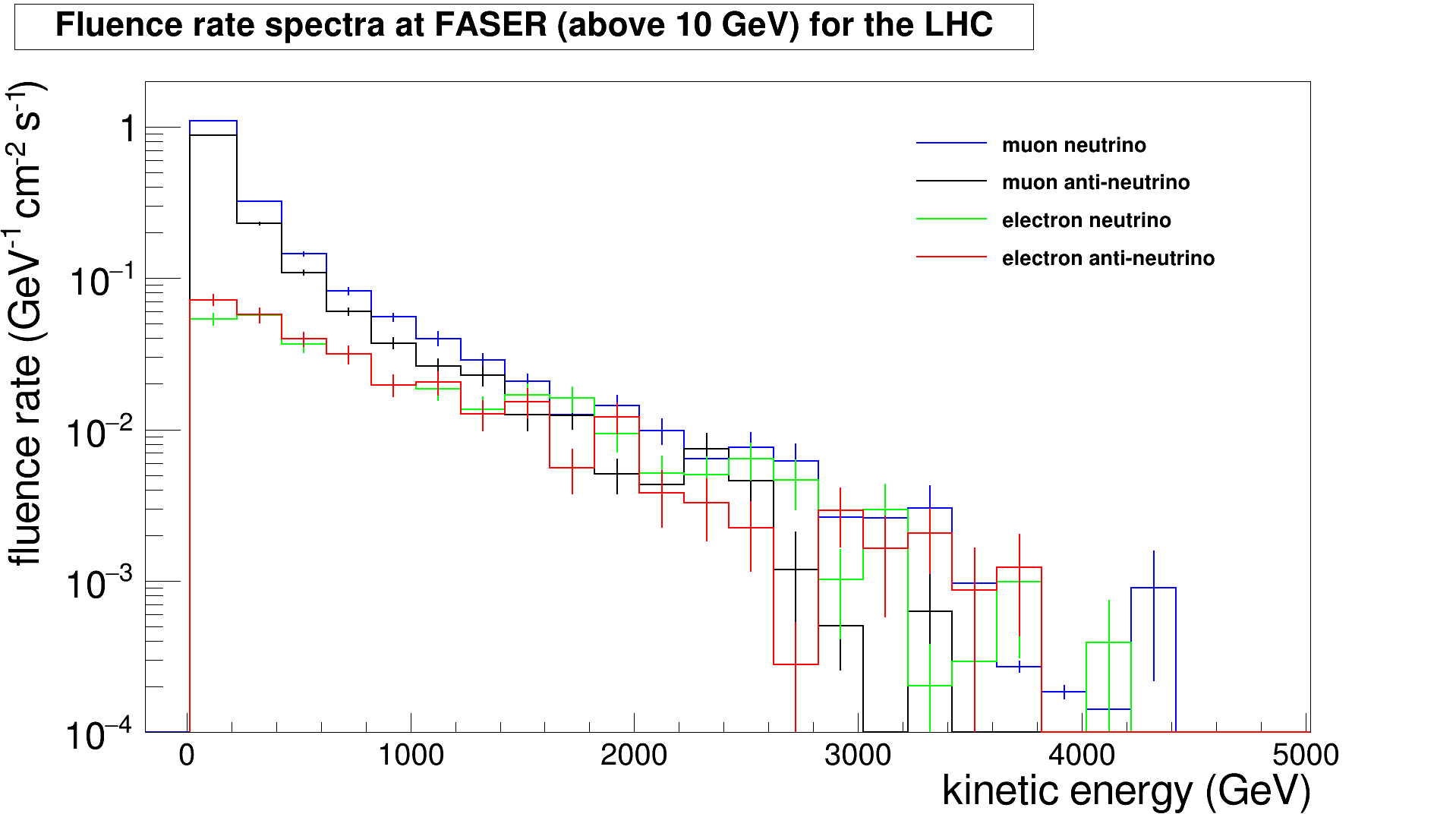}
\caption{FLUKA simulation estimation of the particle flux as a function of energy at the FASER location: (top) for negative and positive muons; (bottom) for different neutrino species. These are normalized to a luminosity of $2 \times 10^{34}~\cm^{-2}~\s^{-1}$. }
\label{fig:particleFlux}
\end{figure}

\begin{table}[tbhp]
  \centering
  \begin{tabular}{|c|c|}
  \hline
	Energy threshold  & Charged particle flux  \\
    $[\gev]$ & [$\cm^{-2}~\s^{-1}$] \\
    \hline
10 &  0.40 \\ 
100 &  0.20 \\ 
1000 &  0.06 \\ 
 \hline
  \end{tabular}
  \caption{The expected charge particle flux at the FASER location from FLUKA simulations for different energy thresholds, normalized to the expected Run 3 luminosity of $2 \times 10^{34}~\cm^{-2}~\s^{-1}$. The rate is entirely from muons.
}  \label{tab:FLUKAflux}
\end{table}

FLUKA does not include the rate of particles produced in neutrino interactions in the rock, but this is estimated to be completely negligible (less than 0.01~Hz).

The FLUKA estimations of the particle flux have an uncertainty of order a factor of a few, dominated by statistical effects in the current simulation samples. With larger samples there will still be sizable systematic uncertainties at the tens of percent level.

\Figref{ParticleFluxMap} shows the muon flux as a function of radial position around FASER. It can be seen that the FASER detector is in a region with reduced particle flux, with significantly higher rates expected 1--2~m on either side of FASER (for positive and negative muons separately), due to the bending of the muons in the LHC magnetic field.

\begin{figure}[bp]
\centering
\includegraphics[width=0.47\textwidth]{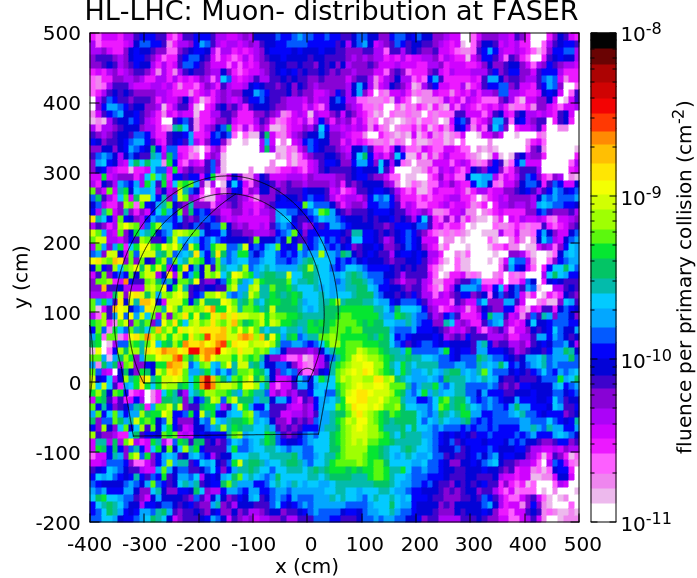}
\includegraphics[width=0.47\textwidth]{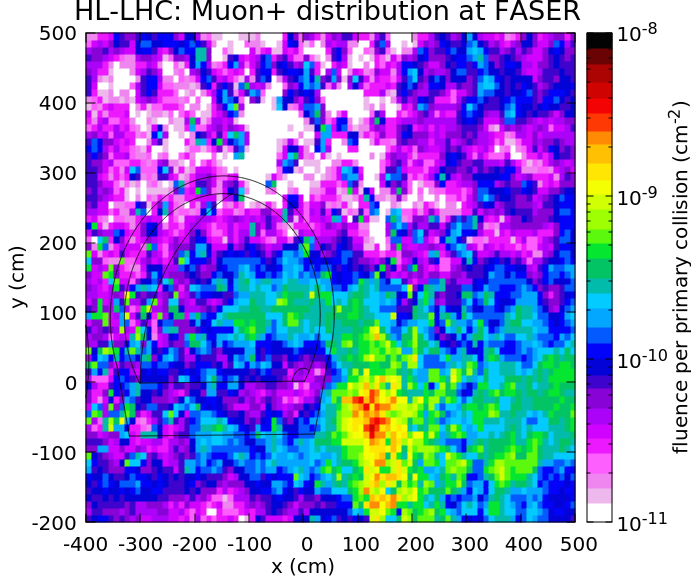}
\includegraphics[width=0.45\textwidth]{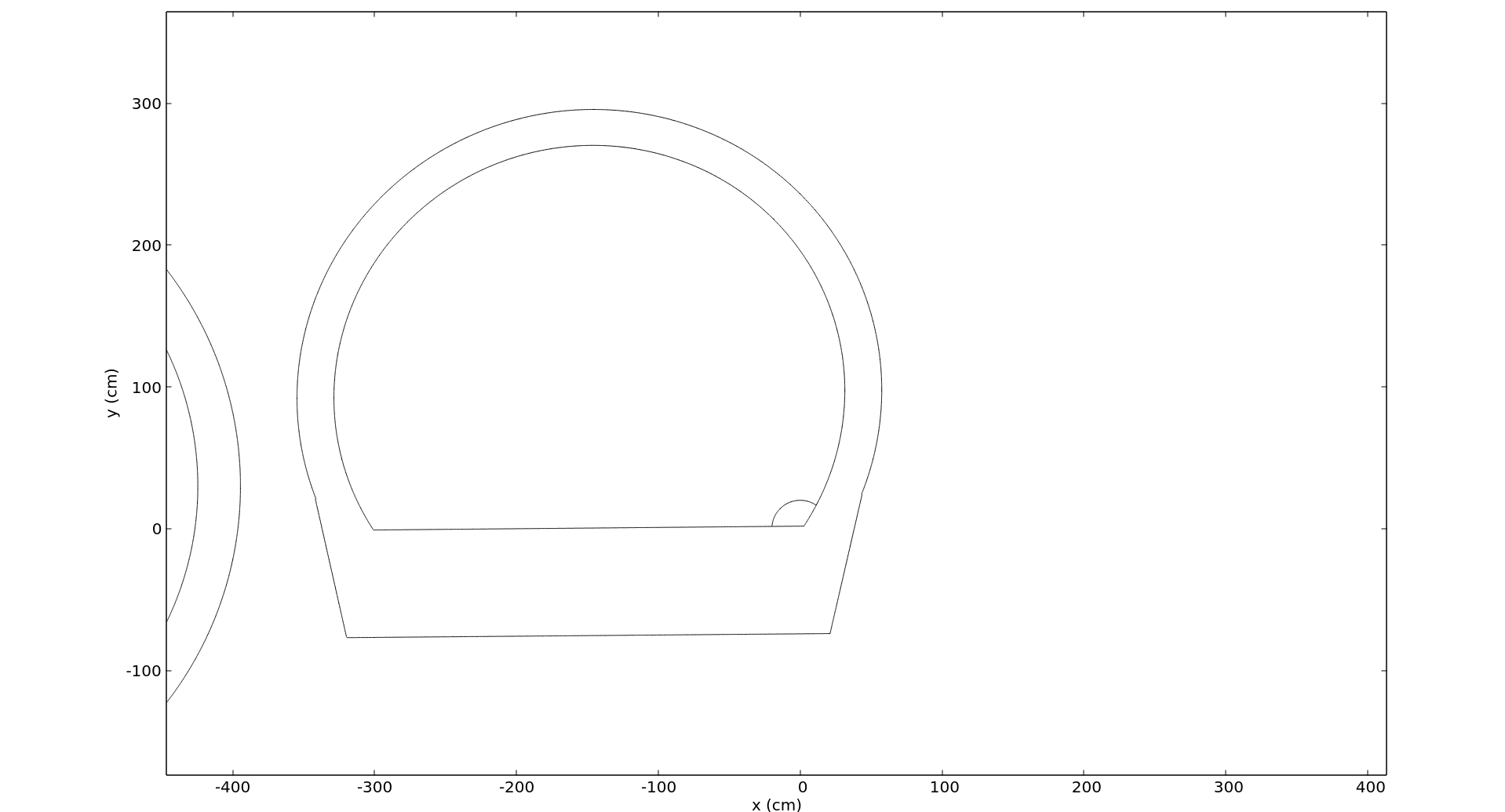}
\caption{FLUKA simulation estimates of negative muon fluxes (top left) and positive muon fluxes (top right) in the transverse plane at the FASER location.  These results assume the TI18 location, 485~m from the IP. The diagram at the bottom shows the geometry used in the simulations.  The FASER detector is visible as a small, partially-cut circle of radius 20~cm at the bottom right of the tunnel. 
}
\label{fig:ParticleFluxMap}
\end{figure}

FLUKA can also be used to estimate the flux of particles entering FASER that are not produced in interactions at IP1. These include particles coming from showers in the LHC beam pipe in the dispersion suppressor (which arise from off-momentum protons following diffractive interactions in IP1, hitting the beam aperture, and causing particle showers), and particles produced in beam-gas interactions in the LHC beampipe. Both of these sources are highly suppressed: the dispersion function close to the FASER location minimizes proton losses in this region, and the excellent vacuum in the LHC means beam-gas interaction rates are extremely small. The FLUKA simulations predict the background of high-energy particles entering FASER from these processes is negligible.

\subsubsection{{\em In situ} measurements}

To measure particle fluxes at FASER's location, emulsion detectors were placed in the TI18 and TI12 tunnels. Photos and maps of the installation and the angular distributions of particles detected are shown in \figref{InSituMeasurement}.  The angular distributions show clear peaks of charged particles entering the detector from directions compatible with the ATLAS IP (the peak population of tracks at close to (0,0) on the figures). The width of the main peak was measured to be 2.3~mrad as shown in \figref{ti_ty}, which is as narrow as the angular resolution of the emulsion films, which is 2~mrad. This implies that the particles in this peak are sufficiently energetic not to be affected by multiple Coulomb scattering through about 100~m of rock and concrete shielding. Note that the expected transverse momentum due to the multiple Coulomb scattering through 100~m of rock is 0.54~GeV, corresponding to a scattering of 2~mrad for 270~GeV particles without including their ionization energy loss.
In addition, the emulsion detector measurements were taken both with and without 0.5-mm-thick tungsten plates, which effectively imposed different energy cutoffs and provided insights about particle energy as a function of angle. As shown in \figref{projection_ti12}, the tracks at the peripheral part are mostly composed of low-energy particles below 1 GeV.

\begin{figure}[t]
\begin{tabular}{lll}
\includegraphics[height=5.5cm]{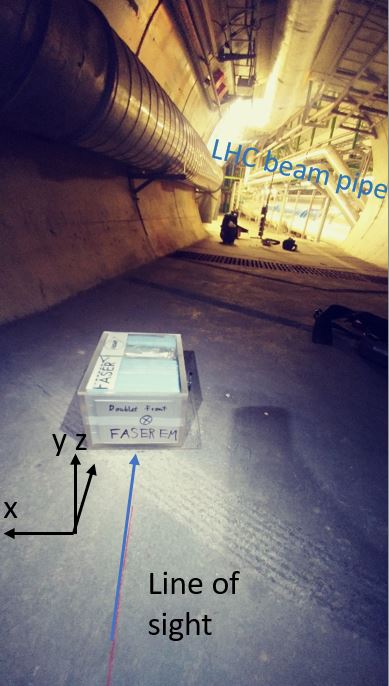} &
\includegraphics[width=5.5cm]{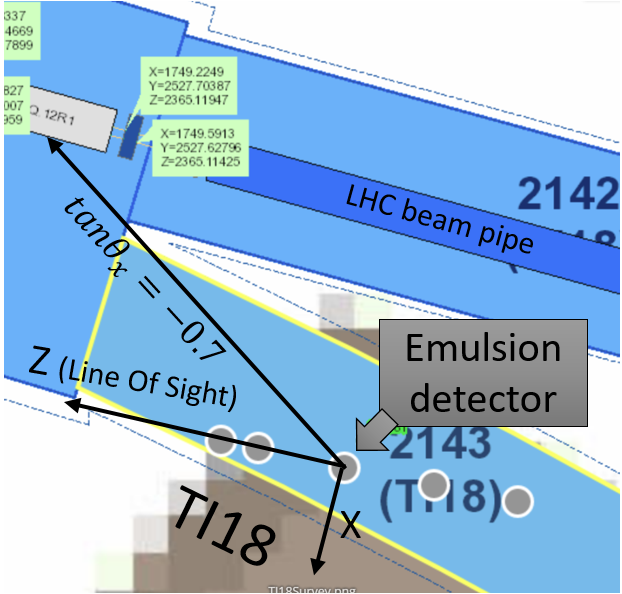}& 
\includegraphics[height=6.cm]{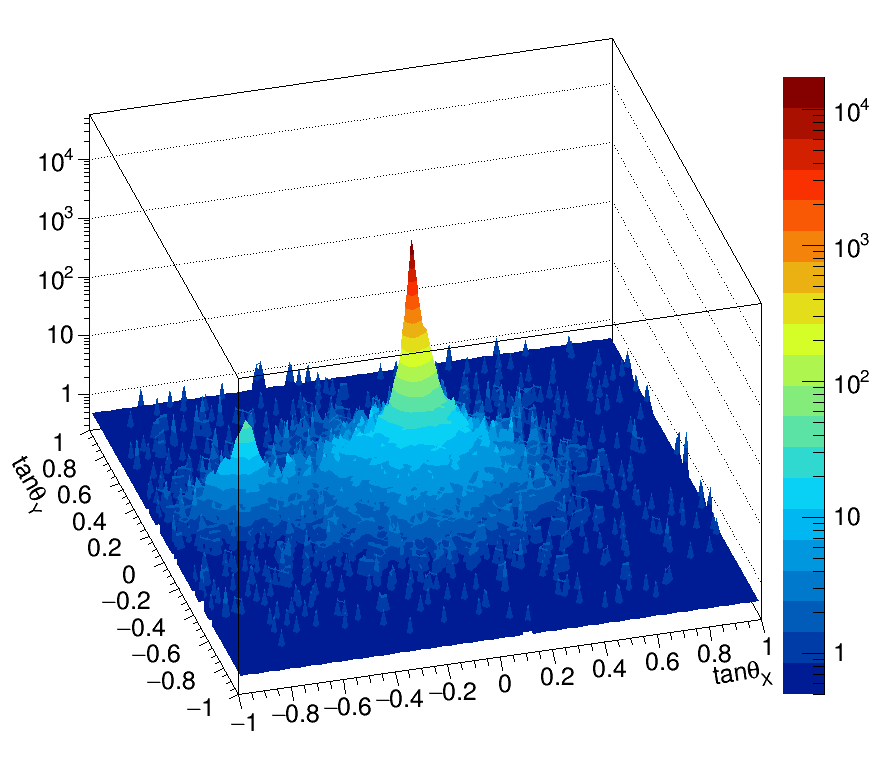} \\
\includegraphics[height=5.5cm]{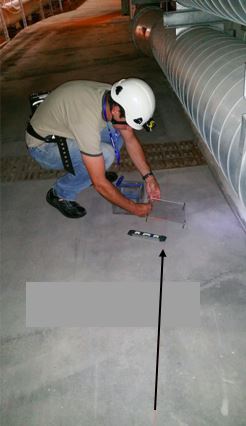} &
\includegraphics[width=5.5cm]{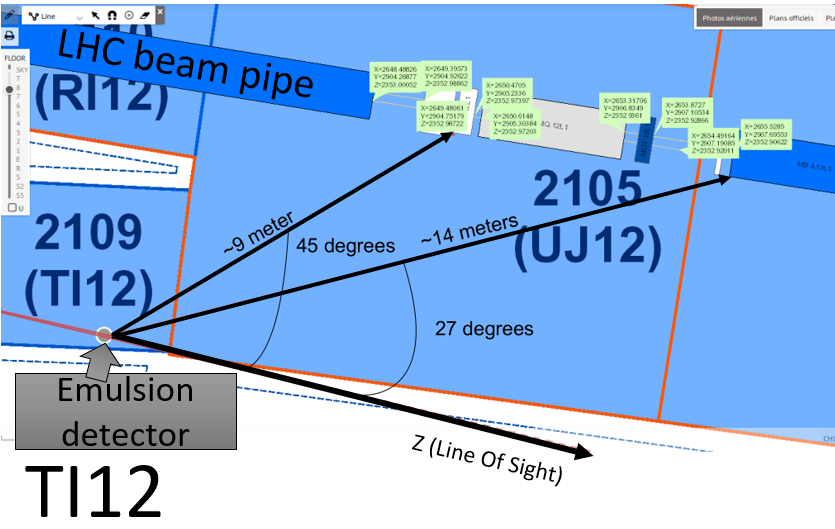} &
\includegraphics[height=6.cm]{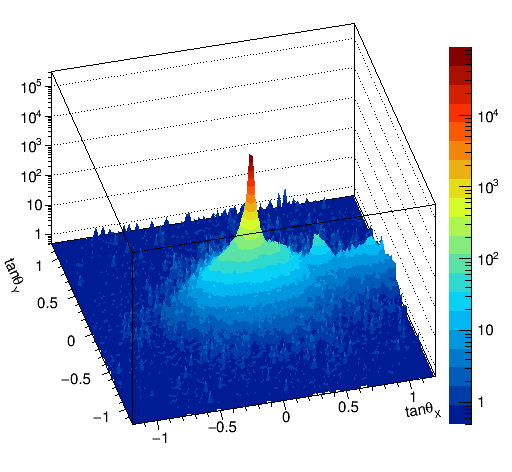}
\end{tabular}
\caption{\textit{In situ} measurements by emulsion detectors at the TI18 location (upper panels) and TI12 location (lower panels). We show photos of the installed detectors (left), maps of the installation locations (center), and angular distributions of the detected particles (right).}
\label{fig:InSituMeasurement}
\end{figure}

\begin{figure}[tbhp]
\begin{minipage}[t]{0.48\textwidth}
\includegraphics[width=\textwidth]{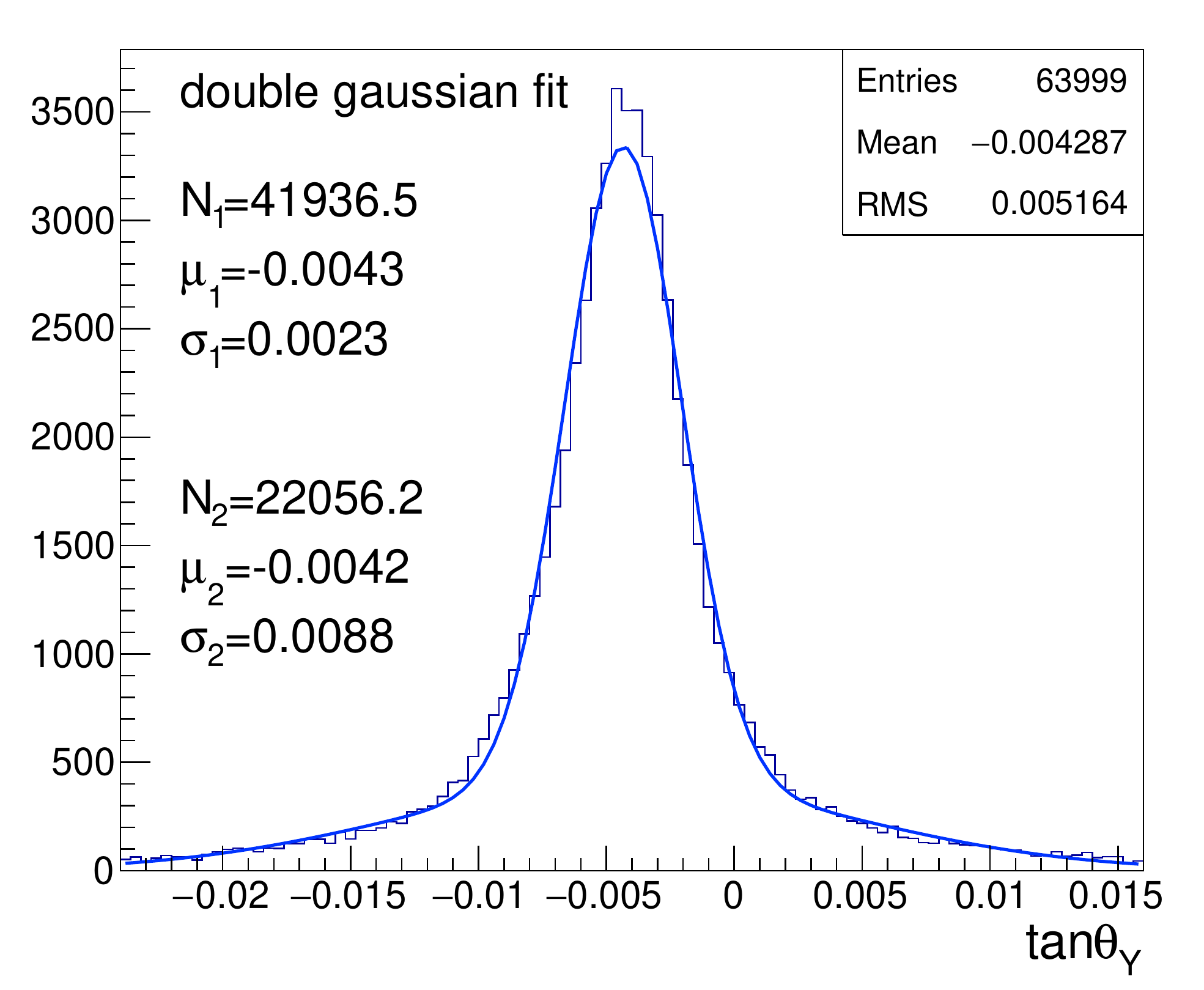}
\caption{A magnified view of the main peak of the angular distribution of particles detected at TI12, projected into the $y$-axis.  The width of the main peak is 2.3 mrad, which is nearly consistent with the emulsion detector's angular resolution of 2 mrad. }
\label{fig:ti_ty}
\end{minipage} 
\hfill
\begin{minipage}[t]{0.48\textwidth}
\includegraphics[width=\textwidth]{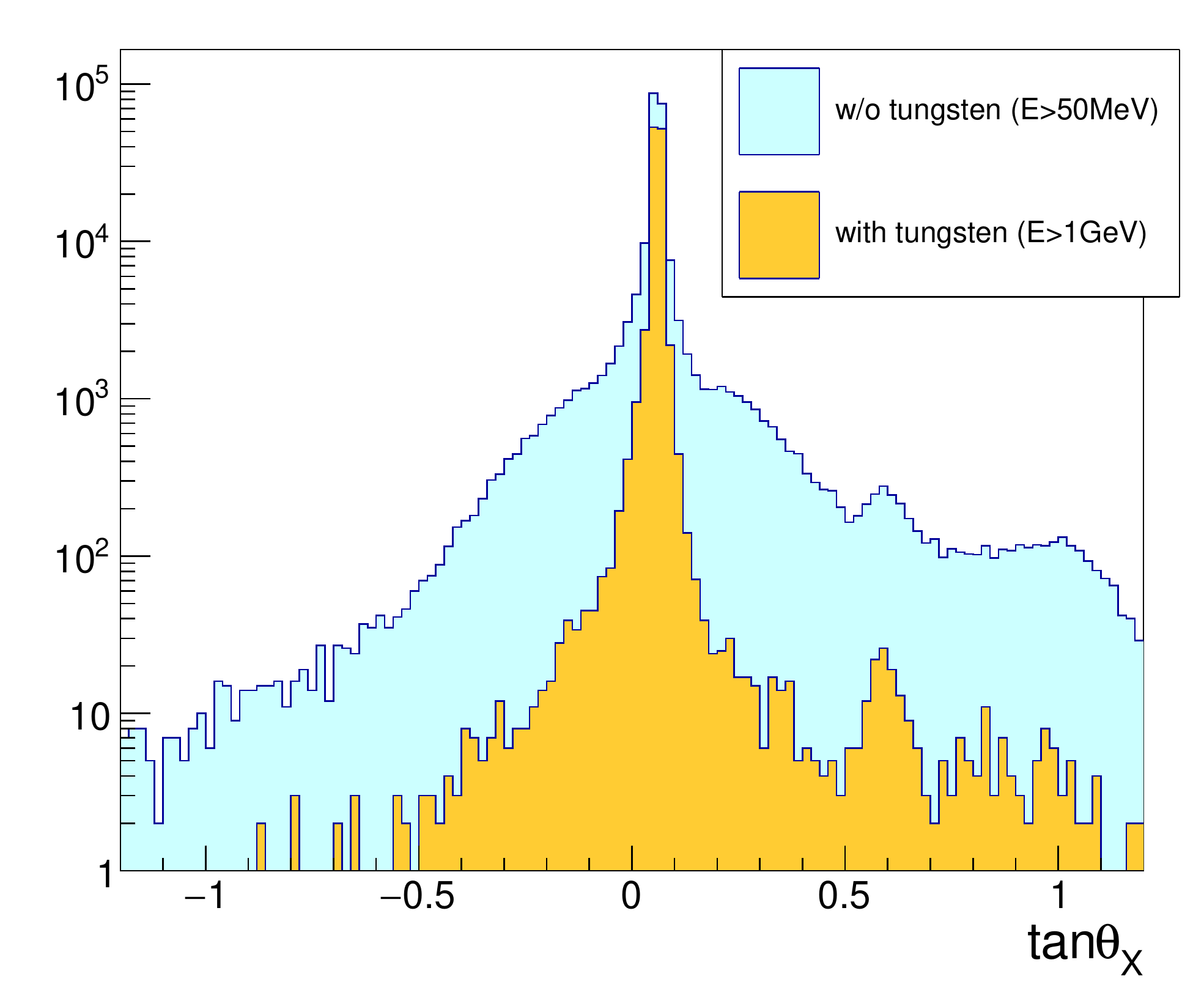}
\caption{Angular distributions, projected into the $x$-axis, by three emulsion films with and without tungsten plates, corresponding to energy cutoffs of about 1 GeV and 50 MeV, due to multiple Coulomb scattering, respectively.}
\label{fig:projection_ti12}
\end{minipage}
\end{figure}

The observed number of charged particles and rates are summarized in Table~\ref{tab:particleflux}.  The particle fluxes in the main peak (within 10~mrad in angular space) were measured to be $(1.2-1.9)\times 10^4~\fb~\cm^{-2}$ when corrected for detection efficiency. These can be compared with the expected flux from the FLUKA simulations of $2.0\times 10^4~\fb~\cm^{-2}$ for particles with the energy above 10 GeV. 
The {\em in situ} measurement therefore agrees well with the simulation given the uncertainties in the detection efficiency and uncertainty in the simulations.

\begin{table}
\begin{tabular}{|c|c|c|c|c|c|}
\hline
& beam & observed tracks & efficiency & normalized flux, all & normalized flux, main peak \\
& $[\ifb]$ & [$\cm^{-2}$] & & [$\fb~\cm^{-2}$] & [$\fb~\cm^{-2}$] \\
\hline
TI18 & 2.86& 18407 & 0.25 & $(2.6\pm 0.7)\times 10^4$& $(1.2\pm 0.4) \times 10^4$\\
\hline
TI12 & 7.07& 174208 & 0.80 & $(3.0\pm 0.3)\times 10^4$& $(1.9\pm 0.2)\times 10^4$\\
\hline
\hline
\multicolumn{4}{|l|}{FLUKA simulation, E$>$100 GeV}  & \multicolumn{2}{|c|}{$1 \times 10^4$}\\
\hline
\end{tabular}
\caption{Measured fluxes from emulsion detector data. The fluxes in the main peak (within 10 mrad) should be compared with the FLUKA simulation.}
\label{tab:particleflux}
\end{table}

An interesting feature in the emulsion detector data is small secondary peaks visible in the angular distributions, for example, at $(-0.75,0)$ in the TI18 data shown in \figref{InSituMeasurement}, with $\sim 1\%$ of the total number of tracks. This corresponds to tracks entering the detector with an angle consistent with originating at the LHC beamline at the bottom of the TI18 tunnel and, therefore, entering the detector without passing through any rock. The coordinate system and the tunnel geometry can also be seen in the center panels of \figref{InSituMeasurement}. The FLUKA simulations do not show such a population of tracks, but for the beam-gas simulations they only included high energy particles with $E> 100~\gev$, whereas the emulsion detectors are sensitive to particles with much lower energies. It is therefore likely that these are low energy particles and will not be problematic for the experiment. A more detailed analysis of the emulsion detector data is ongoing, and it should provide a determination of the rate of high-energy electromagnetic objects. 

An active monitoring device (a TimePix3 Beam Loss Monitor~\cite{TimePix}) was installed in the TI18 tunnel on the LOS during LHC Technical Stop 2. This device has the capability to correlate the rate of detected particles with the beam conditions.  In particular, it can separately determine the particle detection rate during periods with high-luminosity collisions, periods with high-energy beams but no collisions (for example, during the `squeeze' beam process), and periods with no beam in the machine. During periods with high-luminosity collisions, the rate can also be correlated with the instantaneous luminosity at IP1. However the device and the reconstruction algorithm are not calibrated, so it cannot currently provide absolute measurements of the particle flux.  
A first analysis of the TimePix detector data shows a clear correlation between the observed cluster counts and the instantaneous luminosity in IP1. It also shows slightly larger rates when there is beam in the LHC, but with no collisions, compared to no beam in the machine. Example results are shown in Table~\ref{tab:TimePix}.

\begin{table}
  \centering
  \begin{tabular}{|l|c|c|c|}
  \hline
	{\bf Period} & {\bf Luminosity} & {\bf Counting Rate} & {\bf Counting Rate/Luminosity} \\
     & $[10^{34}$~s$^{-1}$~cm$^{-2}]$ & [s$^{-1}$] & [$10^{-34}~$cm$^{2}]$ \\
    \hline
    \hline
No beam & - & 0.16 & -  \\ \hline
Beam (no collisions) & - & 0.55 & - \\ \hline
Collisions & 1.8 & 7.0 & 4.0 \\
Collisions & 1.3 & 4.8 & 3.8 \\
Collisions & 0.8 & 3.3 & 4.2 \\
Collisions & 0.6 & 2.7 & 4.3 \\
Collisions & 0.5 & 2.2 & 4.1 \\ \hline
  \end{tabular}
  \caption{Preliminary results from the TimePix detector installed in TI18, indicating that the main particle rate is proportional to luminosity in IP1. This also shows a small, but significant, increase in rate with non-colliding beam, compared to no beam in the machine. Beam (no collisions) corresponds to a full machine (2556 bunches) at the start of a physics fill, providing a total intensity of 2.7$\times 10^{14}$ protons per beam.} 
\label{tab:TimePix}
\end{table}

\subsection{Radiation Level}

The radiation level in the TI12 tunnel is an important input for determining what electronics can be operated in and close-to the detector. Both simulation studies and {\em in situ} measurements have been carried out to assess the radiation level.

\subsubsection{FLUKA simulation}
The FLUKA setup described above was also used to estimate the radiation level in the TI18 tunnel. For such locations, the radiation field is usually driven by proton showers in the dispersion suppressor, and, as mentioned above, these losses are very low for the LHC cell around FASER. Given this, radiation from showers due to beam-gas interactions in the incoming beam, where the particles from the beam-gas interaction showers can enter FASER without passing through any rock, can also be relevant. The simulations estimate a dose less than $5 \times 10^{-3}$ Gy per year and a 1 MeV neutron equivalent fluence of less than $5 \times 10^7$ per year. These numbers are roughly estimated to be comparable to or smaller than the dose in the ATLAS underground area USA15, where non-radiation-hard commercial electronics are used.

\subsubsection{{\em In situ} measurements}

Four BatMon battery operated radiation monitoring devices were installed in the TI18 tunnel by the EN-SMM CERN group during LHC Technical Stop 1. Two were tuned to measure the high-energy hadron flux, and two to measure the thermal neutron level. Given the expected low radiation in TI18, they were deliberately installed at the entrance to TI18 from the LHC, so closer to the LHC than any FASER electronics would be, as otherwise they may not have been able to make meaningful measurements. This means the measurements should represent a conservative estimate of the radiation field that FASER electronics will be exposed to. The devices were read out after 3~fb$^{-1}$ of 13~TeV $pp$ collision data was delivered to IP1.  The readings show that the high-energy hadron fluence is below the device sensitivity (corresponding to $10^6$~cm$^{-2}$), completely consistent with the expectation from the FLUKA simulation studies. For thermal neutrons the measured flux is $4\times10^6$~cm$^{-2}$, to be compared with the simulation estimate of $3\times10^6$~cm$^{-2}$. In general, the measured radiation level is consistent with the simulations, and it is generally low and dominated by beam-gas interactions in beam-2 (the incoming beam for TI18). The BatMon detectors have now been installed in TI12, but the detector data has not been analyzed yet.

In summary, in general we expect to be able to use non-radiation-hard electronics in FASER without problems.

\subsection{Other Environmental Factors}

A temperature and humidity sensor (of type TAND TR-72wf) was installed in the TI12 tunnel on the LOS as part of the emulsion detector installed there to measure the charged particle flux. The measured temperature and humidity as functions of time are shown in \figref{TI12-tem-humidity}. During this period the temperature in TI12 was constant at 18$^\circ$C, whereas the humidity varied between 40\% and 60\%, with a value around 55\% for most of the time. The figure also shows the variation over a longer timescale of about a year, but using the LHC environmental monitoring system, with the reading for the sensors closest to TI12. The temperature in the tunnel is very stable also over the longer time scale.

Vibrations and ground movements in the LHC tunnel are carefully monitored, as even small movements can cause the beam to be lost. For FASER, movement can effect the detector alignment. Studies by LHC experts~\cite{LHCvibrations} show that, for wavelengths up to 500~m, the tolerable range of seismic wave amplitudes is in the range of a few micrometers or less, with large uncertainties due to the exact wavelength or frequency content of the earthquake and due to its orientation relative to the LHC ring plane. For wavelengths above 500~m, wave amplitudes of a few tens of micrometers are acceptable and should not provoke beam aborts. This suggests that any ground motion would be a problem for the LHC before it becomes problematic for FASER.

\begin{figure}[btp]
\centering
\includegraphics[width=15cm,height=5cm]{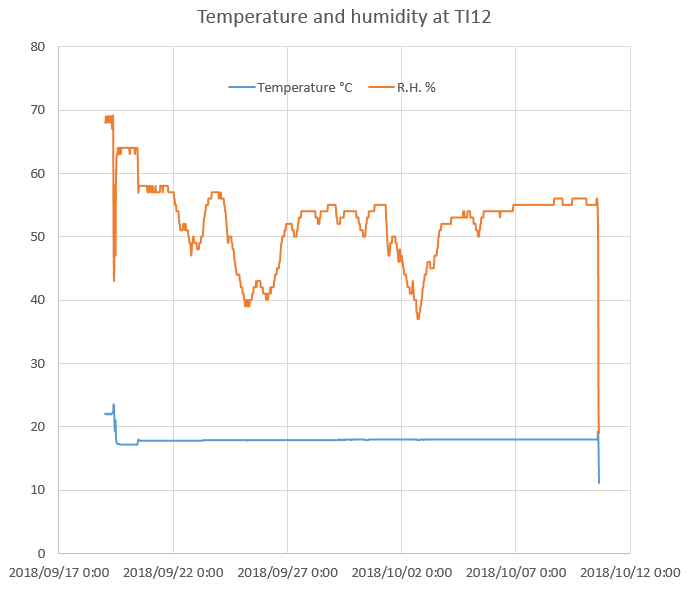} 
\includegraphics[width=15cm,height=5cm]{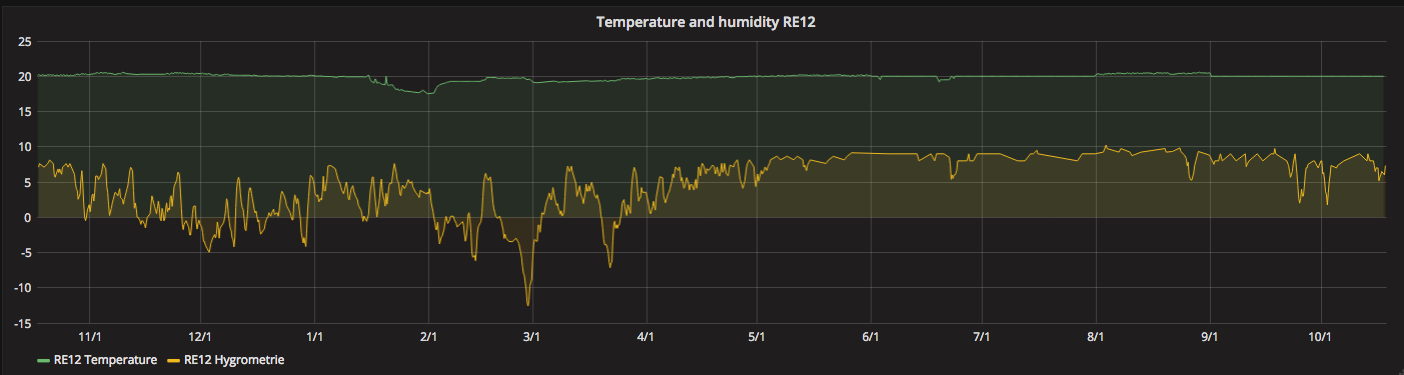} 
\caption{Measured temperature and humidity. Top: during the period when the emulsion detector was installed on the LOS in TI12 (between mid-September and mid-October).  Bottom: in the LHC tunnel close to TI12 (in the UJ12 region) from LHC monitoring, over about 1 year. For the top plot the humidity is shown, whereas for the bottom it is the dew point of the air.}
\label{fig:TI12-tem-humidity}
\end{figure}

\clearpage
\section{Magnets} 
\label{sec:magnet}

\subsection{Requirements and Magnet Design}

The main requirements on the magnets needed for FASER are:
\begin{itemize}
\item The dipole field should be large enough to sufficiently separate pairs of oppositely-charged, high-energy particles originating from a high-energy, low-mass particle decay inside the detector decay volume;
\item The magnet should be sufficiently thin to be able to fit into the tunnel (after lowering the floor by up to 46~cm) to allow a 5~m-long detector centered on the LOS (this suggests a magnet thickness of 15~cm would be best);
\item The services needed for the magnet should be minimized (for example, minimizing power and cooling requirements);
\item The stray field must be small enough that the PMTs used for the scintillator and calorimeter readout can operate correctly;
\item The magnet dimensions and weight must be compatible with transport to the TI12 location (including being lifted over the LHC machine).
\end{itemize}
The above requirements suggest that a permanent magnet with a field of $\approx$0.5~T would be a good solution. For such a magnet there are no services, and the magnet can be thin.
A conceptual design of such a magnet, based on a design made for the NTOF experiment, has been made by the CERN warm magnet section. This is based on a Halbach array of permanent magnets arranged to produce a dipole field. Since the radiation level in TI12 is very low, the NTOF design can be slightly improved for FASER by replacing the magnetic material of SmCo with NdFeB, which allows a slightly higher field of 0.6~T to be achieved.  The magnet design is shown in \figref{magnetDesign}, and Table~\ref{tab:magParams} shows the main parameters.
\begin{figure}[bp]
\centering
\includegraphics[width=0.57\textwidth]{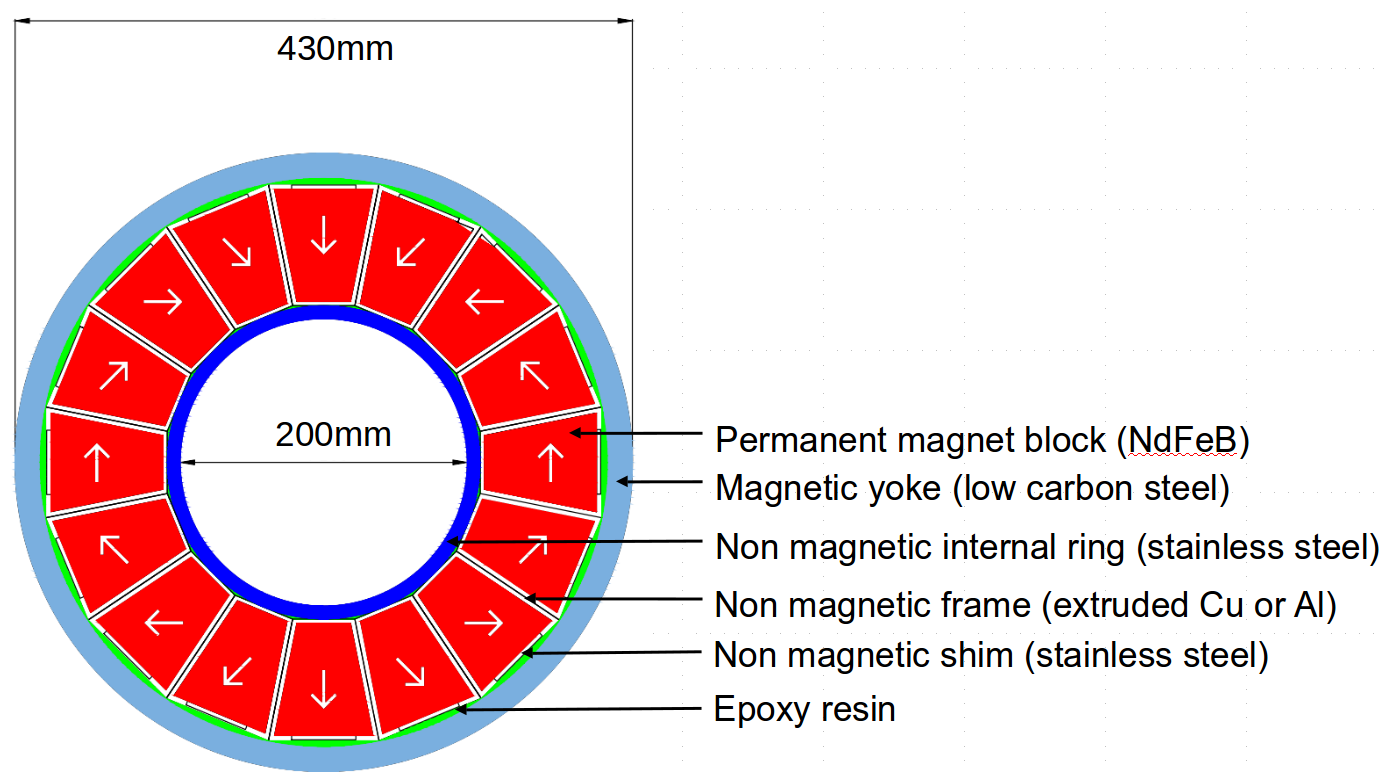} \qquad
\includegraphics[width=0.30\textwidth]{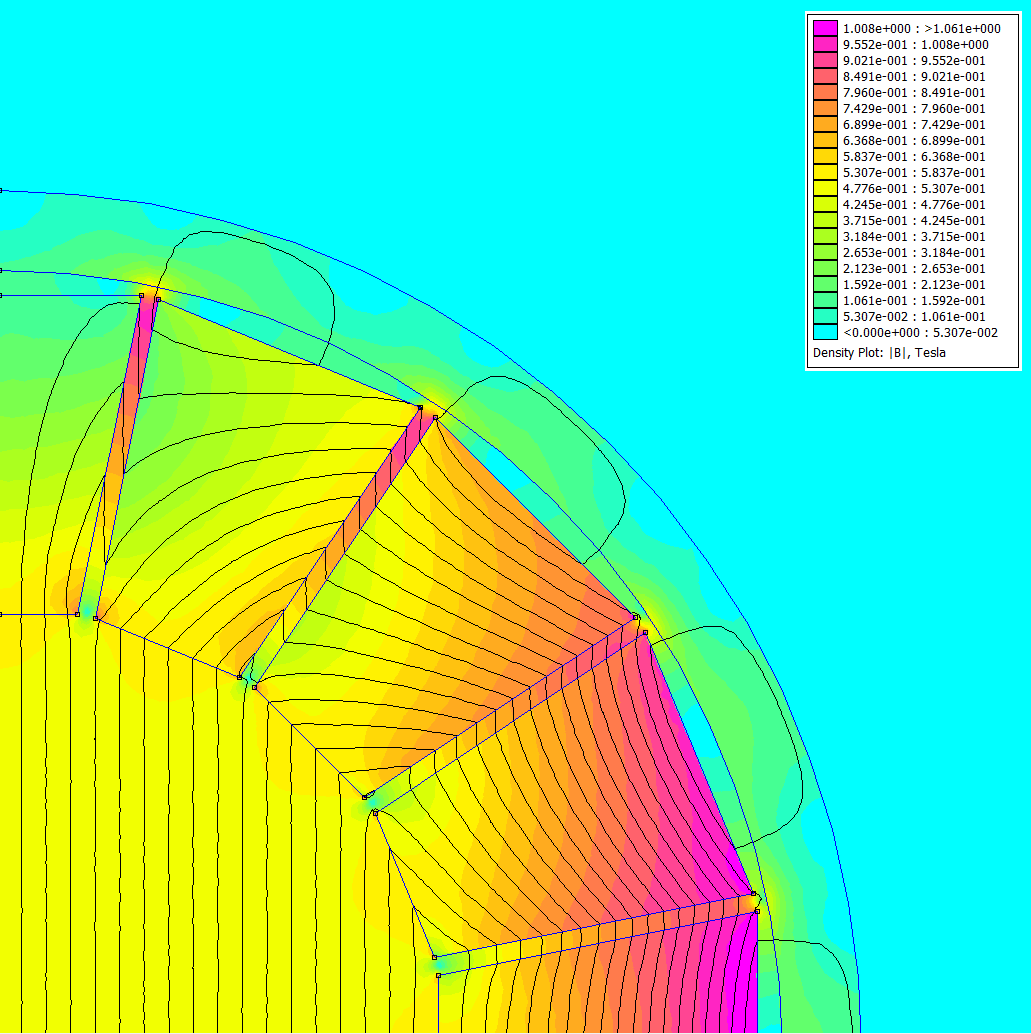}
\caption{Left: a sketch of the magnet design. Right: the field lines in the magnet. 
}
\label{fig:magnetDesign}
\end{figure}

\begin{table}
  \centering
  \begin{tabular}{|l|c|c|}
  \hline
	{\bf Parameter} & {\bf Value} & {\bf Unit} \\
    \hline
Magnetic material & NdFeB& \\ \hline
Central field & 0.6 & T \\ \hline
Aperture & 200 & mm \\ \hline
Outer diameter & 430 & mm \\ \hline
Field homogeneity & $\pm$2 & \% \\ \hline
Temperature dependence & -0.12  & \%/K \\ \hline
Weight/length & $\approx$1000  & kg/m \\ \hline
  \end{tabular}
  \caption{Main magnet parameters for the FASER magnet design.
}  \label{tab:magParams}
\end{table}

The stray field outside the magnet openings can be quite large, as is demonstrated in \figref{strayField}. It maybe that shielding is needed to reduce the field sufficiently to be able to operate the PMTs in the detector. Because of the field between the magnets, the magnets should be separated by at least 20~cm, and no magnetic material should be placed between them.
\begin{figure}[t]
\centering
\includegraphics[width=0.48\textwidth]{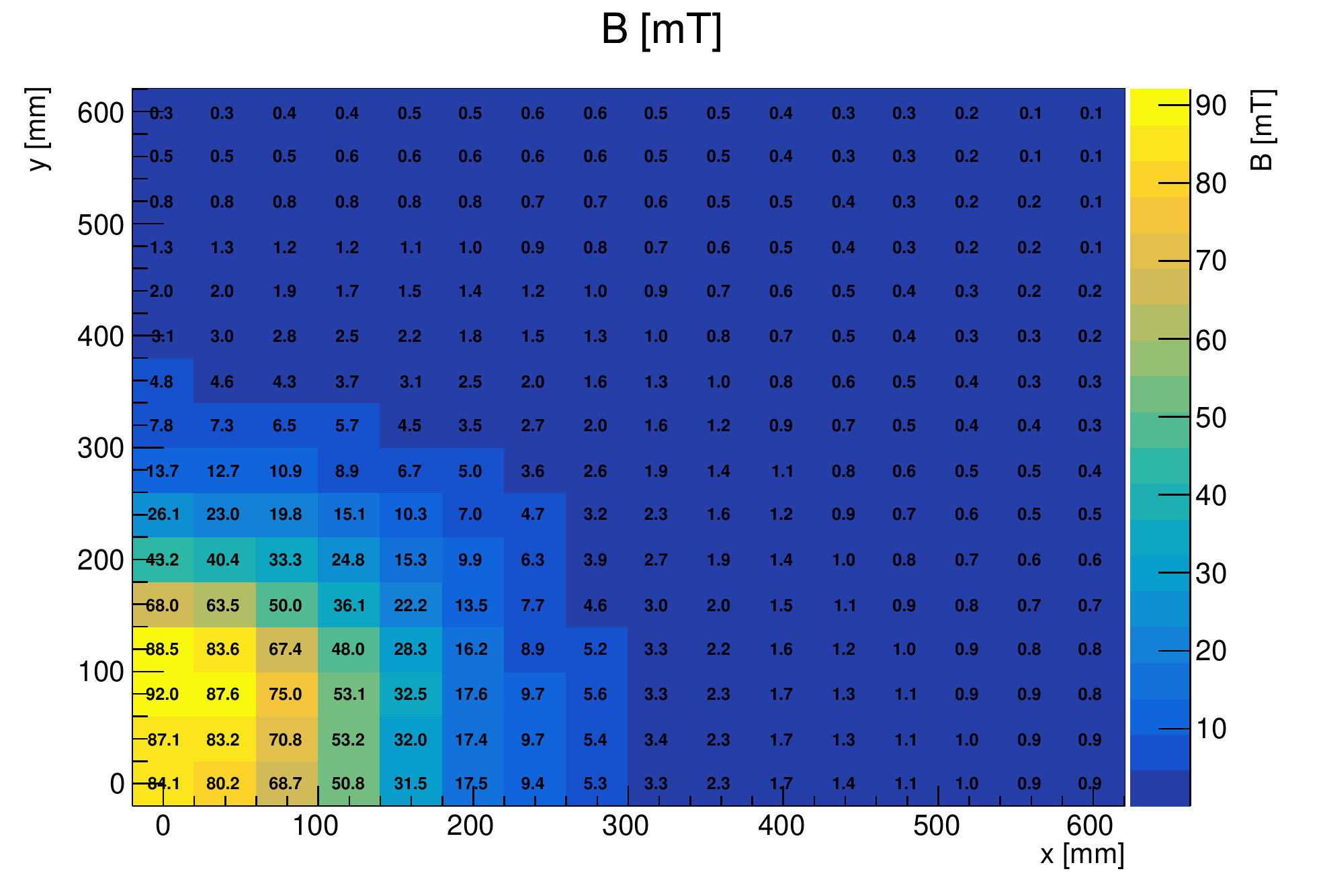} 
\hfill
\includegraphics[width=0.48\textwidth]{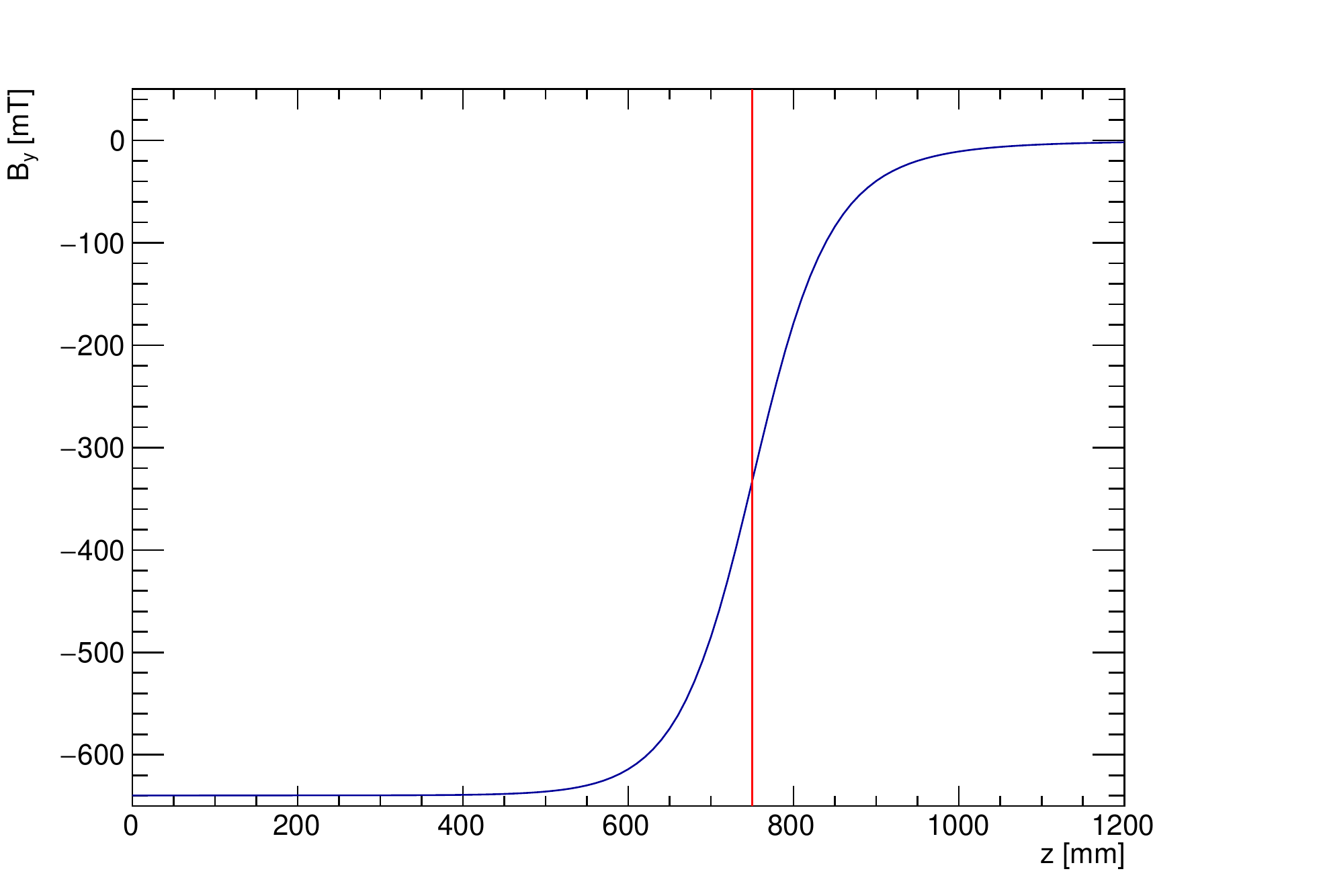} 
\caption{Left) The magnetic field (in mT) outside of the magnet estimated from the magnet design model 10~cm from the magnet end. Right) The vertical field along the center of the magnet and beyond; the red line indicates the end of the magnet. The closest scintillator PMT is located 35cm from center.
}
\label{fig:strayField}
\end{figure}

\subsection{Cost and Schedule}
The cost of producing the FASER magnets has been estimated by the CERN magnet group to be 420~kCHF, including 100~kCHF for personnel costs. The cost breakdown is outlined in \tableref{magnet-budget}.

\begin{center}
\begin{table}[tbp]
\centering
\resizebox{0.75\textwidth}{!}{%
\begin{tabular}{|l|c|}
\hline
\  {\bf 1.5m Magnet Component} \ &  \ {\bf Cost [kCHF]} \  \\ \hline
\ Permanent Magnet Blocks \ &  90 \quad  \\ \hline
\ Machining of Magnetic and Non-Magnetic Parts \ & 30 \quad  \\ \hline
\ Magnet Assembly \ & 10 \quad  \\ \hline
\  {\bf 1.5m Magnet Total (1 Unit)} \ & {\bf 130}  \quad  \\ \hline 
& \\ \hline
\  {\bf 1.0m Magnet Component} \ &  {\bf Cost [kCHF]} \\ \hline
\ Permanent Magnet Blocks \ & 60 \quad  \\ \hline
\ Machining of Magnetic and Non-Magnetic Parts \ & 25 \quad  \\ \hline
\ Magnet Assembly \ & 10 \quad  \\ \hline
\  {\bf 1.0m Magnet Total (1 Unit)} \ &  {\bf 95} \quad  \\ \hline 
& \\ \hline 
\  {\bf Magnet Engineer} \ &  {\bf 100} \quad  \\ \hline 
& \\ \hline 
\ {\bf Total (1 1.5m Magnet $+$ 2 1.0m Magnets $+$ Engineer)} \ & {\bf 420}  \quad   \\ \hline
\end{tabular}
}
\caption{Budget for magnet construction for the FASER experiment.}
\label{table:magnet-budget}
\end{table}
\end{center}

The timeline for producing the FASER magnets has been estimated by the CERN magnet group to be 15~months, with the breakdown outlined in \figref{magnet-timeline}.
\begin{figure}[tbp]
\centering
\includegraphics[clip, trim=0.5cm 10.8cm 0.5cm 0.5cm,width=0.98\textwidth]{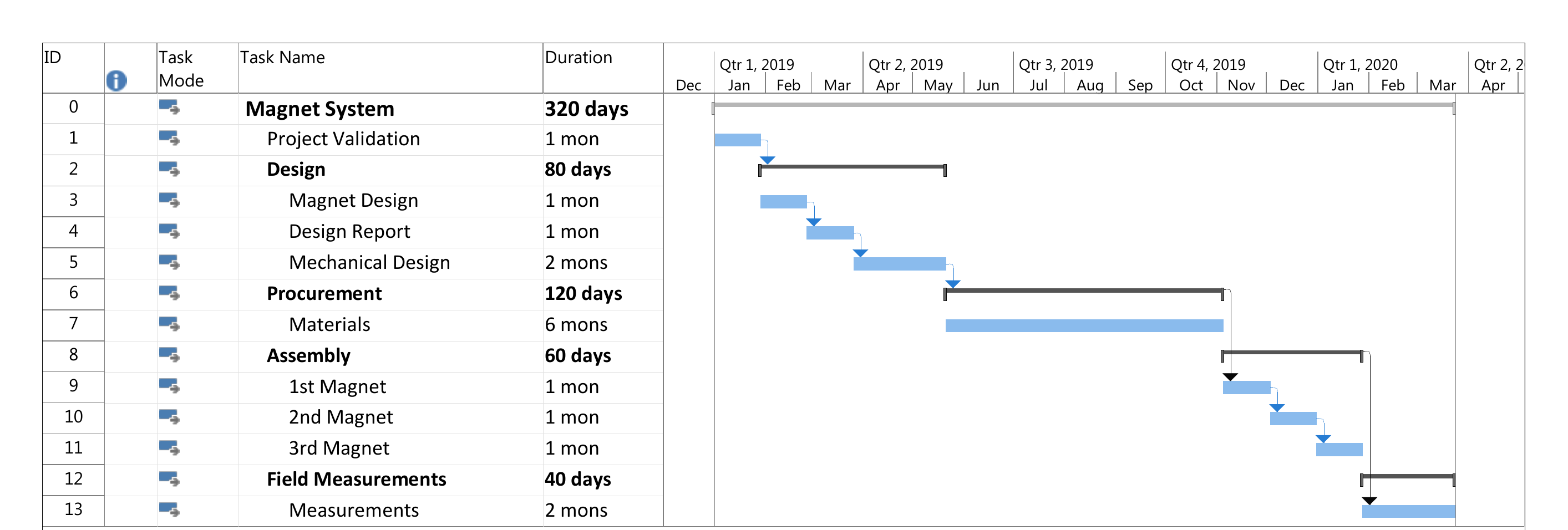}
\caption{
Timeline for magnet construction for the FASER experiment.
}
\label{fig:magnet-timeline}
\end{figure}

\clearpage
\section{Tracker}
\label{sec:tracker}

\subsection{Overview}

The FASER tracker's purpose is to reconstruct the trajectories of energetic charged particles from the direction of ATLAS. It must have sufficient hit resolution to efficiently and reliably identify the signature of two closely-spaced, oppositely-charged tracks from decay of a very high-energy, low-mass particle, where the tracks are separated by the detector's magnetic field. The Letter of Intent~\cite{Ariga:2018zuc} showed that requiring tracks separated by at least $300 \, \mu\text{m}$ results in good physics performance. 

Given the detector length and magnetic field that can be practically achieved, the high momentum of signal particles, and the uncertainty in the relative alignment of tracking stations in the magnetic bending plane, precise momentum measurement is not expected, and is not required by FASER's searches for new physics. Nevertheless, the detector is designed to achieve the best momentum resolution possible, which will be important for understanding backgrounds and for any auxiliary physics measurements (for example, neutrino studies with FASER). 

For cost and schedule reasons, the tracker will be constructed from spare modules of the ATLAS semiconductor strip tracker (SCT)~\cite{ATLAS:1997ag, ATLAS:1997af}. The SCT Institute Board has kindly agreed to allow FASER to use 80 of the available spare barrel modules~\cite{Abdesselam:2006wt, Abdesselam:2007ec}. \Figref{SCT} shows a barrel module with 6 on-detector ASICs per side, which are integrated into the module. These ``ABCD'' ASICs~\cite{Campabadal:2005rj} are the first stage of the detector readout and control the module's hit threshold and overall configuration. The SCT modules have proven their reliability in ATLAS during LHC Runs 1 and 2, operating without problems and delivering excellent physics performance. 

The basic tracker design is three tracking stations, located at the front, center and back of the spectrometer, where each station contains three planes of SCT modules. An SCT module comprises two pairs of silicon strip detectors glued back-to-back to a central thermal pyrolytic graphite (TPG) baseboard. The modules will be oriented to optimize measurement precision in the magnetic bending plane; a stereo angle of $\pm 20 \, \hbox{mrad}$ between the two sides allows track position measurements in the non-precision direction as well. The strip pitch is 80~$\mu$m allowing track position resolution in the precision coordinate of order 17~$\mu$m per plane (assuming 4 SCT planes, and perfect detector alignment). The resolution in the non-precision coordinate is about 580~$\mu$m.~\cite{Aad:2014mta} 

The number of tracking planes per station is chosen to give high efficiency and sufficient redundancy, while limiting the amount of material inside the tracking volume. To simplify the system, all three tracking stations have an identical design. The design is also matched to the number of available SCT spare modules. Each SCT module is 6~cm $\times$ 12~cm and the tracking plane is made of 8 modules, giving a square of about 24~cm $\times$ 24~cm covering the full active area of the detector (a 10~cm-radius circle determined by the magnet aperture). Each SCT module has 768 readout channels per side, giving the full tracker system a total of 111 $\times 10^3$ channels.

With the low radiation levels in the TI12 tunnel, the silicon itself does not require cooling to give acceptable efficiency and noise, but the on-detector ASIC chips (12 chips per module) must be cooled. Based on experience during commissioning of the ATLAS SCT, the silicon should be operated at a temperature below 30$^\circ$C. A water cooling system with inlet water temperature of 5 to 10$^\circ$C is sufficient to achieve this, but the tracking station will need to be enclosed in dry air with a dew point low enough to prevent condensation.

The low trigger rate (expected to be $\sim 600 \, \hbox{Hz}$, as described in \secref{tdaq}), the very small occupancy of the tracking detector (generally a few hits per plane from a single muon traversing the detector, along with a small number of noise hits), and the low radiation level all combine to greatly reduce the challenges of reading out SCT modules in FASER, compared to ATLAS. A simpler readout architecture than ATLAS, based on shipping the data to a custom made board via flex cable, has therefore been chosen. The readout board contains a single FPGA, which carries out simple data processing and error handling before sending the data to an event builder process running on the surface. Whether one readout board per tracker plane (nine total) or per tracking station (three total) will be required is still under study. Either way, the readout system consists of only a small number of boards; these need to be situated about 1~m from the detector to minimize data cable lengths, while remaining out of the stray magnetic field. 

The readout system must also collect calibration data from the SCT modules during beam-off periods, when all strips are read to measure noise and efficiency and to tune thresholds. Preliminary calculations show that the proposed architecture can complete the most bandwidth-intensive calibration for the full detector in short downtimes ($\approx$ 30 minutes). 

\begin{figure}[tbp]
\centering
\includegraphics[width=0.6\textwidth]{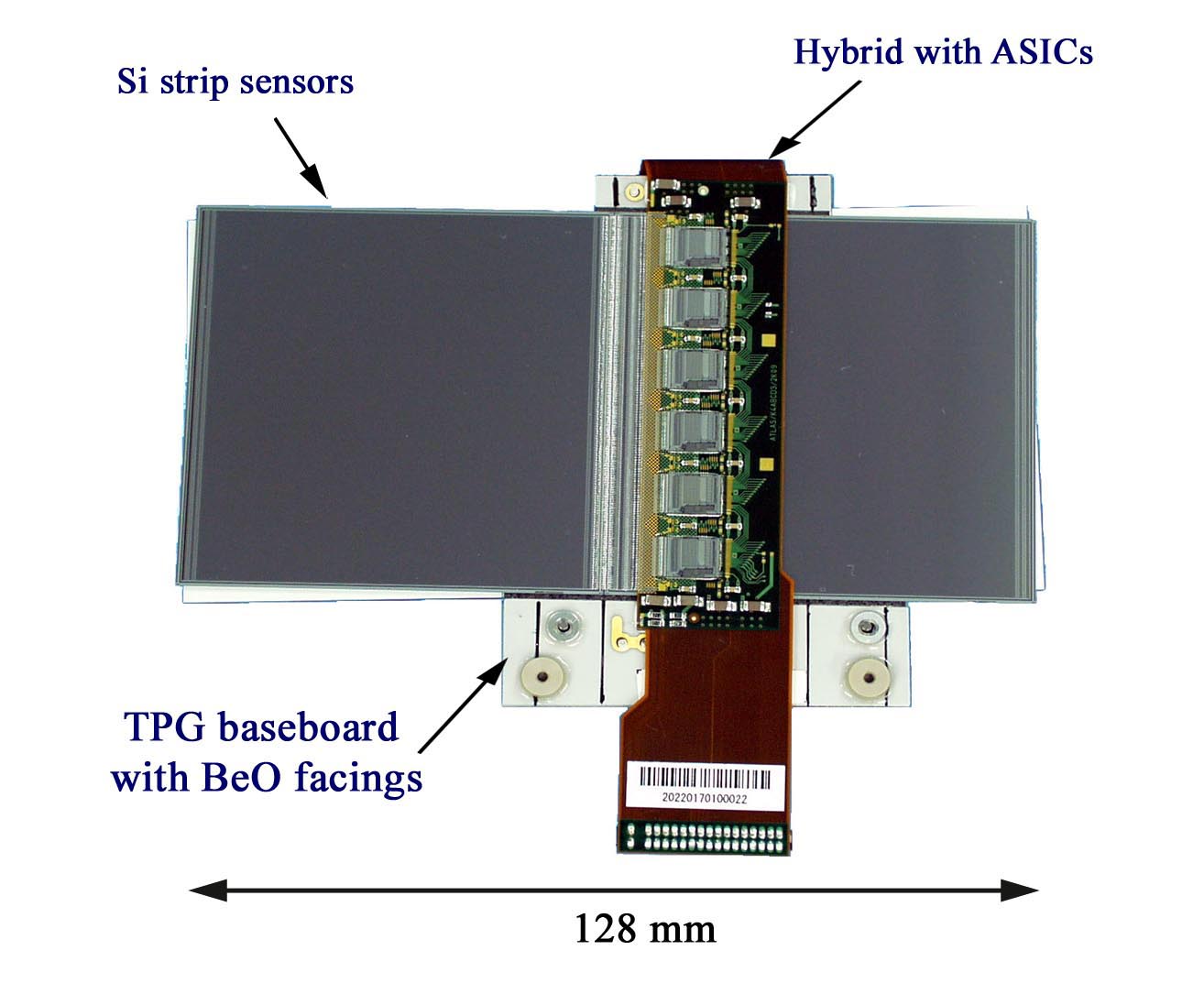}
\caption{
Illustration of the ATLAS SCT barrel module to be used in FASER.
}
\label{fig:SCT}
\end{figure}

\subsection{Module Selection and Quality Assurance}
\label{sec:QA}

SCT modules for the FASER tracker will be selected based on a series of Quality Assurance (QA) tests. For module QA, a so-called Chimaera digital board attached with a Tengja trigger card will be used (see \figref{cambridge_board}).  These were originally developed for a test system of the LHCb RICH upgrade at Cambridge University and were later used for cosmic muon tomography with eight spare SCT modules~\cite{Keizer:2018nju}. 

The Chimaera digital board is FPGA-based. It sends commands and reads data to/from the module, operated by a PC connected via Ethernet. The Tengja card is an interface between the digital board and an adapter card for the SCT module. The trigger board is directly connected to the digital board via an AUX connector and provides the trigger signal to issue a calibration pulse to the ABCD readout ASICs on the SCT module.

\begin{figure}[tbp]
\centering
\includegraphics[width=0.6\linewidth]{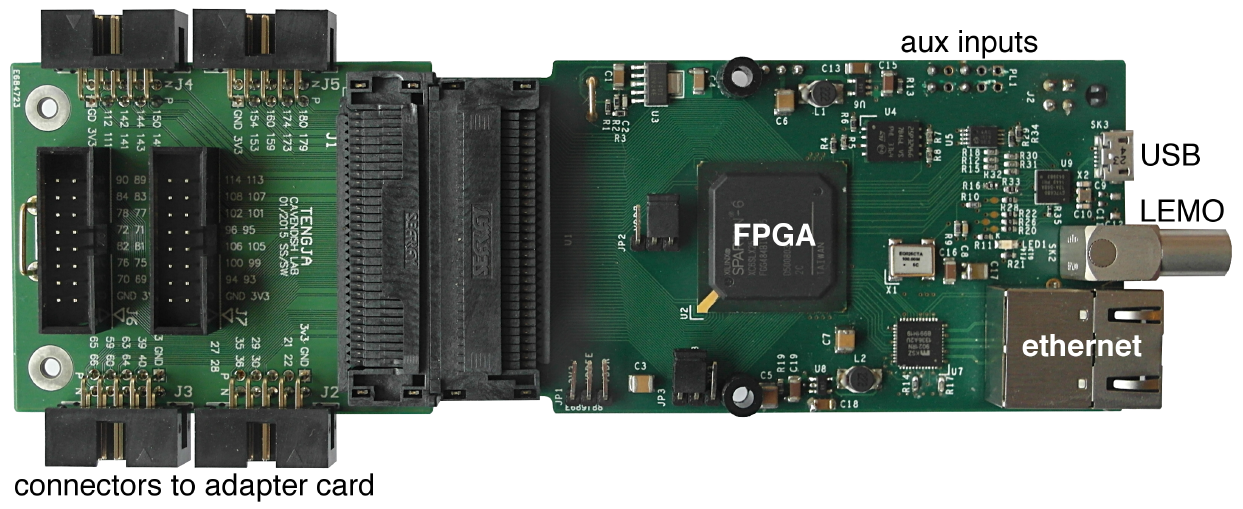}
\caption{Chimaera digital board (right) with the Tengja card (left).}
\label{fig:cambridge_board}
\end{figure}

The adapter card connects to the SCT module via a 36-pin connector and provides electrical lines for the clock, command, data, thermistor readout and several voltages to the SCT module. A low-voltage supply is connected with the adapter card to power the ABCD chips, while high voltage for the sensors is supplied via a LEMO connector on the card. 

The digital board and trigger board are operated via a GUI-based operating software. The user can set the scan parameters via the GUI window.

The SCT module will be operated at room temperature during the QA and placed on a cooling jig connected with a chiller. The temperature on the module is monitored by using a Negative Coefficient Thermistor (NTC) on the module.

During the QA, the performance of the modules will be checked by a set of tests that includes scans of threshold and trigger latency for qualification of the ABCD chips, and an HV scan for qualification of the sensors themselves. Modules will be ranked according to the test results and those with the highest scores selected for use in the FASER tracker.

The QA setup was integrated at CERN as shown in Fig. \ref{fig:qa_setup}, and it is possible to communicate with the SCT module and read data. The operation software is being optimized, and a list of the tests is under preparation. The full QA will start at the beginning of 2019.

\begin{figure}[bp]
\centering
\includegraphics[width=0.45\linewidth]{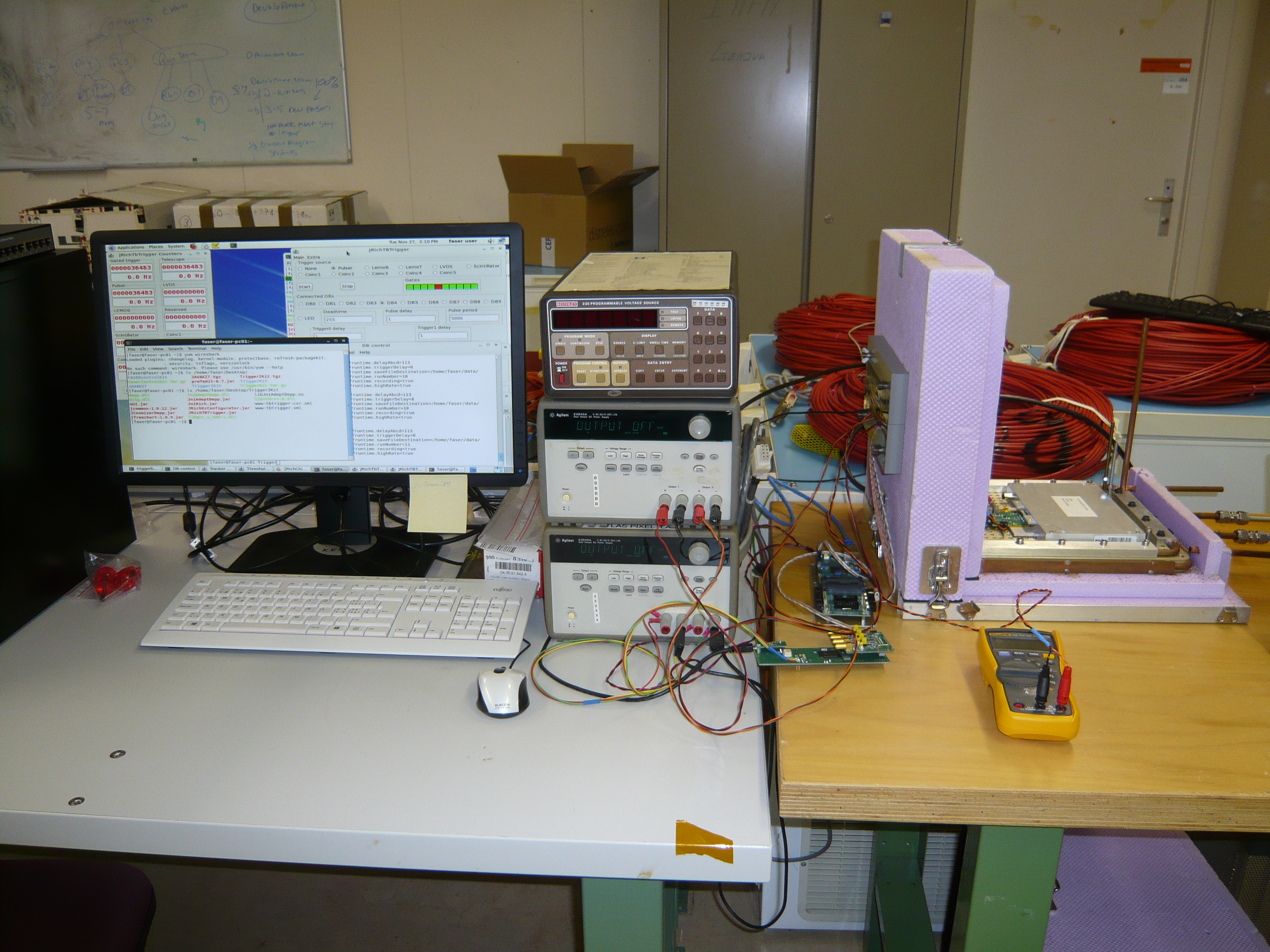}
\includegraphics[width=0.45\linewidth]{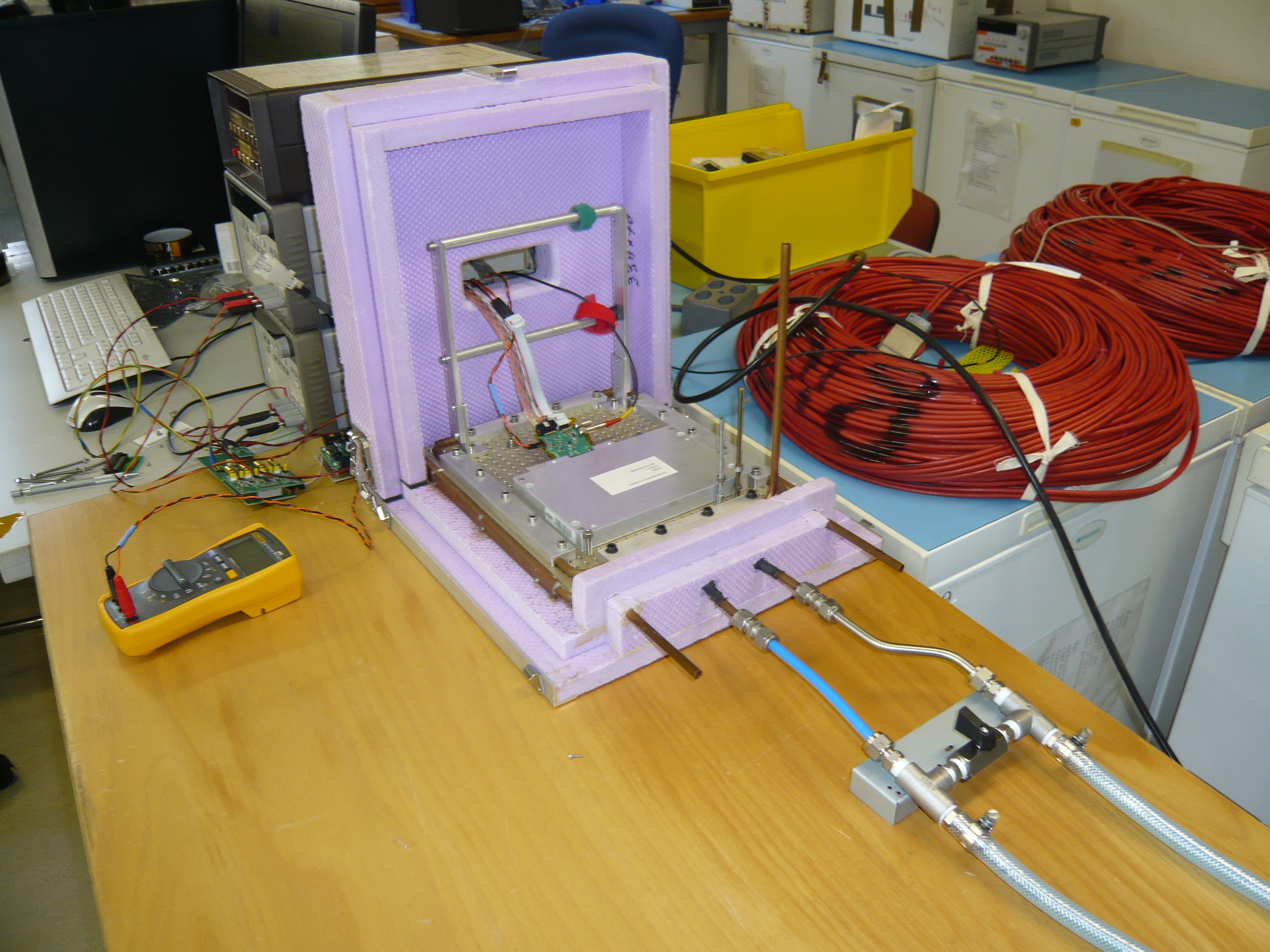}
\caption{Module QA setup at CERN.}
\label{fig:qa_setup}
\end{figure}

\subsection{Mechanics}

The tracker consists of three stations deployed between magnet sections. A station comprises three planes of eight SCT modules, arranged over two rows as shown in \figref{layer_schema}. The tracker as a whole thus uses 72 of the 80 SCT modules provided by ATLAS, with the remaining 8 available as spares.

\begin{figure}[tbp]
\centering
\includegraphics[width=0.8\linewidth]{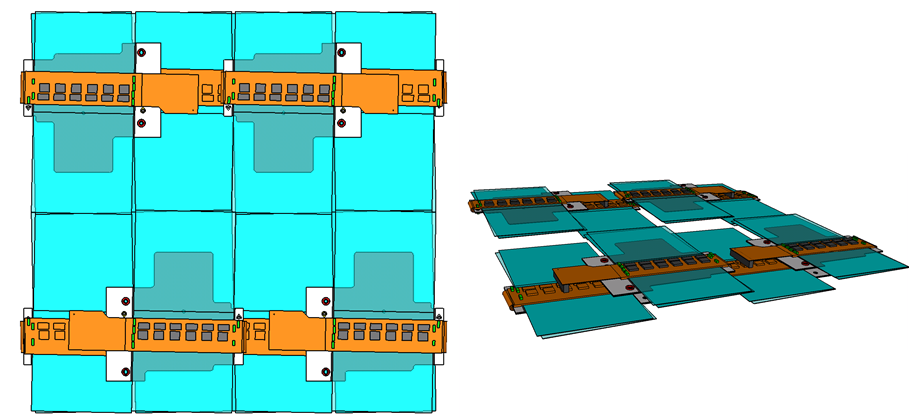}
\caption{Left: A full FASER tracker plane, with two rows of four SCT modules. Right: To allow the edges to overlap, modules are staggered in and out of the plane by several millimeters.}
\label{fig:layer_schema}
\end{figure}

\subsubsection{SCT modules}

Mechanical assembly of tracker modules into planes re-uses most of the existing parts of an SCT module (\figsref{module_schema}{module_pic}). The aim is to retain the module's primary assembly (two tilted silicon planes + TPG backbone + hybrid) apart from the cooling pipe. Attachment will be made through the three existing kinematics mounts (shown in \figref{module_schema}) so that no major modifications to the SCT modules themselves are necessary, saving time.

\begin{figure}[tbp]
\begin{centering}
\includegraphics[width=0.8\linewidth]{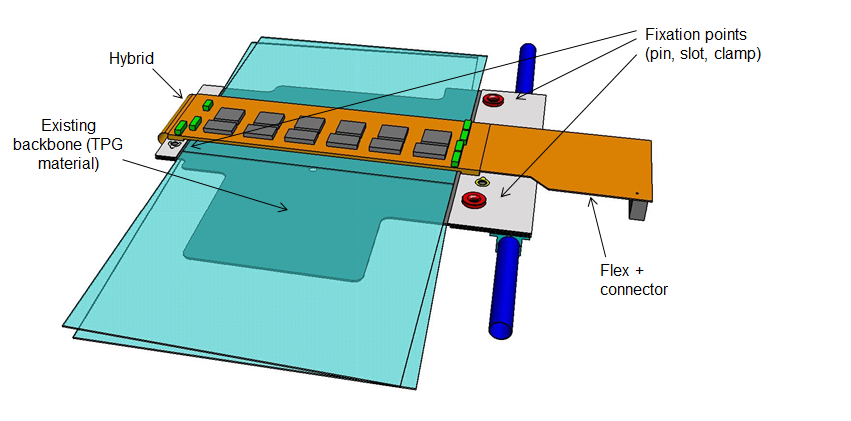}
\caption{SCT module with its cooling pipe (dark blue).}
\label{fig:module_schema}
\end{centering}
\end{figure}

\begin{figure}[tbp]
\centering
\includegraphics[width=0.8\linewidth]{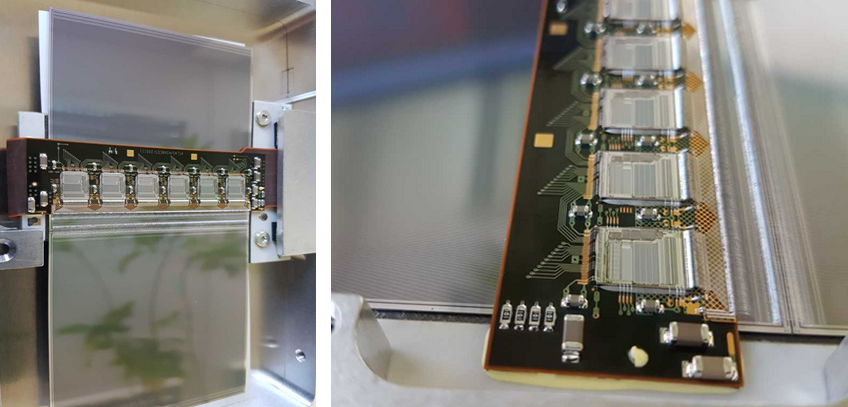}
\caption{An SCT module in its holding frame and close-up of the hybrid part.}
\label{fig:module_pic}
\end{figure}

Overlap between adjacent modules (a minimum of five silicon strips) is optimized to guarantee full geometric efficiency for near-normal tracks over the entire area of a plane. The arrangement of the two rows is chosen to minimize the material in the central region (no hybrids, connectors, TPG…). This mechanical design also allows replacement of any individual module of a plane, if necessary.

\subsubsection{Module frame}
Tracker planes are mounted in a module frame (see \figref{layer_schema2}) that serves several purposes:

\begin{itemize}
\item Provide stable attachments for the kinematic mounts of each of the  eight modules in a plane,
\item Provide attachment and alignment within 30 microns for each of the three planes within a station,
\item Provide the main thermal path for module cooling.
\end{itemize}

Aluminum is a candidate material for the module frame.

\begin{figure}[htbp]
\centering
\includegraphics[width=0.8\linewidth]{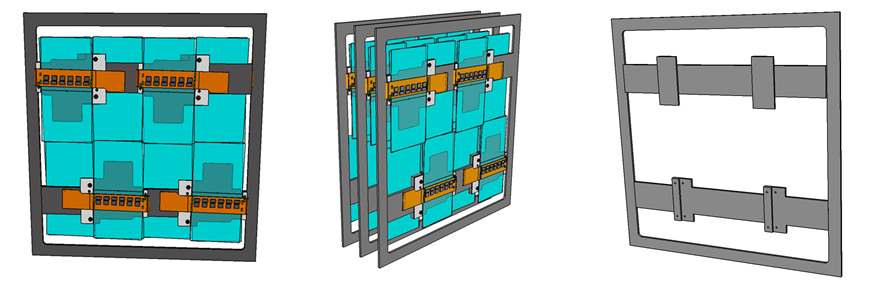}
\caption{Left: Module frame with a tracker plane mounted inside. Center: Three module frames assembled into a tracking station. Right: The module frame itself (holes for alignment and attachment are not shown).}
\label{fig:layer_schema2}
\end{figure}

The spacer of each module frame (\figref{sct_station}) determines the separation between planes and allows connections for cooling. It is also the main interface to the outer box that allows each station to be aligned with respect to the others and isolated from the tunnel environment. All service connections (power, data, cooling water, dry air flushing) are located on the same side of a station due to tunnel access constraints.

\begin{figure}[tbhp]
\centering
\includegraphics[width=0.8\linewidth]{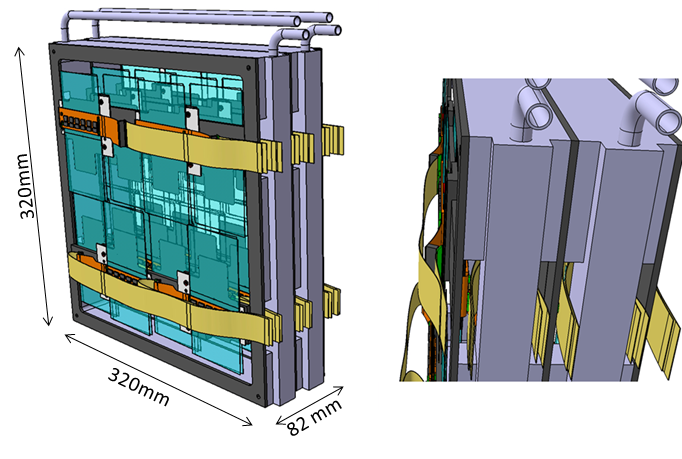}
\caption{A tracking station with module frame spacers, showing cooling and electronic connectors.}
\label{fig:sct_station}
\end{figure}

\subsection{Cooling and Humidity Control}
\label{sec:cooling}

Each tracking station is housed in an outer box (\figsref{sct_station2}{sct_station3}), which provides:
\begin{itemize}
\item An attachment for the enclosed planes,
\item A light-tight and air-tight environment for the modules inside, to isolate them from the detector tunnel, 
\item An interface for services (cables, water pipes, dry air pipes),
\item An interface to the FASER support system (with an adjustable mechanism for alignment).
\end{itemize}

The outer box has a patch panel for connectors and fittings. The material will be aluminum to enhance EMI performance by providing a Faraday cage of sorts. It also has fiducial points to establish the position of the station as accurately as possible via survey.

\begin{figure}[tbhp]
\centering
\includegraphics[width=0.8\linewidth]{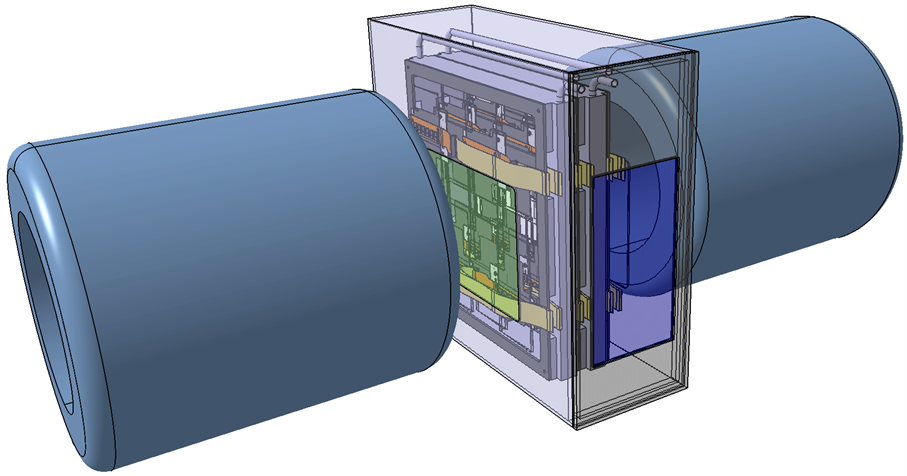}
\caption{A tracking station in its outer box between two magnet segments.  Nearby scintillators are not shown.}
\label{fig:sct_station2}
\end{figure}

\begin{figure}[htbp]
\centering
\includegraphics[width=0.8\linewidth]{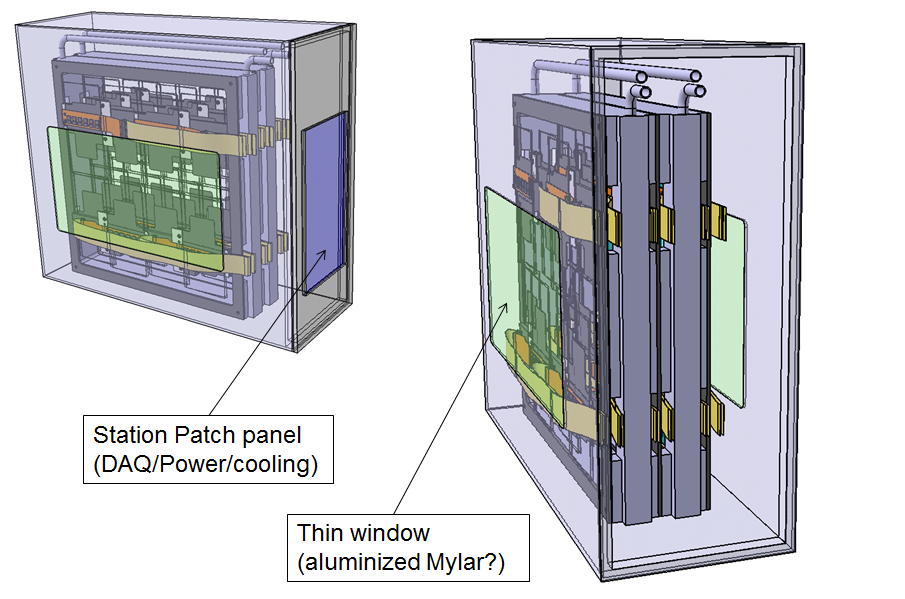}
\caption{The tracking station outer box showing the locations of the patch panel and aluminized mylar windows (to minimize material and multiple scattering).}
\label{fig:sct_station3}
\end{figure}

The tracking stations will be cooled by chilled water at an inlet temperature between 5 and 10$^\circ$C. To provide this chilled water, the CERN cooling and ventilation group (EN-CV) recommends a small chiller installed close to the experiment in the TI12 tunnel.  An SMC chiller (model HRS030-AN-020 or equivalent) with a 5-liter water tank can provide cooling capacity up to 1500~W (\figsref{Chiller1}{Chiller2}) at 25$^\circ$C ambient temperature. The total cooling capacity required for the three tracking stations is about 600~W. The chiller would be controlled and monitored using an RS485-to-Ethernet converter (\figref{Chiller3}). Some insulation (Armaflex) will be required along the water lines to the tracking stations (including the patch panels for the stations). The chiller will be placed in the tunnel at least 30~cm from any obstacle, to allow free air ventilation, but limiting the length of the cooling lines as much as possible.

\begin{figure}[tbp]
\centering
\includegraphics[width=0.80\linewidth]{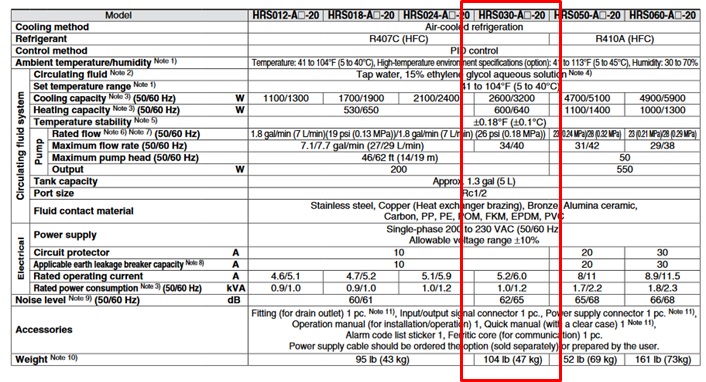}
\caption{Chiller data sheet.}
\label{fig:Chiller1}
\end{figure}

\begin{figure}[tbp]
\centering
\includegraphics[width=0.85\linewidth]{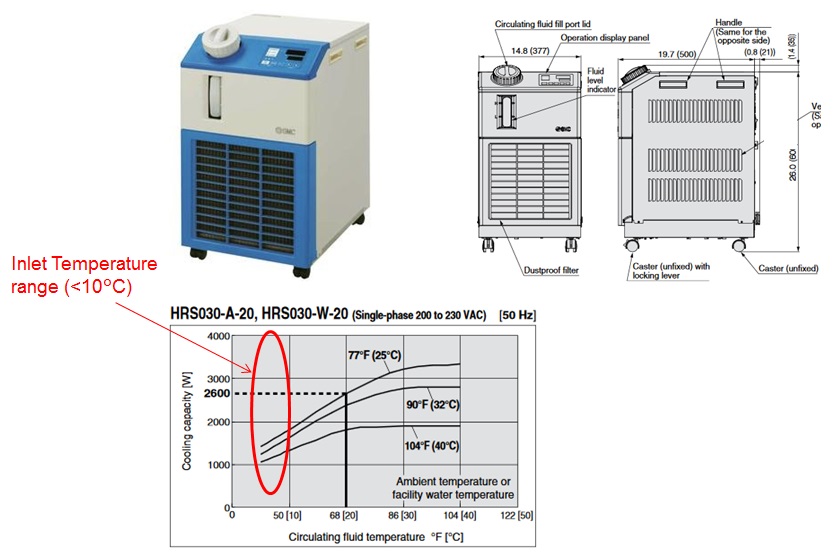}
\caption{HRS030 with cooling capacity and dimensions.}
\label{fig:Chiller2}
\end{figure}

Humidity inside the tracker stations will be controlled using compressed air provided by the EN-CV group, which can be installed during LS2.  This air has a guaranteed dew point below $-40^\circ$C and thus meets FASER's requirements perfectly.

\begin{figure}[h!]
\centering
\includegraphics[width=0.9\linewidth]{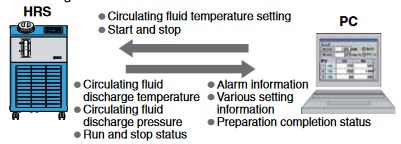}
\caption{HRS030 Monitoring via PC.}
\label{fig:Chiller3}
\end{figure}

\subsection{Power, Control and Interlocks}

Each SCT module requires three separate power lines. For a non-irradiated module the requirements are typically 900~mA at 3.5~V for the analogue low voltage, 400~mA at 4~V for the digital low voltage, and 150~V reverse bias voltage with less than 0.5~$\mu$A leakage current. Several powering schemes are under consideration; the most conservative (baseline) design would provide dedicated power supply channels for each module, i.e., one 3.5 V line, one 4 V line, and one 150 V channel per module. Having separate powering for each module minimizes the impact on the tracker performance of a power supply channel or module failure. It also allows for individualized monitoring and tuning of each module for best performance and reliability.

This powering system can be implemented using eighteen CAEN A2517A 8-channel floating low voltage power supply modules and three A1542 24-channel 500~V floating high voltage boards. These can be housed in two SY4527 power supply crates and controlled via Ethernet. The PS modules should allow for individual channels to be enabled/disabled via an external input signal. Both analogue and digital low voltages are sensed (four wires) for a precise control of the supply voltages. 

One of the most problematic issues for the SCT modules is overheating.  This could occur due to a cooling failure (e.g., chiller failure, chiller power-cut, cooling circuit clogging), and therefore temperature monitoring is essential. 

The tracker's detector control system (DCS) will control and monitor the power supplies, chiller, module temperatures, and environmental sensors inside the stations.  Each SCT module has two (one per hybrid side) NTC thermistors ($R25=10\,\rm{k\Omega} \pm 1\%)$ that provide two temperature measurements. In addition, the environmental conditions will be monitored by other, redundant sensors. Each tracking station will have at least four NTCs for temperature measurements (two located at the exhaust pipes to measure the temperature of the spent cooling water and two to measure the air temperature inside the station) and two humidity sensors (e.g., Honeywell) for relative humidity measurements. The complete tracker will include 144 hybrid thermistor channels and 18 environmental sensors to be monitored. The DCS system monitors all supplied low and high voltages and return currents, hybrid temperatures, and environmental conditions. The power supply crate and chiller are controlled by the DCS software, allowing for remote operation of the power and cooling systems. All conditions data will be  regularly archived in a database. 

Ensuring safe operation of an underground experiment with very limited access is mandatory. Several safety mechanisms at both software and hardware levels will be implemented. The power supply cards contain programmable current and voltage protections, so that a given output channel will be automatically switched off if it exceeds a safe maximum  value. The DCS software will also be programmed to take automated protective actions based on monitored sensor readings (e.g., HV-channel ramped-down in case the measured leakage current goes above a certain threshold). Finally, a purely hardware-based interlock system (IS) will provide a final line of protection for the detector modules in the event that software-based protection mechanisms fail. The IS acts as an ``ultimate emergency switch,'' and as such it must be autonomous, simple, and reliable (without relying on software). The baseline is to have custom circuit boards with simple analogue circuitry perform a comparison of temperatures, voltages, and currents to fixed values. Some logic circuits (e.g., OR-ing the comparator outputs for a given module) will be implemented. If thresholds are exceeded, a logical signal (typically TTL-logic) is sent to the interlock inputs of the PS cards to trip the corresponding channel. User intervention would be required to re-enable the channel through the DCS software after diagnosing the fault.

\subsection{Readout}

The hardware readout architecture of the FASER tracker is illustrated in \figref{sct_readout_architecture}. 

Each plane is composed of 8 SCT modules and has 2 patch panel boards, one for the top and bottom row, respectively. The connection between the patch panel board and the four modules of a row is made with a module flex circuit having one flex for the two modules with the connectors on the front and another for the two modules with connectors on the back. The patch panel board merges signals and preserves the light-tightness of the tracking station, having only one connector on the external face of the station. The SCT modules will be controlled and read out using a ``UniGe USB3 GPIO'' board described in \Secref{UniGeGPIO}. This board requires an additional interface board (the ``FASER station GPIO'' board) which will be designed to allow communication with the SCT modules. The architecture of this interface board is relatively simple: it is composed of LDVS drivers and receivers for the clock, command and two data signals of each module connected (eight per plane) and is connected to the two patch panel boards per plane by cables. 

\Figref{sct_readout_architecture} presents an optimized architecture with one ``Unige USB3 GPIO'' board per station, each driving three planes. If the firmware cannot handle three planes but only one, then nine ``FASER station GPIO'' boards and nine ``Unige USB3 GPIO'' boards will be required in the readout box for the overall experiment. The detailed functionality of the readout boards is described in \Secref{trackerReadout}.

\begin{figure}[tbp]
\includegraphics[width=0.95\linewidth]{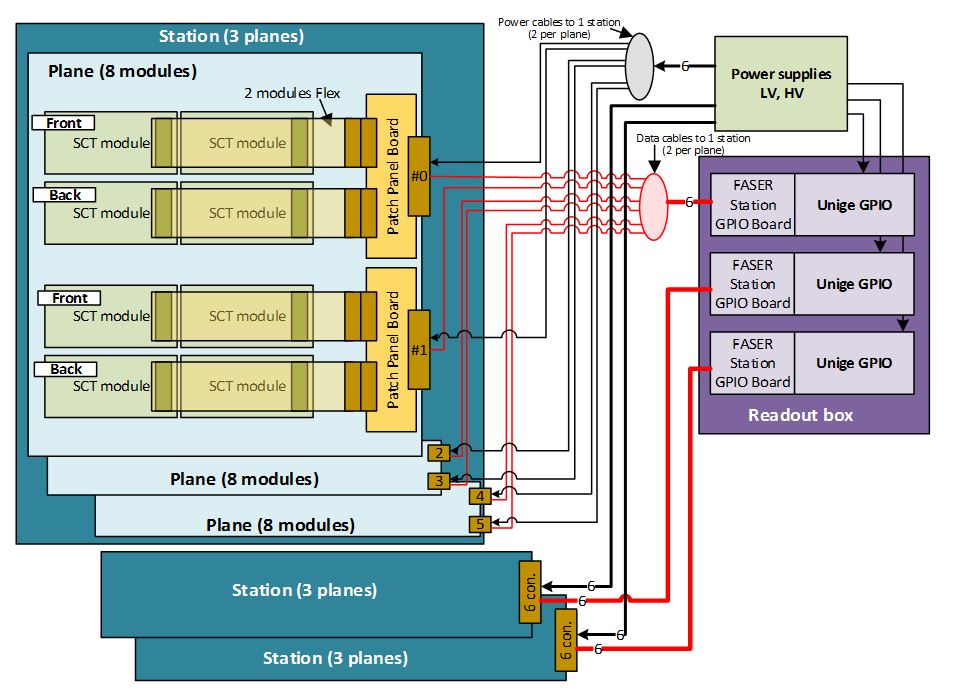}
\caption{Tracker readout hardware architecture.}
\label{fig:sct_readout_architecture}
\end{figure}

\subsection{Alignment}

Preliminary studies of detector performance have assumed perfect alignment, but the precision with which the positions and orientations of the 72 tracker modules can be determined will be an important limitation on the detector's momentum resolution. 

The modules in each tracking station will be measured carefully when assembled, and surveyed upon installation. A crucial tool for constraining the module positions will be the sizable flux of high-energy muons during proton-proton collisions (see Section~\ref{sec:particleFlux}). The overlapping geometry of the modules in each station will facilitate precision measurement of their relative positions with these nearly straight tracks.

Without, for instance, a detectable resonant particle whose mass can be used for absolute momentum calibration, certain ``weak modes" of misalignment~\cite{ATLAS:2010nca} that cannot be constrained by geometrical single track fits may remain and would limit the momentum resolution achievable. Use of calorimeter measurements to help resolve these degeneracies will be explored. Fortunately, FASER's ability to discover a dark photon signal does not rely on precision momentum measurement.

\subsection{Calibration}

Calibration of the SCT modules is performed to optimize the threshold and voltage settings and ensure stable performance by measuring the properties of the sensors and their electronics. The calibration procedure also includes basic tests to identify and flag dead or noisy channels.

The most important chip parameter probed by the calibration is the threshold of the discriminator. Each front-end chip has an 8-bit DAC, which allows the threshold to be set globally across that chip. To assure uniformity of threshold across the chip, each channel has its own 4-bit correction DAC (TrimDAC). The DAC steps can themselves be set to one of four different values. This means that uniformity of threshold can be maintained even with uncorrected channel-to-channel variations. Each channel also has an internal capacitor used to inject test charges. By injecting various known charges the analogue properties of each channel can be measured.

In normal operation, the target threshold is set to 1 fC, which should be compared to the 3.6 fC deposited by a minimum ionizing particle at normal incidence. With this threshold, the noise occupancy should be below $5 \times 10^{-4}$. 

Since radiation damage is not a serious issue in FASER, it should be unnecessary to tune the thresholds frequently. Calibrations will likely be performed during technical stops and machine development periods. 

\subsection{Cost and Schedule}
The expected cost for the tracking mechanics is detailed in Table~\ref{table:trackerMechanicsBudget}, and the costs for other Tracker related items are details in Table~\ref{table:trackerBudget}. 
To efficiently deal with hardware failures, we will need a spare chiller, a spare power supply crate, one spare HV PS and two spare low-voltage power supplies for a total cost of spares of 22~kCHF. 
The schedule for the production of the Tracker is shown in \figref{FAS_MP_Nov2018-2}.

\begin{table}[tbp]
\centering
\begin{tabular}{|l|c|}
\hline
\  {\bf Item} \ &  \ {\bf Cost [kCHF]} \  \\ \hline
\ 9 Tracking plane frames \ &	15 \quad \\ \hline
\ 6 Spacers frames  \ &		20 \quad \\ \hline
\ 3 Tracking Station Outer boxes  \ &		20 \quad \\ \hline
\ 3 set of cooling piping  \ &		1 \quad \\ \hline
\ 3 Fixation to the base frame \ &		10 \quad \\ \hline
\ {\bf Total} \ & {\bf 66}  \quad   \\ \hline
\end{tabular}
\caption{Budget for tracker station mechanical assembly.}
\label{table:trackerMechanicsBudget}
\end{table}

\begin{figure}[bp]
\centering
\includegraphics[clip, trim=0.5cm 3.7cm 0.5cm 0.5cm,width=1\textwidth]{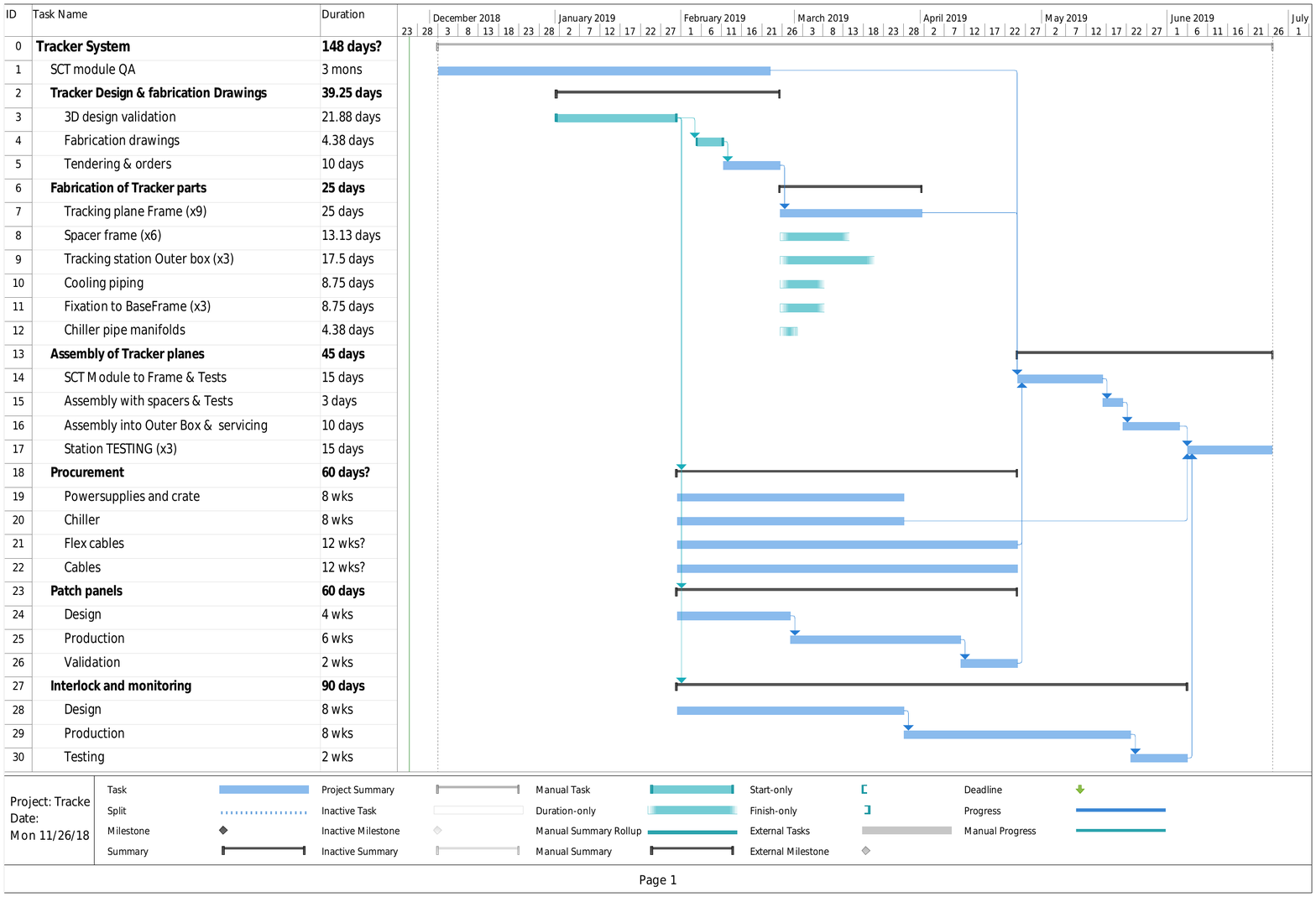} 
\caption{Schedule of Tracker stations}
\label{fig:FAS_MP_Nov2018-2}
\end{figure}

\begin{table}[tbp]
\centering
\begin{tabular}{|l|c|}
\hline
\  {\bf Item} \ &  \ {\bf Cost [kCHF]} \  \\ \hline
\ Chiller \ &	7\quad \\ \hline
\ Connection to EN-CV compressed air  \ &		1 \quad \\ \hline
\ Power supply crate  \ &		10 \quad \\ \hline
\ HV Power supplies  \ &		17 \quad \\ \hline
\ LV Power supplies  \ &		36 \quad \\ \hline
\ Power cables  \ &		6 \quad \\ \hline
\ Flex cables  \ &		10 \quad \\ \hline
\ Patch panel and connectors  \ &		8 \quad \\ \hline
\ Interlock and environmental monitoring \ &		10 \quad \\ \hline
\ {\bf Total} \ & {\bf 105}  \quad   \\ \hline
\end{tabular}
\caption{Budget for other tracker-related hardware.}

\label{table:trackerBudget}
\end{table}

\clearpage
\section{Scintillator Veto and Trigger Layers}
\label{sec:sci}

The FASER experiment has four scintillator stations as shown in \figref{DetectorLayout}. These serve several different purposes in the experiment. The stations are constructed from one or two scintillator layers.

\subsection{Veto Stations}

The first two stations, the veto stations located in front of the  dipole magnets, are primarily used to suppress events with incoming particles, mostly high-energy muons. To avoid energetic photons from muon bremsstrahlung before the detector entering undetected, an absorber block of lead is placed between the two veto stations, which will either contain the photons completely or generate a shower that is detectable by the second station. The block is foreseen to be 20 radiation lengths (about 11 cm) thick. High-energy muons passing through the absorber will also radiate photons, but in this case the muons will be detected by the front station. To fully suppress background related to muons from the interaction point, each station should detect more than 99.99\% of the incoming muons. 

\begin{figure}[bp]
\centering
\includegraphics[width=0.42\textwidth]{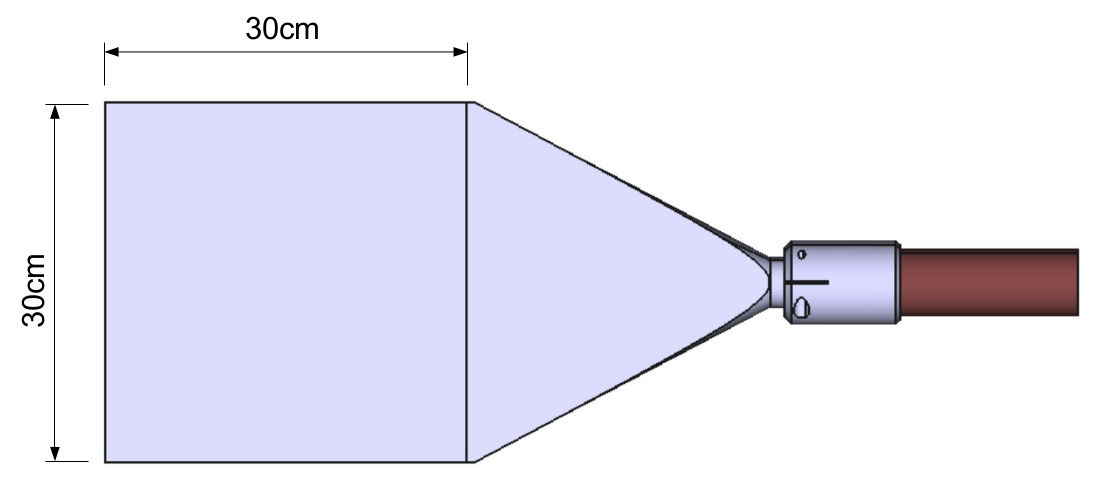} 
\includegraphics[width=0.55\textwidth]{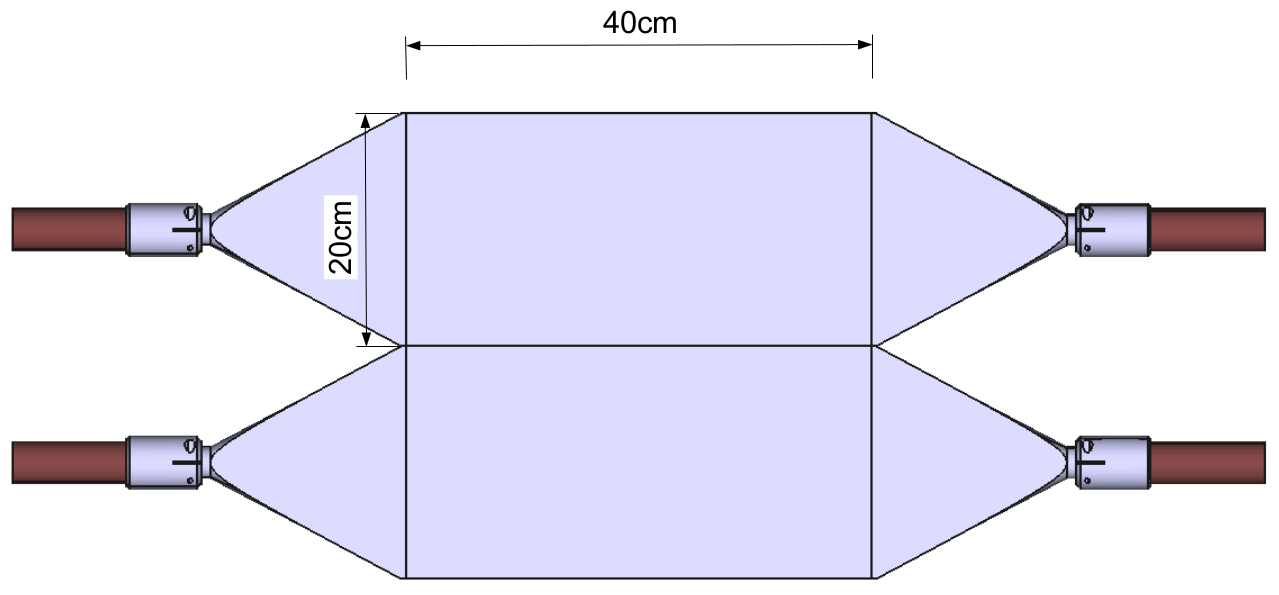} 
\caption{Design of the two types of scintillator layers.}
\label{fig:Scintillators}
\end{figure}

Each of the two veto stations consists of two identical scintillator layers. The basic design of each layer is shown on the left in \figref{Scintillators} and consists of a $2~\cm$-thick, $30~\cm \times 30~\cm$ plastic scintillator (Bicron BC 408) connected through a light guide to a PMT (for example a Hamamatsu H3178-51). The PMT will be encased in mu-metal to ensure its operation is not affected by the magnetic fringe fields (up to 5~mT). To minimize the required space in the horizontal direction, the two layers will be installed with a $90^\circ$ angle between them. The transverse size of the scintillator layer is larger than the magnet aperture to further ensure no charged particles can enter undetected, and the $2~\cm$ thickness is chosen to provide a very high single layer detection efficiency (well above 99\%). The two independent layers provide redundancy and ensure a very high veto efficiency, which can easily be measured {\em in situ}, as there should be no correlated inefficiencies. 

\subsection{Trigger and Timing Station}

The trigger/timing station located after the first magnet and first tracking station is used to detect the appearance of a charged particle pair from the decay of a LLP in the decay volume of the first dipole magnet. It provides the primary trigger signal and will also be used to precisely measure the arrival time of the signal with respect to the $pp$ interaction at the IP. The precision should be better than 1~\ns to suppress any non-collision backgrounds. Additional signal to background suppression is provided by measuring the light yield since the prospective signal has two charged particles unlike most background types, but this is not expected to be a decisive discriminator due to the large event-to-event Landau fluctuations in the light yield. Since this layer is in the tracker volume, the material should be minimized while still maintaining an efficiency above 98\%. Finally the active area of this station will also be made large enough to cover most of the magnet front surface. This is to detect muons coming in at an angle, missing the veto stations and the first two tracker stations, but causing an electromagnetic shower in the magnet material and thus a detectable shower in the downstream calorimeter that mimics a photon-only signature. 

The trigger/timing station consists of a single scintillator layer made from two $20~\cm\times 40~\cm$ scintillator blocks with a thickness of 1~\cm{} (2.5\% radiation length) shown on the right in \figref{Scintillators}. The layer is split in to two blocks to reduce the size of the light guide, keep high detection efficiency, and improve the timing precision by minimizing the vertical time-walk. The horizontal time-walk is compensated by using the average signal time of the two PMTs, hence the requirement to have a PMT on each side of the scintillator layer. With this setup, the timing resolution will be limited by the precision of the readout electronics. Having two blocks will imply a gap of about 1.5~\mm between the active parts of the scintillators. To avoid an inefficiency, the two blocks will be offset along the line-of-sight so that a small overlap (about a mm) can be introduced.

\subsection{Preshower Station}

The final trigger/preshower station is located just in front of the calorimeter. This station provides an additional trigger signal which, if needed, can be used in a coincidence with the first trigger station to reduce the rate of non-physics triggers. It will be preceded by a thin layer of radiator (about 2 radiation lengths of tungsten or lead) to create a simple preshower detector that with high efficiency can detect a physics signal of two close-by energetic photons. The preshower helps distinguish this physics signal, which would otherwise leave only large energy deposition in the calorimeter, from deep inelastic scattering of high energy neutrinos in the calorimeter. This is needed because the calorimeter does not have any longitudinal segmentation; see \secref{calo}.

To reduce the number of different scintillator configurations, the preshower station will use two layers in the same configuration as the veto stations with half of the radiator material placed in front of the first layer and the second half in between the layers. A double arm-layer as used in the timing station, which would have provided another precise timing measurement, is not possible due to physical constraints from the tunnel wall on one side of the detector.

\subsection{Powering}

Each PMT requires 1.5-2.5~kV. To avoid correlated inefficiencies, each PMT should be supplied by its own power supply channel. The baseline is to provide this with two 6-channel high-voltage power supplies (CAEN V6533). These will be installed in the VME crate that is also used by the PMT readout described in \secref{PMTReadout}. An alternative option is a separate high voltage power supply module in the power supply crates used for the tracker stations. 

\subsection{Cost and Schedule}

The cost of the different scintillator components apart from the readout is estimated in Table~\ref{table:scintillatorBudget}. One spare PMT and a spare HV power supply module at a total cost of 6~kCHF is also foreseen to be purchased.

\begin{table}[tbp]
\centering
\begin{tabular}{|l|c|}
\hline
\  {\bf Item} \ &  \ {\bf Cost [kCHF]} \  \\ \hline
\ 8 scintillator plates \ &	4 \quad \\ \hline
\ 10 light guides  \ &		8 \quad \\ \hline
\ Assembly  \ &		5 \quad \\ \hline
\ 10 PMTs  \ &		20 \quad \\ \hline
\ 2 HV power supply  \ &		8 \quad \\ \hline
\ HV Cables	  \ &	1\quad \\ \hline 
\ Lead absorber block \ &		1.5 \quad \\ \hline
\ 2 Tungsten plates \ &		4.5 \quad \\ \hline
\ {\bf Total} \ & {\bf 52}  \quad   \\ \hline
\end{tabular}
\caption{Budget for scintillator system for the FASER experiment.}
\label{table:scintillatorBudget}
\end{table}

The expected schedule for the construction of the FASER scintillator system is shown in \figref{ScintillatorTimeLine}. The schedule assumes that the light guides and PMT mounts are produced by an outside company, while the assembly is done by the CERN scintillator lab. 

\begin{figure}

\centering
\includegraphics[clip, trim=0.5cm 11.4cm 0.5cm 0.5cm,width=0.95\textwidth]{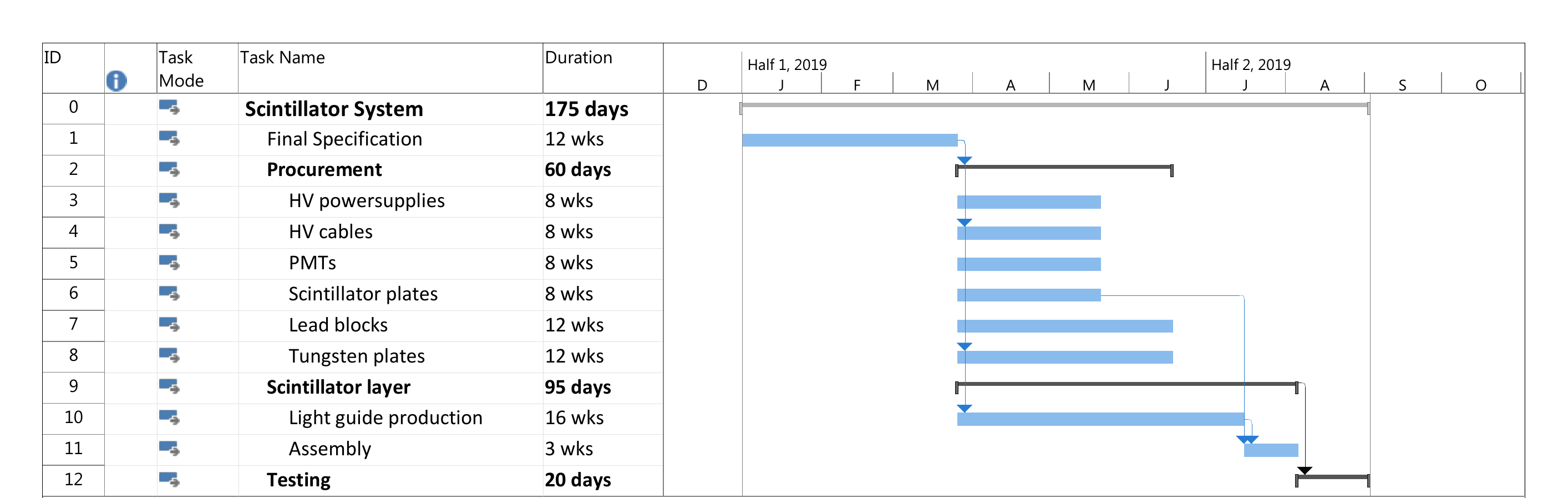} 
\caption{\label{fig:ScintillatorTimeLine}
Timeline for the construction of the scintillator system.}
\end{figure}

\clearpage
\section{Electromagnetic Calorimeter}
\label{sec:calo}

\begin{figure}[bp]
\centering
\includegraphics[width=0.95\textwidth]{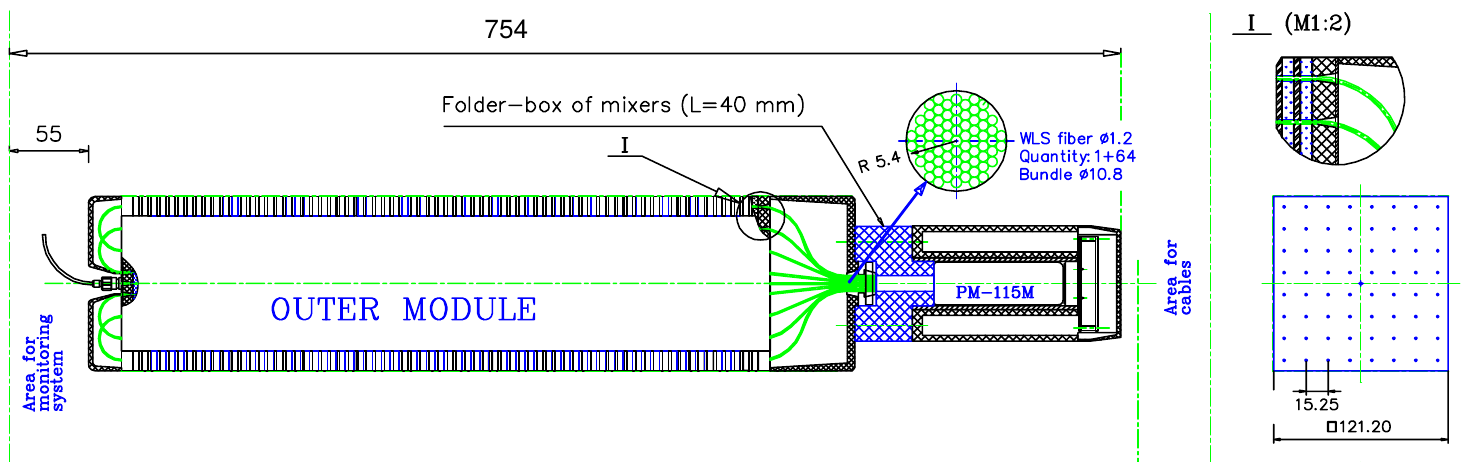} 
\caption{
Design of the LHCb outer ECAL modules~\cite{LHCB:2000ab}. 
}
\label{fig:Calorimeter}
\end{figure}

The electromagnetic calorimeter provides strong identification of high-energy electrons and photons over muons and hadrons and allows to measure their energy. Since for most signal events the $e^+e^-$ or photon pair is separated by less than a few millimeters, it is not feasible to measure the individual particle energies.  The main calorimeter requirement is, therefore, to measure the total electromagnetic energy with good accuracy for multi-TeV deposits in a compact detector. 

\subsection{LHCb ECAL Modules}

The calorimeter will use four spare LHCb outer ECAL modules~\cite{LHCB:2000ab}, shown in \figref{Calorimeter}. The LHCb Collaboration has kindly agreed to allow FASER to use six of these modules on indefinite loan.  These are so-called Shashlik-type calorimeters with interleaved scintillator and lead plates. Each module contain wavelength shifting fibers penetrating the full calorimeter, which extract the signals to a single PMT on the back. The modules are 754~\mm-long, including the PMT, and have transverse dimensions of $121.2~\mm \times 121.2~\mm$. The full FASER acceptance is therefore covered with these four modules. The calorimeter contains 66 layers of $2~\mm$ lead and $4~\mm$ plastic scintillator, for a total depth of 25 radiation lengths. The baseline is to use the same PMT as LHCb (Hamamatsu R7899-20), but it might be necessary to use either a different model or attenuate the light in order to not saturate at the highest energy. The high voltage for the PMTs will come from the same type of power supply as for the scintillator system and the same readout will be used as well. The energy resolution for TeV deposits in such a calorimeter is expected to be around 1\%, although this will be degraded at the highest energies, since 25 radiation lengths will not fully contain all such showers. It is not foreseen to have a leakage detector behind the calorimeter as the impact on the signal efficiency is expected to be negligible, and there is very limited space.

A short test beam run for calibration using high energy electrons in the SPS (up to 300~\gev) is being considered and if needed a request for test beam time will be submitted. This would only be able to take place in 2021 or later and therefore would have to be done with a spare module and PMT. 

\subsection{Cost and Schedule}

The cost of the different calorimeter components, apart from the readout, is estimated in Table~\ref{table:calorimeterBudget}. One spare PMT, with a cost of 2~kCHF, is planned, but no additional HV modules are planned, as the spare is shared with the scintillators. As noted above, six calorimeter modules will be on indefinite loan from the LHCb collaboration.

\begin{table}[tbp]
\centering
\begin{tabular}{|l|c|}
\hline
\  {\bf Item} \ &  \ {\bf Cost [kCHF]} \  \\ \hline
\ 4 PMTs  \ &		8 \quad \\ \hline
\ HV power supply  \ &		4 \quad \\ \hline
\ HV Cables	  \ &	1\quad \\ \hline 
\ {\bf Total} \ & {\bf 13}  \quad   \\ \hline
\end{tabular}
\caption{Budget for the calorimeter system for the FASER experiment.}
\label{table:calorimeterBudget}
\end{table}

The expected schedule for the preparation of the FASER calorimeter system is shown in \figref{CalorimeterTimeLine}. Tests of each module will be done in collaboration with LHCb ECAL experts. 

\begin{figure}[tbp]
\includegraphics[clip, trim=0.5cm 13.2cm 0.5cm 0.5cm,width=0.98\textwidth]{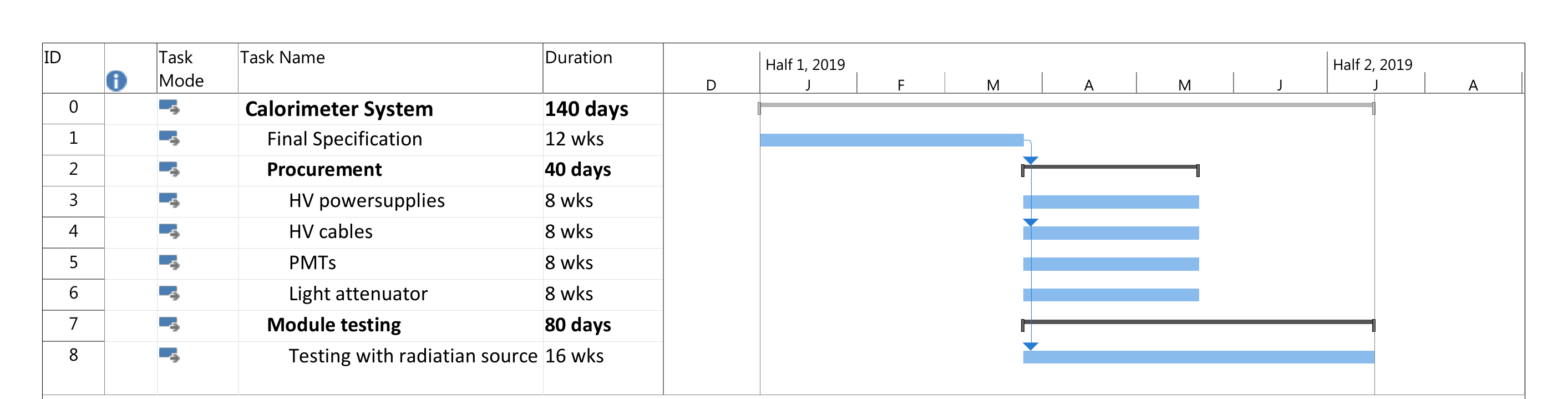} 
\caption{\label{fig:CalorimeterTimeLine}
Timeline for the preparation of the calorimeter system. }
\end{figure}

\clearpage
\section{Detector and Magnet support}
\label{sec:support}

The detector is mechanically supported by a common structure (the detector support structure or main cradle), which also allows fine alignment of the different components. 
At this stage a preliminary conceptual design has been carried out, which will be refined over the next months including adding the sub detectors
specific needs (for the calorimeter, scintillators, and Tracker stations).
\subsection {Requirements}
The main requirements for the detector support structure are:
\begin{itemize}
\item Providing a common mechanical support for the detector components and the magnet segments (limiting also the number of interfaces needed for the detectors with respect to the floor);
\item Providing a guaranteed safe distance between the magnets of 20~cm;
\item Allow for fine alignment between the components {\it in situ};
\item Provide stable alignment between the components (especially the tracking stations) as precisely as possible given the detector design. A placement alignment precision of 100~$\mu$m and stability of the alignment of 100~$\mu$m would be desirable to provide good spectrometer momentum resolution for high energy charged particles.
\end{itemize}

\subsection{Design}
The main support structure as shown in \figref{FAS_BaseFrame2} is a conceptual design to describe the key features. It is a common cradle which positions each part independently and acts as the main interface to perform the necessary alignment fine  tuning with respect to the theoretical beam axis. 

\begin{figure}[h!]
\centering
\includegraphics[width=0.9\textwidth]{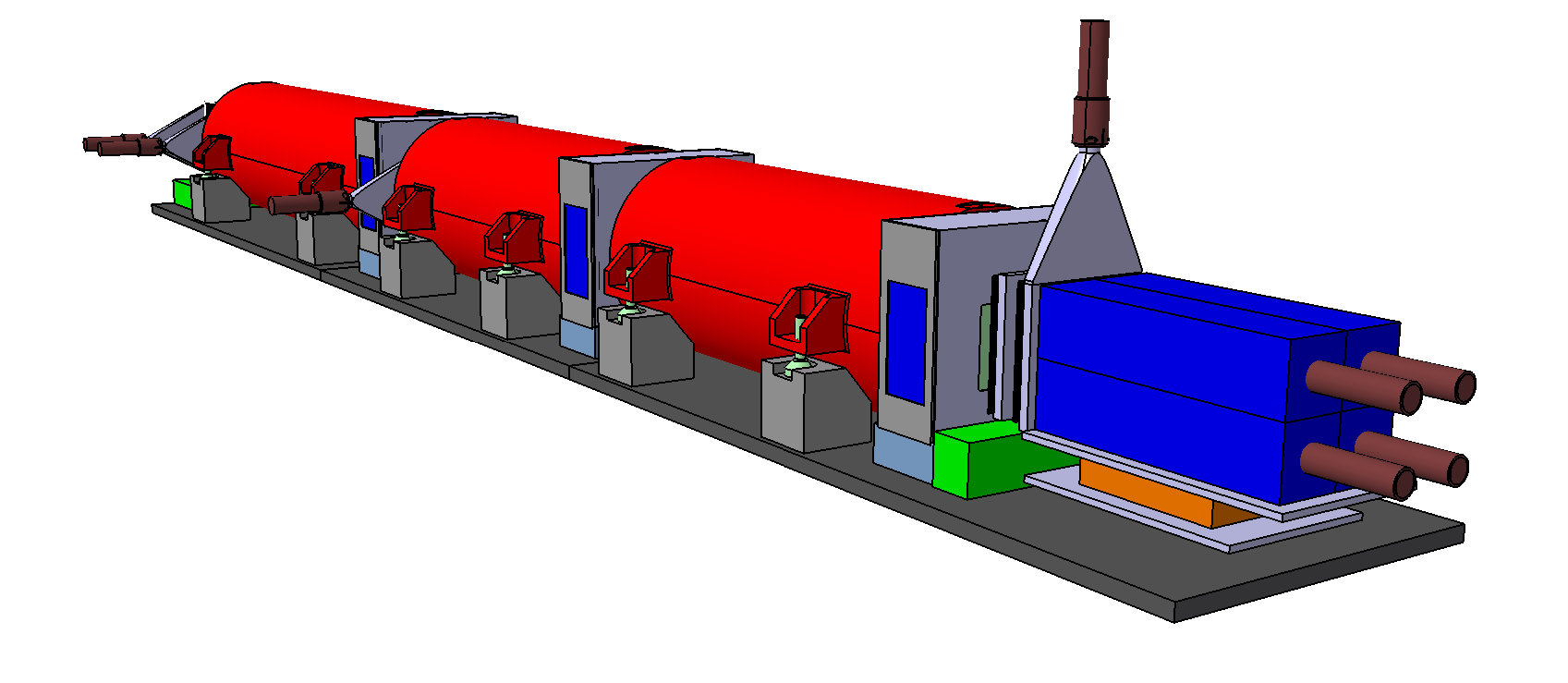} 
\caption{Detector support overview}
\label{fig:FAS_BaseFrame2}
\end{figure}

The base plate (or base frame) shown in \figref{FAS_BaseFrame6} will be the first element to be installed in the tunnel prior to any detectors assembly. It will be surveyed in 3-dimensions and aligned by means of jack screws (or equivalent) before being clamped to the ground to provide a stable interface for the detectors. The conceptual approach is to limit the civil engineering work by removing individual fixations per detector. Instead the fixation is only done once at the main support structure level. Due to size limitation (in particular for CNC machining, and transport to TI12), this base plate or frame has to be build in segments to be precisely re-assembled using calibrated pins when installed in the tunnel. 
The overall precision (both for machining and assembly) that can be expected here is at the order of 100-200 microns (along the beam axis).The lateral precision over 50~cm (plate width) could be lowered to 50 microns at best. This strongly depends on the CNC machine type.
The in-plane alignment between plate segments is performed and frozen by the machining itself (no fine tuning). Only the out-of-plane alignment (perpendicular to the base pate) is needed to provide a common theoretical axis to be aligned with the beam axis. This will be carried out with the CERN survey group by means of a laser system and targets attached to the base frame. The precision of the measurement is of the order of 100 microns (to be confirmed with CERN survey group).

\begin{figure}[tbp]
\centering
\includegraphics[width=0.7\textwidth]{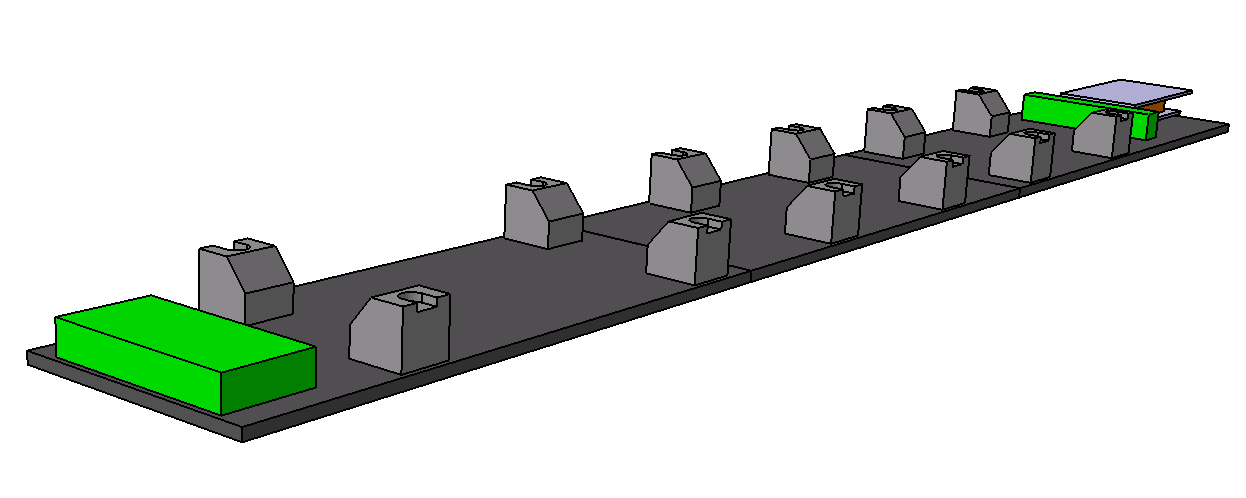} 
\caption{Base plate with detector resting blocks.}
\label{fig:FAS_BaseFrame6}
\end{figure}

The material choice is based on stability requirements, magnetic compatibility and cost (procurement, and fabrication). Stainless steel is a good candidate since it is less sensitive to temperature changes (as compared to aluminum). As an example, over a 5~m long steel part (the overall base plate length), and for 1 degC change in temperature, the length variation is of the order of 70 microns. 
The design is still open and will be driven mostly by the final clearance between the Magnet outer cylinder and the floor.
\begin{figure}[h!]
\centering
\includegraphics[width=0.70\textwidth]{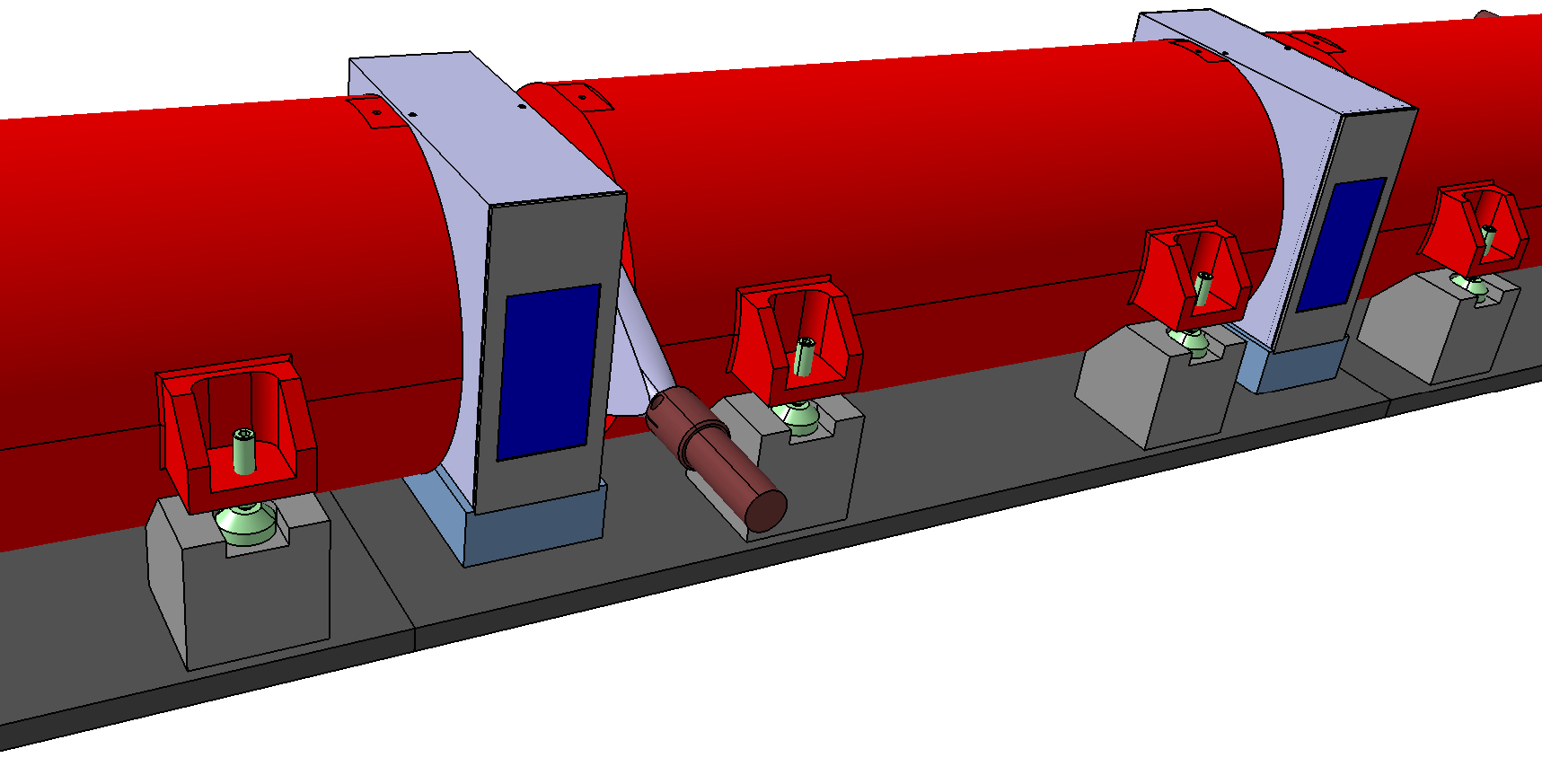} 
\caption{Close up of magnet and Tracker station alignment.}
\label{fig:FAS_BaseFrame4}
\end{figure}

\Figref{FAS_BaseFrame4} shows the basic principle of the magnet vertical tuning by means of jack screws (4 per magnet segment). The in-plane alignment is done by inserting the jack screw into machined resting blocs (acting as a pocket), but such as system cannot guarantee any better than 1~mm precision. If a greater precision is requested from the physics requirements (to be confirmed), lateral jack screws could be added to solve this issue. An example of the support structure for a Tracker station is shown (with the block in light blue), that could either be a precisely machined block or some linear stages (in 2 axes, perpendicular to the beam axis) to allow for fine tuning. A similar support structure would be used for each detector component. All of the interfaces to the base plate are aligned (in-plane) precisely by the CNC plate machining and will comply with the above mentioned level of precision.

\subsection{Cost and Schedule}

The expected cost estimate for producing the detector support structure is presented in Table~\ref{table:SupportStructure}, based on the preliminary conceptual design.

\begin{table}[tbp]
\centering
\begin{tabular}{|l|c|}
\hline
\  {\bf Item} \ &  \ {\bf Cost [kCHF]} \  \\ \hline
\ Stainless steel Plate or frame \ &	30 \quad \\ \hline
\ Detector interfaces  \ &	5 \quad \\ \hline
\ Magnet adjustable feet  \ &	25 \quad \\ \hline
\ {\bf Total} \ & {\bf 60}  \quad   \\ \hline
\end{tabular}
\caption{Budget for Detector support for FASER experiment.}
\label{table:SupportStructure}
\end{table}

\begin{figure}[tbp]
\centering
\includegraphics[width=0.82\textwidth]{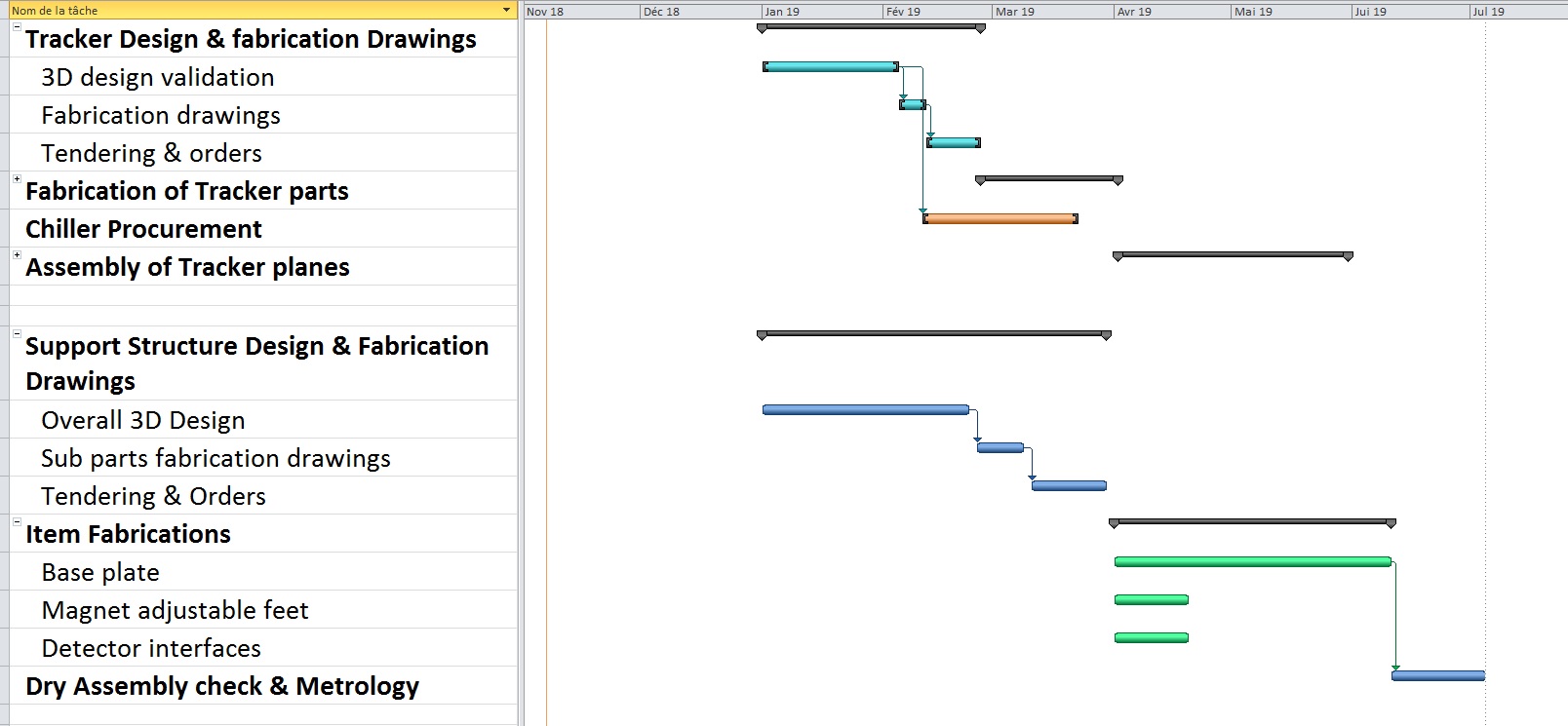} 
\caption{Schedule for FASER Support structure.}
\label{fig:FAS_MP_Nov2018-1}
\end{figure}

\Figref{FAS_MP_Nov2018-1} shows the schedule for producing the FASER support structure. This is inter-connected with the other sub-system timelines, for instance the design completion of the Tracker stations. Such interconnection will apply also for the other sub-parts since the design will be a common effort. The starting date is given as an example.

\clearpage
\section{Trigger and Data Acquisition}
\label{sec:tdaq}

The FASER detector is located in a low-background area of the LHC. The aim is therefore to trigger the read out of the full detector on every detectable high energy particle passage. For redundancy and to measure the detector efficiencies, all scintillators layers and the calorimeter will provide trigger signals. The scintillator trigger threshold will be below that of a single minimum ionizing particle, while the calorimeter threshold will be set to trigger on any significant electromagnetic shower. The expected trigger rate per source is shown in \tableref{TriggerRates} under the assumption that the rates will be completely dominated by high-energy muons, as shown by simulation studies and {\em in situ} measurements (see \secref{environment}). The rates are therefore highly correlated, and the total physics rate is expected to be approximately equal to the highest rate source. In case the rate of a trigger source is too high for the available readout bandwidth, the trigger system will allow triggering only on combinations of certain (anti-)coincidences, such as a calorimeter signal or signal in preshower scintillators with no signal in the veto scintillators. In this case a fraction of the individual trigger sources will still be recorded in the form of ``pre-scaled" triggers for calibration and alignment purposes.

\begin{table}[bp]
\centering
\begin{tabular}{|l|c|}
\hline
{\bf Source} & \quad {\bf Rate [Hz]} \quad \\ \hline 
Veto scintillators & 360  \\ \hline
Timing scintillators & 640  \\ \hline
Preshower scintillators & 360  \\ \hline
Calorimeter ($E>100\gev$) & $<5$ Hz \\ \hline
Random trigger & 10  \\ \hline
{\bf Total} & {\bf 650} \\ \hline
\end{tabular}
\caption{Expected trigger rate per trigger source at a instantaneous luminosity of $2 \times 10^{34}~\cm^{-2}~\s^{-1}$. Note that since the primary source of triggers is high-energy muons passing through the full detector, the rates are highly correlated. The random trigger rate is meant for pedestal calibration and noise monitoring.}
\label{table:TriggerRates}
\end{table}

To be able to identify a rare non-SM physics signal, as much information as possible about each triggered event will be read from the detector. For the tracker, the information available is limited by the pre-existing readout chip to three 25~\ns-wide time bins (hit/no hit) per strip. The trigger signal will be adjusted so that signals originating from highly relativistic particles from the IP will be seen in the central bin. For all PMTs, the full waveform will be read out in a much wider time window (about 1~\mus) to accurately reconstruct the charge and time of each pulse and to ensure there is no anomalous activity near the time of a possible physics signal. Basic information from the trigger system, such as the source of the trigger firing, event number, etc., will also be read out. In order to measure non-collision backgrounds and minimize the complications of the system, it is not foreseen to require the trigger signals to come in time with the collisions at IP1 and therefore no use of the bunch patterns will be made online. Instead the in-time and out-time event selection will need to be done offline.

\begin{table}[tbp]
\centering
\begin{tabular}{|l|c|}
\hline
{\bf Source} & \quad {\bf Size [kB]} \quad \\ \hline 
PMTs & 20 \\ \hline
Tracker & 5 \\ \hline
Trigger Logic Board & $<1$ \\ \hline
{\bf Total} & {\bf 25} \\ \hline
\end{tabular}
\caption{Average expected event sizes. The PMT size assumes a 1~\mus readout window (500 samples of 2~\ns).} \label{tab:EventSize}
\end{table}

A diagram of the trigger and date acquisition system is given in \figref{TDAQOverview}. The system is designed to minimize the amount of equipment underground where access is highly limited, while having the minimal number of signal cables coming into or out of the experiment, as those have to be more than 500~m long. The trigger system will run synchronously to the 40.08~\MHz clock of the LHC. All of the PMTs will be read out by a high-speed digitizer, which also provides a trigger signal when the signal on a pair of PMTs exceeds a preset threshold. The trigger signals are combined in the Trigger Logic Board (TLB) to form the final trigger signal (L1A) after possibly applying coincidence and/or veto logic, which combines information from the scintillator and calorimeter PMTs. The L1A is sent to the PMT digitizer and the tracker readout boards to initiate the readout of all of the detectors. The readout is done over Ethernet to a PC (the Event Builder) located on the surface. The expected event size is listed in Table \ref{tab:EventSize}; given the trigger rate, a single Gbit/s connection is more than sufficient. The event builder merges the data fragments into events that are buffered and undergo basic data quality checks on a second PC before being copied to CASTOR for permanent storage. The various Detector Control Systems (DCS) in the experiment will be connected over the same Ethernet connection to a separately-controlled PC on the surface, which will also be used to monitor and control the trigger and readout system. The Ethernet connection to the surface will use a single optical fiber pair connecting the TI12 experimental area to the SR1 surface building, close to the ATLAS control room. 

\begin{figure}[tbhp]
 \centerline{\includegraphics[width=0.8\linewidth]{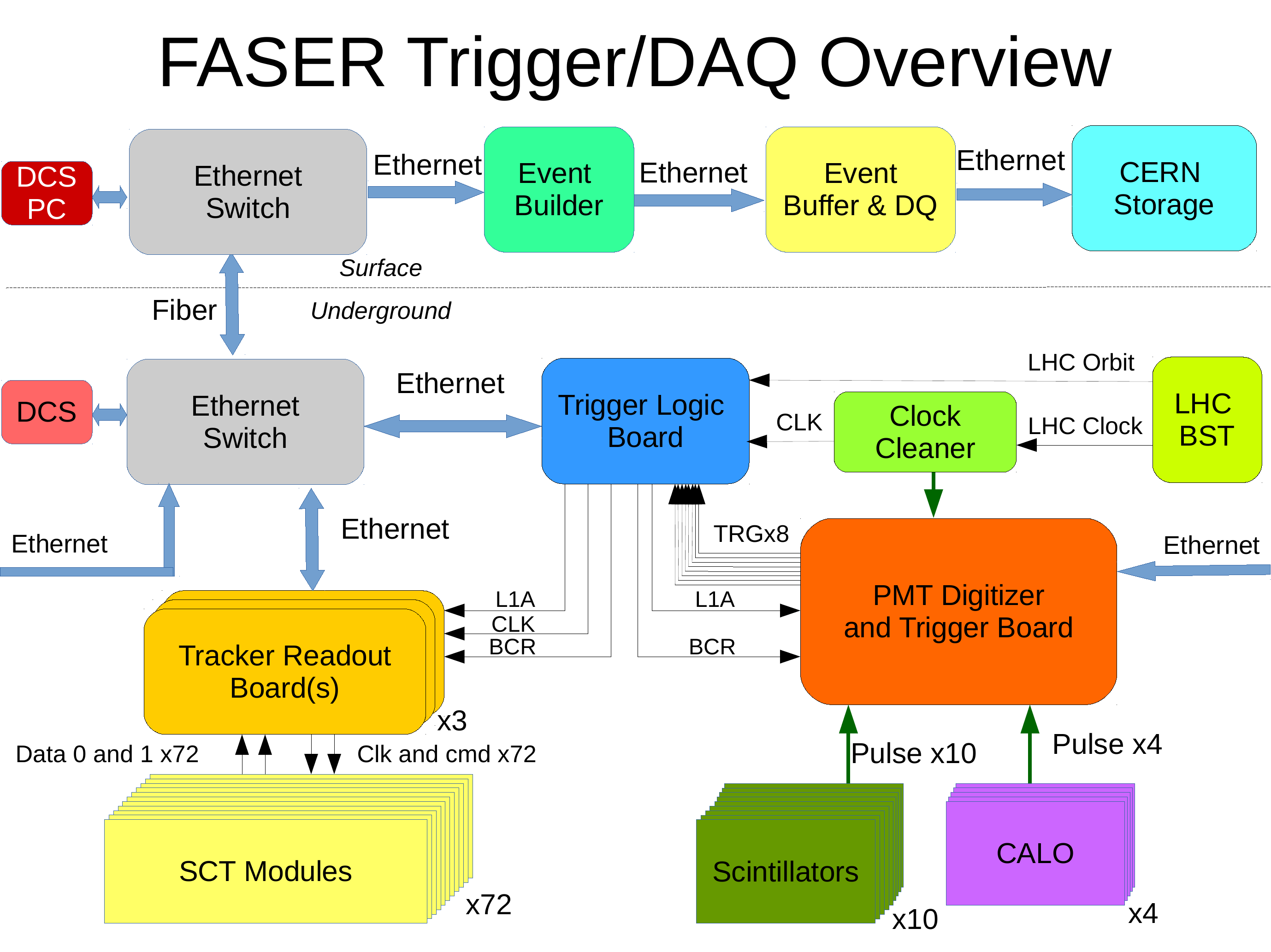}}
 \caption{Schematic diagram of the Trigger and DAQ system.}
 \label{fig:TDAQOverview}
\end{figure}

\subsection{LHC Signals}

The LHC clock (40.08~\MHz) and orbit signal (11.245~\kHz) are part of the Beam Synchronous Timing (BST), transmitted to beam instrumentation equipment around the LHC over optical fibers using the TTC system~\cite{TTC}. For FASER, this signal will be received over a single optical fiber by a BOBR VME card~\cite{BOBR} from the LHC beam instrumentation group, which extracts the LHC clock and orbit signal to two front panel LEMO connectors. It also provides additional information, such as the global time and accelerator mode over the VME bus, but this is not foreseen to be used. 

The LHC clock provided by the BOBR has a non-negligible jitter, changes during the ramp of the LHC, and is not guaranteed to be continuous. The clock will therefore pass through a Si5344 jitter attenuating clock  multiplier chip~\cite{Si5344} mounted on an evaluation board in the VME crate. This provides a ultra low-jitter clock (90~\fs), which tracks the LHC clock at a programmable speed, provides a holdover clock in case the input clock disappears, and provides a zero delay mode that ensures the clock is always at the same delay with respect to the beam interactions. Daily and seasonal variations of up to several \ns{} in the phase of the clock is expected, due primarily to ground temperature variations, and the absolute timing offset will therefore need to be monitored and calibrated with high-energy muons from the IP.

The orbit signal provides a 25~ns pulse once per turn of the LHC. It will be sent directly to the input of the TLB and used to reset the bunch counters in all of the readout.

\subsection{Scintillator and Calorimeter Trigger and Readout}
\label{sec:PMTReadout}

The readout and triggering of the scintillator and calorimeter  PMT signals will be done by a single, 16-channel CAEN digitizer VME board, VX1730~\cite{VX1730}. The digitizer samples each channel at 500~\MHz using 14-bit ADCs and stores the samples in a circular buffer of programmable length. For FASER, a $1~\mus$ long buffer (500~samples) will be used to capture the full PMT waveform and preceding data for offline analysis. The length of the readout might later be reduced to just cover the full PMT waveform (1-200~\ns).

Trigger thresholds can be set separately on the ADC value of each channel and a trigger signal generated for each pair of input channels. For the FASER scintillators, the threshold will be set well below that of a MIP, but both PMTs in a station will be required to be above threshold to keep noise triggers to a negligible level while maintaining more than 98\% trigger efficiency.  For the calorimeter, a larger signal is required, but any of the four channels
can raise a trigger signal. The board supports using a digital constant fraction discriminator logic for generating the trigger signal, but since the rise time of the PMT is shorter than the 2~\ns sampling time, this is not foreseen to be used. The trigger logic runs at 125~\MHz and provides eight trigger signals transmitted as 32~\ns-wide pulses to the TLB. The 32~\ns-wide pulse is needed since the signals are not synchronized to LHC clock and therefore could arrive at the TLB in separate bunch-crossings and not trigger the coincidence logic. The digitizer trigger signals will not on their own trigger the digitizer readout, which instead is triggered on the signal from the TLB. 

The digitizer does not run synchronously with the LHC clock. Instead it provides a trigger time in units of 16~\ns since the last LHC orbit signal. 
This does not provide sufficient precision to determine if any signal is in time with the collisions, which have an intrinsic spread of around 180~\ps. Therefore the LHC clock from the clock cleaner is also sampled on one of the digitizer input channels with 2~\ns sampling time. This will allow the relative timing with respect to a bunch crossing to be measured at a precision better than the 1~\ns precision targeted for the timing scintillator station. The absolute timing offset will be calibrated using high-energy muons.

The VX1730 board is able to buffer up to 1024 events while waiting to be read out. The board does not support direct readout over Ethernet, but instead has to be read out over  the VME backplane.  The maximum readout speed over the VME exceeds 50~MB/s, well above the needs of FASER. The readout to Ethernet will be done with an SIS3153  Ethernet to VME interface~\cite{SIS3153}, eliminating the need for a single-board computer in the VME crate.

\subsection{``UniGe USB3 GPIO" Board}
\label{sec:UniGeGPIO}

The Trigger Logic Board and the Tracker Readout Boards require custom firmware in an FPGA interfaced to the hardware signals and Ethernet readout/control. For both cases, this will be based on a general purpose board developed by the University of Geneva for slow control and readout of particle physics ASICs, detectors tests, and qualification. This board, named the ``Unige USB3 GPIO'' board, can also be used for small experiments using several detectors like the FASER experiment.

For each application, a dedicated interface board needs to be developed to interface to the hardware. This interface board will be connected with the connectors of the analogue and/or digital inputs/outputs (the latter only, for the case of FASER).

\Figref{board_logic} describes the architecture of the ``Unige USB3 GPIO'' board, and a photo of the board is shown in \figref{gpio}.  The core of the board is composed of a CYCLONE V A7 FPGA having 150K Logic Elements or 56.5K Adaptive Logic Modules (1 ALM = 8 inputs LUT and 4 registers), i.e., 226K registers, 6.86Mbits + 836Kbits of RAM, 156 double 18x18 DSP blocks and 7 PLLs. The FPGA is driven by a 40MHz local clock or an external clock thanks to a LVDS clock input through a dedicated 2 pin LEMO connector.

\begin{figure}[tbp]
\begin{center}\includegraphics[width=0.8\linewidth]{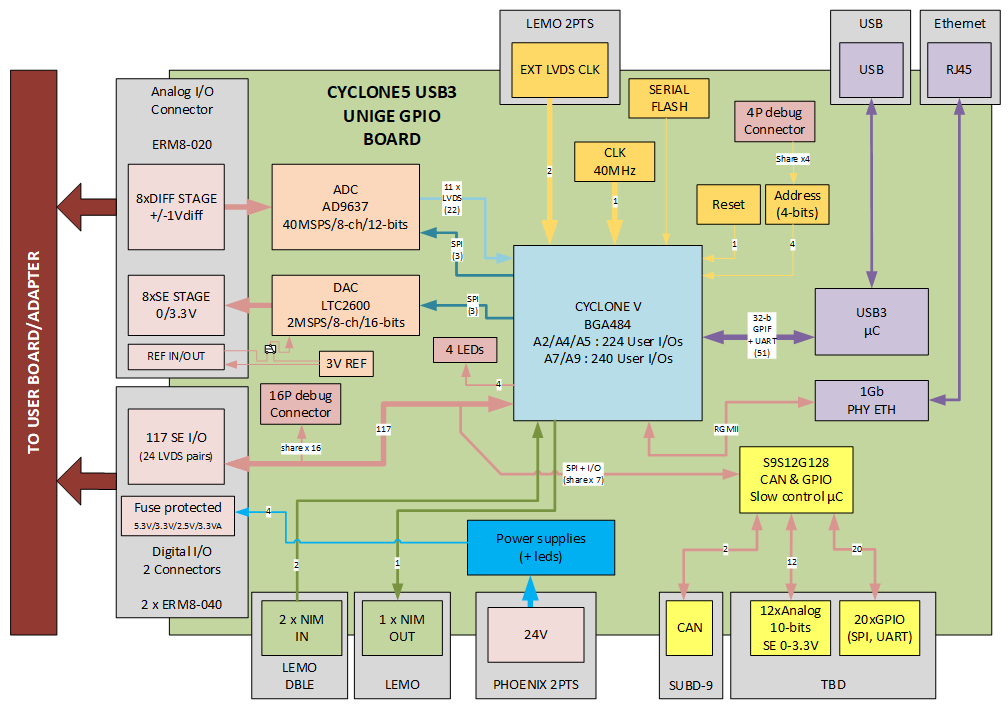}
\caption{Architecture of the ``UniGe USB3 GPIO'' board.}
\label{fig:board_logic}
\end{center}
\end{figure}

\begin{figure}[tbp]
\begin{center}
\includegraphics[width=0.5\linewidth]{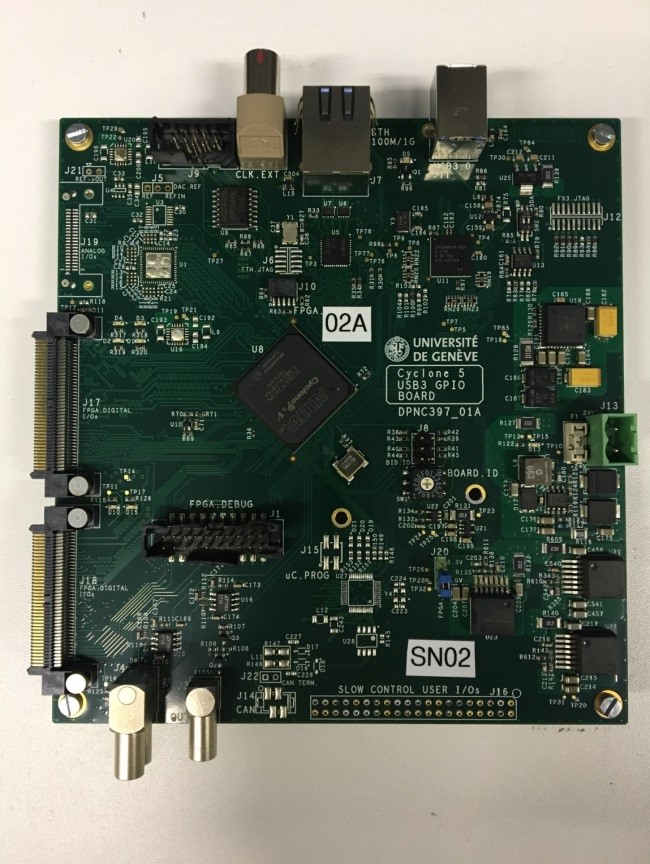}
\caption{The ``UniGe USB3 GPIO'' board.}
\label{fig:gpio}
\end{center}
\end{figure}

Concerning the experiment interface of the board (located on the left side of \figref{board_logic}), the FPGA is connected to a 40MSPS 12-bits 8 channels ADC and 2MSPS 16-bits 8 channels DAC, providing a versatile analogue interface through a dedicated 20-pin connector.  For the digital part, two connectors dispatch 117 single-ended inputs/outputs, which can be configured through the FPGA firmware.  In addition, two NIM inputs and one NIM output are available through LEMO connectors.

For the readout and the slow control, two interfaces can be used. Already fully implemented is USB3, where the communication is handled by a FX3 Cypress micro-controller providing a 32-bit parallel interface clocked at 100MHz with the FPGA having a maximum potential data throughput of 3.2Gb/s, but finally limited to 1.8 to 2.5Gb/s depending on the host PC performance when connected with the board. A full and very versatile FPGA library for the readout and the slow control is already available for this USB3 link; see below. Additionally an Ethernet PHY IC is also embedded on the board and should be able to provide a 10M/100M/1000Mbps connection. This interface is currently under development (Ethernet MAC and UDP integration into the FPGA firmware library) and should be available in the 1st quarter of 2019.

Finally a S9S12G64, G96 or G128 (for 64, 96 or 128KB of Flash memory) 16-bit micro-controller can also be mounted on the board. This device is typically used for slow control measurements (e.g., temperature, humidity, or voltage monitoring). This micro-controller has a CAN bus interface, 12 10-bits channels ADC, and a couple of digital I/Os which can also be configured in SPI or UART mode. All of those I/Os are accessible through a 2x20-pin header connector. One single 24~V power supply input is required to supply the board and 5.3~V, 3.3~V and 2.5~V is accessible through the user connectors for the interface board. 

\subsubsection{DAQ framework}

\Figref{sct_readout_logic} describes the architecture of an experiment DAQ using the ``Unige USB3 GPIO'' board, which should be based on the existing generic UniGe Library and have two specific application parts: a VHDL application firmware code in FPGA using the UniGe VHDL library, and a Host application code using the C\# monodev library for Windows/Linux.

The UniGe Library is composed of a VDHL firmware library in FPGA, the generic UniGe FX3 Firmware for the USB3 Cypress FX3 µC, the Windows/Linux FX3 cypress USB driver and the UniGe Windows/Linux C\# monodev library in host PC.

As an example, the Baby-Mind experiment (neutrino experiment installed in January 2018 at J-PARC, Tokai Japan) uses the UniGe Library and works with 4000 channels connected on 48 boards, 6 boards chained to 1 USB link, having 1 DAQ PC and 8 USB links for the overall experiment.

\begin{figure}[tbp]
\includegraphics[width=0.8\linewidth]{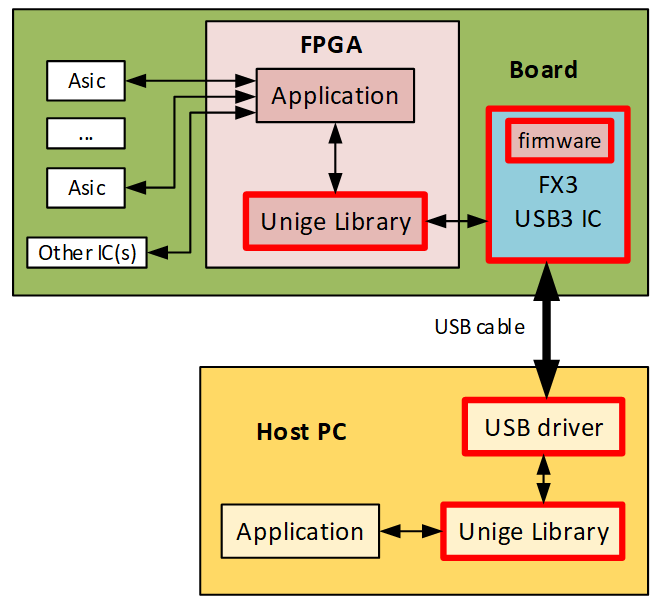}
\caption{DAQ framework for USB3 GPIO board.}
\label{fig:sct_readout_logic}
\end{figure}

\subsubsection{VHDL library}

The Unige VHDL library (Fig.~\ref{fig:fpga_logic}) is ‘linked’ with the C\# Host Library for the slow control protocol handling, for status, read and write parameters, for the usage of Structured configuration objects defined by a JSON file (hardware descriptor, no C\# code to be written), for the USB3 Readout and is compatible with single or multiple boards in a chain.

This means that USB3 is transparent from both sides, providing 1 bi-directional slow control channel and 1 fast readout FPGA to PC channel. A similar interface is planned for the Ethernet-based interface.

\begin{figure}[tbp]
\includegraphics[width=0.9\linewidth]{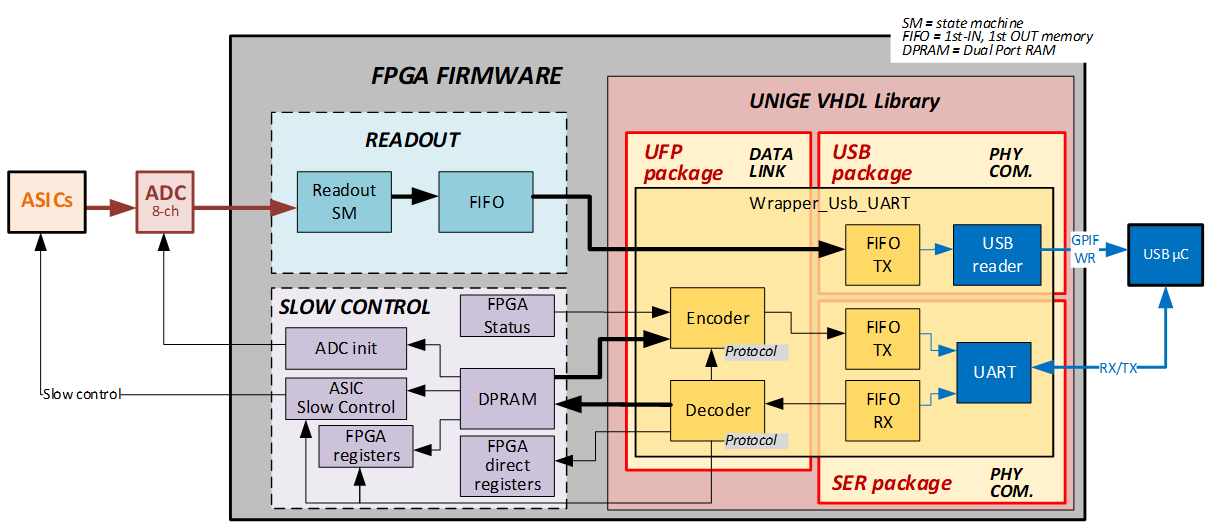}
\caption{UniGe VHDL library and application firmware overview.}
\label{fig:fpga_logic}
\end{figure}

\subsection{Trigger Logic Board (TLB)}

The TLB is the central triggering module of the FASER TDAQ system. It receives trigger signals from the scintillators and the calorimeter, digitized in the CAEN board. The TLB processes these signals to create the trigger decision (L1A) which is transfered to the detector readout boards. The TLB provides the FASER detector readout with a common clock (CLK), envisaged to be synchronous with the LHC clock, as well as bunch counter reset (BCR) signals on every LHC orbit signal. Finally, the TLB receives slow control signals (such as system reset) and sends out trigger data and monitoring information over Ethernet. 

The main functionality of the TLB blocks, shown in \figref{TLB}, is described in what follows. The input trigger lines coming from the CAEN digitizer board are aligned to the 40~\MHz clock, which is either coming from the LHC clock cleaner or generated internally. The eight synchronized trigger lines are merged to four trigger items\footnote{There could be more than four items, but from initial evaluations, four appear to be sufficient.} based on user-defined coincidence logic. User-settable prescales can be applied to each of these trigger items, as well as two additional trigger items resulting from a pseudo-random generator (providing triggers up to 1kHz) and a software trigger accept send by slow control. The logical OR of the six trigger items is created and unless vetoed, the signal becomes the L1A. A simple dead time requirement ensures that detector read-out is vetoed if it comes within 5-10~\mus from the previous one; the dead time requirements are driven by the tracker readout transmission time (more detail is given in \secref{trackerReadout}) and are user-settable. An event counter will provide a count of the total number of accepted events and a bunch counter will calculate the number of bunch crossings since the last LHC orbit signal. Additional counters will provide monitoring for the board: Monitoring will consist of counters for each trigger item before prescale, after prescale and after veto. The counters will be pushed out over Ethernet and the counters reset at a fixed rate of the order of 1~Hz.  

The output data from the TLB in normal running will be very small. Besides the event and bunch counters, the TLB data fragment to be sent to the Event Builder will contain just a copy of the trigger lines at the different stages of the trigger logic. 

\begin{figure}[tbp]
 \includegraphics[width=0.8\linewidth]{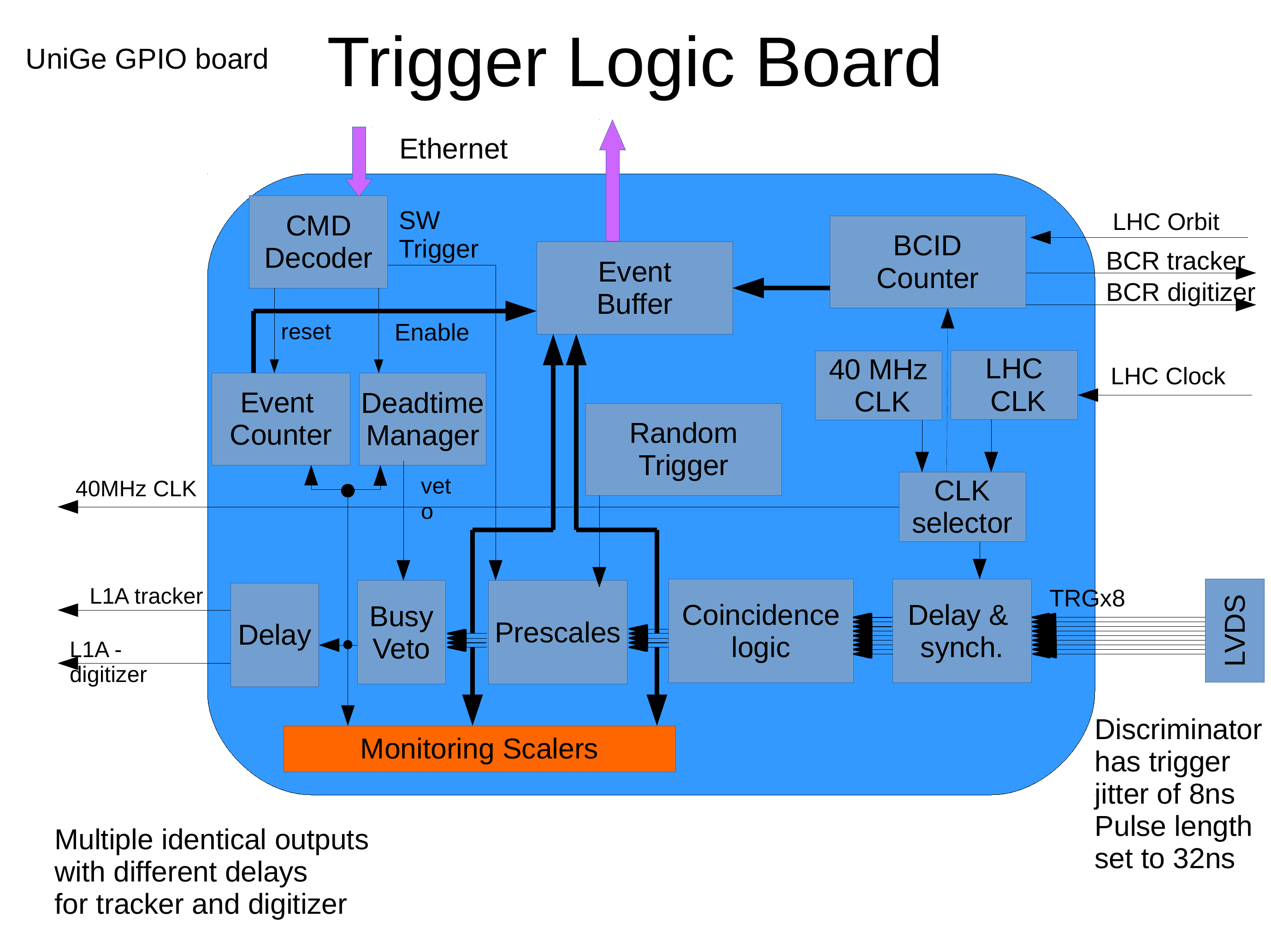}
 \caption{Schematic diagram of the functionality implemented in the Trigger Logic Board}\label{fig:TLB}
\end{figure}

The TLB will be implemented using one ``UniGe USB3 GPIO board'' described in \secref{UniGeGPIO}. The board has sufficient I/O lines (LVDS and Ethernet) and the functionality will be implemented by firmware in the Cyclone V FPGA. A detailed evaluation of the required FPGA resources to implement the above described trigger system still has to be done. However, the FPGA contains more logic cells than the combined sum of the FPGAs used to implement the original ATLAS CTP~\cite{Ask:2008zz} which was significantly more complex (160 incoming trigger lines and 256 trigger items). An adapter board will be built to provide the needed input and output connectors as well as holding the board in place in the VME crate next to the scintillator digitizer.

\subsection{Tracker Readout}\label{sec:trackerReadout}

The tracker readout is largely defined by the pre-existing ABCD chips mounted on the tracking modules. Each module requires a minimum of two input lines, a clock and a command line, and two output data lines (one per side of the module), all running synchronized to the 40.08~\MHz LHC clock. On receiving a L1A over the command line, the location of all strips with a hit are read out to a buffer and transmitted over the data lines. For FASER, a strip is considered hit if the strip fired in the previous, current or next bunch crossing with respect to the L1A, though the trigger timing will be adjusted so that for collision signals the L1A should be aligned to the current bunch crossing. Besides the hit data, the output data contains the lowest 4 bits of the event counter and 8 bits of the bunch crossing counter as seen by the module which allows for synchronization checks at the module level. The minimum data size is 228~bits if no hit is present while a typical size of around 260~bits is expected for most events in FASER. The typical readout transmission time is therefore about 6.5~\mus. Up to eight events can be buffered in a module while it is read out, allowing for close-by triggers.

One ``UniGe USB3 GPIO'' board has sufficient logic cells and I/O connections to handle the readout of at least one out of the three planes located in the tracker station and possible all three planes. Each tracker plane has eight modules. Nine (or three) such boards will therefore be used for the three tracker stations. A special interface card will be constructed to provide the connections from the ``UniGe USB3 GPIO'' board to the tracker stations and the TLB. To keep the cables short and maintain a good signal integrity, the boards will be located in a small mini-crate about one meter away from the center of the middle tracker station, just outside the fringe field of the dipoles. The clock, L1A and BCR from the TLB will be received over a Cat6 cable and distributed from a fan-out to the nine read out boards. These will convert the L1A and BCR signals to the appropriate commands for the tracker modules. Additional commands for configuring and calibrating tracker modules can be received over the Ethernet connection and transmitted to selected modules. The data received by the board from the different modules will be merged and appropriate identifying headers, event and bunch counters added, before being transmitted to the Event Builder over Ethernet. No detailed processing of the data in the readout board is foreseen.

The resource requirements for implementing the tracker readout can be estimated by comparing the resources needed in the Chimaera board used in the module QA in \secref{QA} with those of the ``UniGe USB3 GPIO'' board. The Chimaera board has a different, smaller FPGA (Spartan-6 XC6SLX25 FPGA vs Cyclone V A7), but only reads out four modules and has somewhat less functionality than planned for the FASER readout. A schematic of the circuit blocks is given in \figref{CambridgeSchematic} and the comparison of resources is shown in \tableref{cyclone}.

\begin{table}[htb]
\resizebox{0.98\textwidth}{!}{%
\begin{tabular}{|p{5cm}|p{4cm}|p{4cm}|p{4cm}|} \hline
Resource & Chimaera design (4 SCT modules) Spartan-6 XC6SLX25 & UniGe USB3 GPIO board Cyclone V A7 & Ratio (UniGe/Cambridge) \\ \hline
Logic cells/ Logic elements & 24k & 150k & \\ \hline
Equivalent registers & 30k & 226k & 7.5 \\ \hline
RAM & 936 kB & 6860 kB + 836 kB & 8.2 \\ \hline
$18 \times 18$ DSP blocks & 38 & $156 \times 2$ & 8.2 \\ \hline
PLLs & 2 & 7 & 3.5 \\ \hline
Chimaera design usage (4 modules) & \begin{tabular}{l}45\% Logic\\ 50\% RAM \end{tabular} & \begin{tabular}{l} 6.0\% Logic \\ 6.1\% RAM \end{tabular} & \\ \hline
UniGe USB3 library & & \begin{tabular}{l} 1\% Logic\\ 0.5\% RAM \end{tabular} & \\ \hline
Total for 24 modules + USB3 library & & \begin{tabular}{l} 36.0\% Logic \\ 37.1\% RAM \end{tabular} & \\ \hline
\end{tabular}
}
\caption{Comparison of Xilinx XC6 and Intel Cyclone V. The usage of Logic and RAM was estimated from ratio between XC6 and Intel Cyclone V.}
\label{table:cyclone}
\end{table}

\begin{figure}[tbp]
 \includegraphics[width=0.99\linewidth]{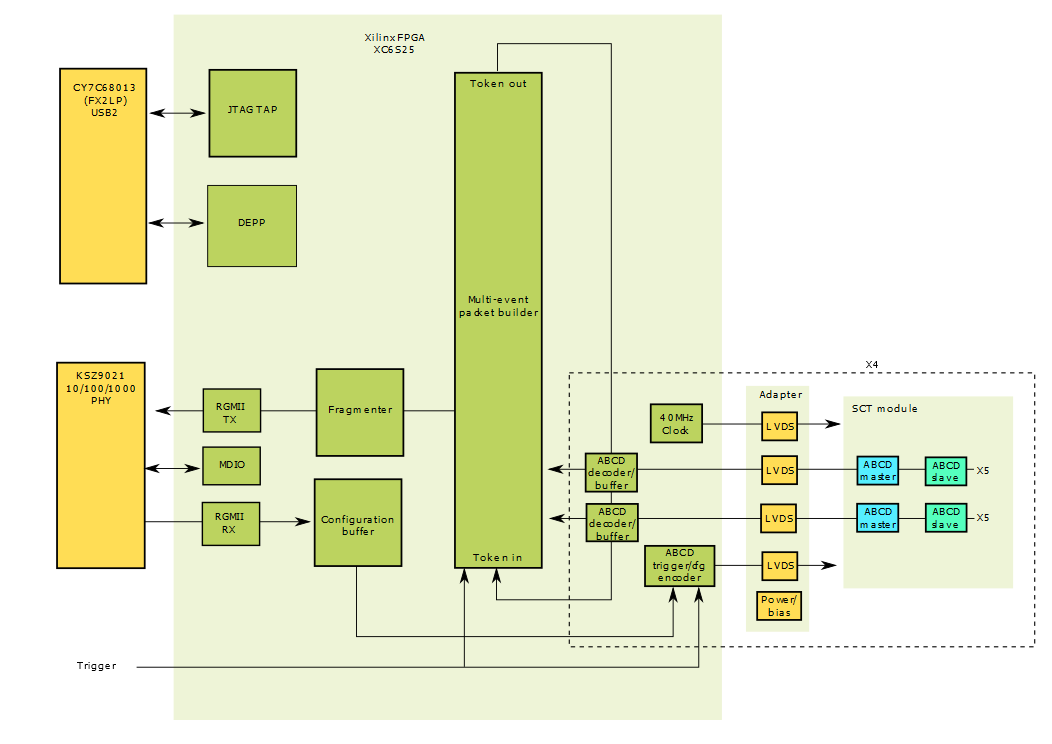}
 \caption{Schematic diagram of the functionality implemented in the Chimaera board for reading out four SCT modules.}\label{fig:CambridgeSchematic}
\end{figure}

Table \ref{table:cyclone} shows that the readout of 24 modules would consume less than 50\% of the Cyclone V resources. This leaves 50\% of the resources for the Ethernet PHY + UDP implementation. While this indicates that a single ``Unige USB3 GPIO'' can control the three planes of a tracker station, the additional requirement for the FASER readout could put a strain on the FPGA resources. Until the preliminary design phase has been completed successfully, it will be assumed that one board will be needed per plane.

\subsection{Event Building, Data-Quality and Storage}

The Event Builder collects the data fragments from the TLB, the tracker readout boards and the PMT digitizer. The first two will push their fragments to the event builder when receiving the L1A, while for the digitizer, the event builder will have to pull the data whenever it receives a fragment from the TLB. The fragments are merged into a single event based on their event counter number. A time-stamp and a run-number will be added and the full event sent to the Event Buffer. The Event Builder will be implemented as a process on a commercial Linux PC located on the surface. 

The same process also controls the overall data acquisition system and will send the configuration to all of the readout boards, enable the trigger and also reconfigure the system in case data corruption is detected. Data corruption is checked by comparing the bunch counter values in the different fragments and inside each tracker module sub-fragment. In case of continuous mismatch or missing data fragments, the system will reconfigure itself and in case this does not help, it will contact the expert on-call. The continuous rate of random triggers ensures that the system is maintained in good running condition even when there are no collisions and therefore no physics triggers.

The Event Buffer is a separate process, normally on a second PC, which collects events with the same run number into files on disk with a convenient size for archiving and analysis (about 1 GByte per file). As the event size is dominated by the PMT readout and most of that data should just be pedestal values given the large readout window, each event will be compressed using for example the Lempel-Ziv algorithm~\cite{Ziv:2006:UAS:2263324.2269180}. This should reduce the event size by at least a factor 2, allowing one file to hold about 5 minutes of data. The files will be copied asynchronously to the CERN CASTOR storage area and a backup at another FASER institute.

The latest event received by the Event Buffer will also be available for sampling by debugging and data quality tools running on the same PC as the Event Buffer. The data-quality will be monitored by a fast offline-based event reconstruction which produces a small set of
monitoring histograms for expert monitoring and automatic checks against reference histograms. 

For redundancy and to minimize the risk of interference, the Event Builder and the Event Buffer runs on two separate Linux PCs located on the surface. In case of a hardware failure it will be made possible to run both processes on a single PC. Each PC will be server-grade rack-mounted PCs with redundant power supplies, at least 16~GB of memory, at least eight cores and 2~TBytes of raid-storage. The latter provides enough storage for one week of data taking at the nominal trigger rate. 

Design of the data-acquisition and detector control software has not yet begun, but it is expected a simple web-based interface will be used for day-to-day monitoring and control.

For security reasons, the readout and control system will be on a private network, where only the PCs will be connected to the CERN general purpose network (GPN) and there will be no direct connection to the wider internet. The connection to CASTOR will be over the GPN as well.

\subsection{Detector Control System (DCS)}

The DCS is responsible for controlling and monitoring all of the different power supplies and electronics crates, as well as monitoring temperatures and humidity in the different detector parts. All devices will be connected through Ethernet to a dedicated DCS Linux PC running the standard LHC SCADA system WinCC OA (formerly known as PVSS).

\subsection{Cost and Schedule}

The cost of the different components is estimated in Table~\ref{table:tdaqBudget}. At least one spare card of each module of the system, one spare VME crate, a spare readout PC, and network switch will be needed in case of a failure. These spares will also be used to form a small test-stand for testing new firmware and software developments. The cost of the spares is estimated to be about 28~kCHF.

\begin{table}[tbp]
\centering
\begin{tabular}{|l|c|}
\hline
\  {\bf Trigger system} \ &  \ {\bf Cost [kCHF]} \  \\ \hline
\ VME Crate \ &  5 \quad  \\ \hline
\ Digitizer \ &  9 \quad  \\ \hline
\ BOBR and fiber \ &  0.5 \quad  \\ \hline
\ Clock Cleaner \ & 0.5  \quad  \\ \hline
\ Trigger Logic Board \ & 1.5 \quad  \\ \hline
\ Trigger fan out \ &   1\quad  \\ \hline
\ Trigger Cables \ &   1\quad  \\ \hline
\  {\bf Trigger system Total} \ & {\bf 18.5}  \quad  \\ \hline 
& \\ \hline
\  {\bf Data acquisition system} \ &  \ {\bf Cost [kCHF]} \  \\ \hline
\ Mini Crate \ &  2 \quad  \\ \hline
\ Tracker readout boards \ &  11 \quad  \\ \hline
\ Data cables to tracker \ &  1 \quad  \\ \hline
\ VME to Ethernet converter \ &  3 \quad  \\ \hline
\ LV power supply \ &  2 \quad  \\ \hline
\ Three readout and control PCs \ &  9 \quad  \\ \hline
\  {\bf DAQ Total} \ & {\bf 28}  \quad  \\ \hline 

& \\ \hline
\  {\bf Networking} \ &  \ {\bf Cost [kCHF]} \  \\ \hline
\ 2 Ethernet switches \ &  6 \quad  \\ \hline
\  {\bf Networking Total} \ & {\bf 6}  \quad  \\ \hline 

& \\ \hline 
\ {\bf Total} \ & {\bf 52.5}  \quad   \\ \hline
\end{tabular}
\caption{Budget for trigger/DAQ system for the FASER experiment.}
\label{table:tdaqBudget}
\end{table}

The expected schedule for the construction and commissioning of the FASER trigger and DAQ system is shown in \figref{TDAQTimeLine}. 

\begin{figure}[tbhp]
\centering
\includegraphics[clip, trim=0.5cm 8.0cm 0.5cm 0.5cm,width=0.95\textwidth]{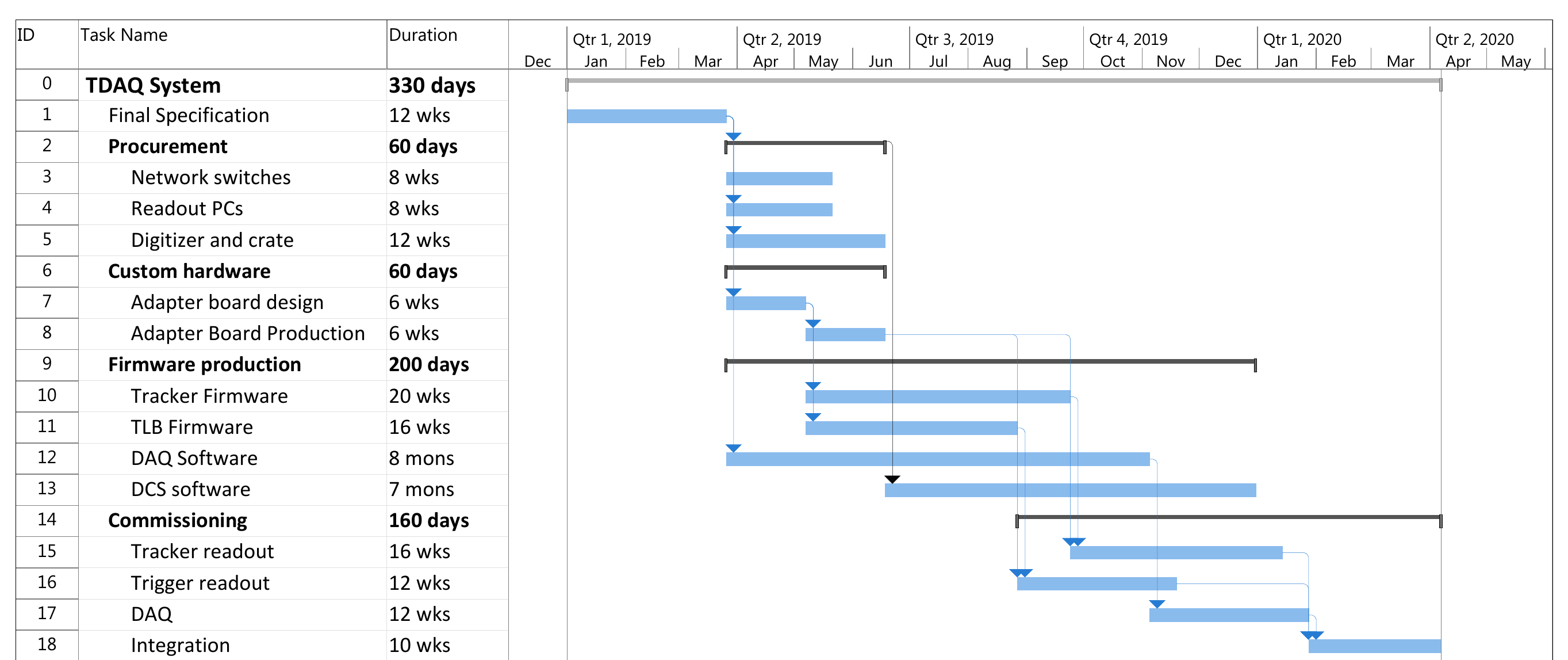} 
\caption{\label{fig:TDAQTimeLine} Timeline for the Trigger and DAQ construction and commissioning.}
\end{figure}

\clearpage
\section{Civil Engineering}
\label{sec:civil}

The experiment is to be located within TI12, a former injection tunnel that connected the Super Proton Synchrotron (SPS) to the Large Electron--Positron (LEP) Collider at CERN's Meyrin site just outside Geneva. TI12 is located just off the LHC, entirely within Switzerland as shown in \figref{Map}. The tunnel is currently disused due to the greater radii required for beam injection into the LHC.  To accommodate the experiment within the chosen location in TI12, a certain amount of civil engineering (CE) work will be required. This section details the existing structure, the civil engineering requirements, the design process, the  planned works methodology, and further work recommended to implement FASER. 

\subsection{Existing Structure}

\begin{figure}[bp]
\centering
\includegraphics[width=0.95\textwidth]{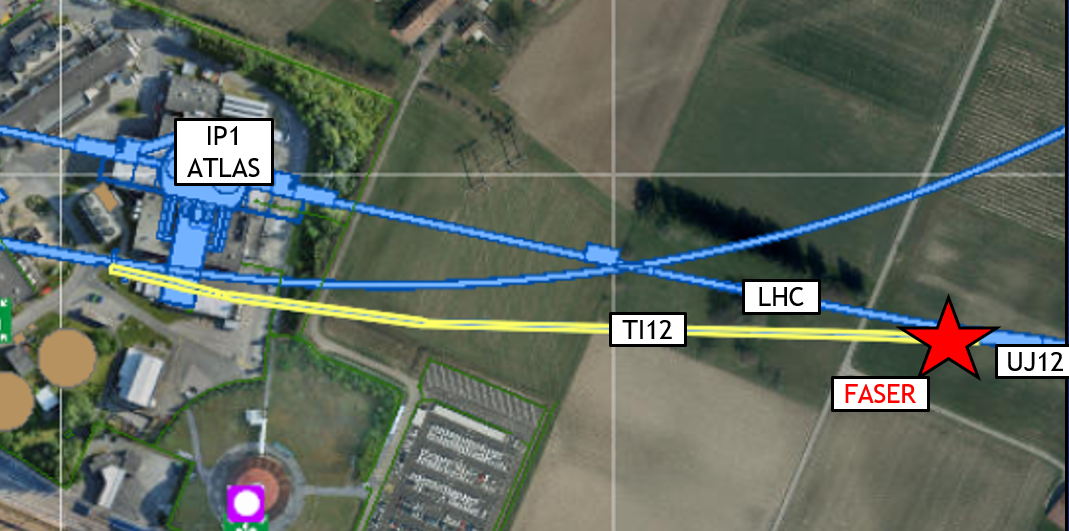} 
\caption{Location plan showing the proposed situation of FASER in TI12}
\label{fig:Map}
\end{figure}

TI12 is a mined tunnel with a horseshoe shaped tunnel cross-section. The tunnel is 2.94~m in width at floor level and is 2.9~m tall from floor to crown. The tunnel structure consists of a 100~mm-thick spray applied shotcrete lining, rock bolts projecting into the surrounding Molasse bedrock, as well as a 250~mm-thick cast {\em in situ} invert and secondary lining as shown in \figref{ABXSec}.

\begin{figure}[tbp]
\centering
\includegraphics[width=0.95\textwidth]{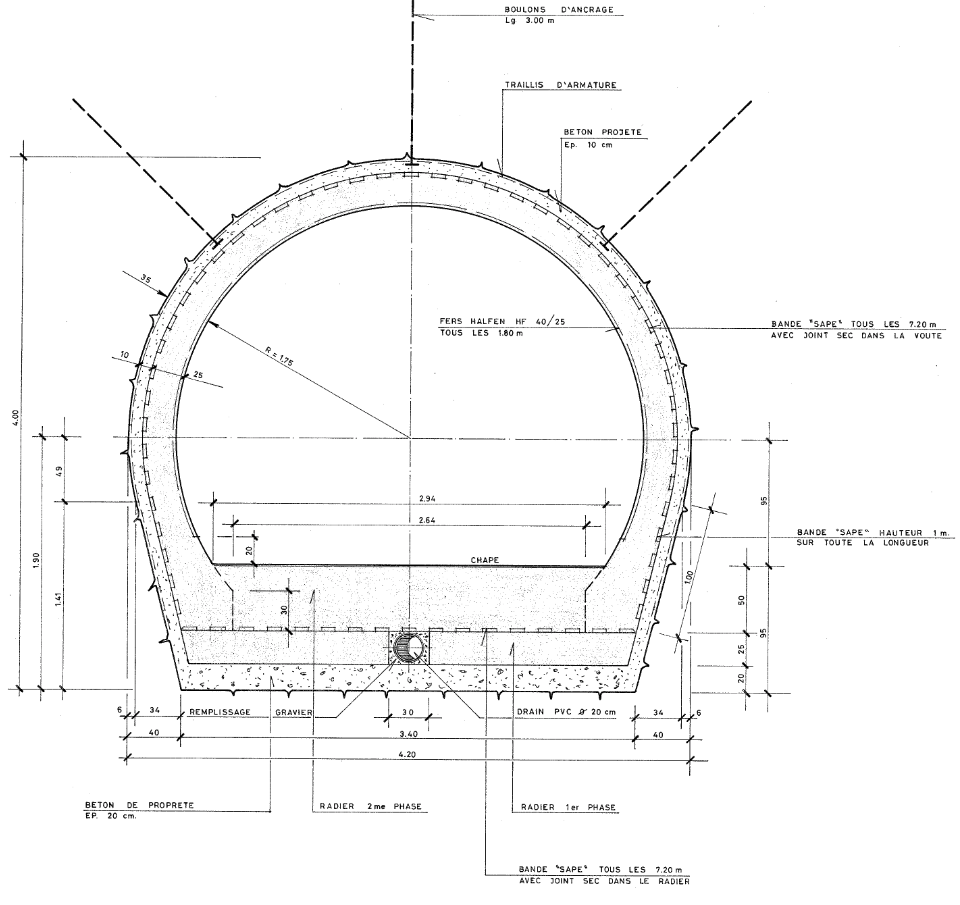} 
\caption{As-built typical section of tunnel TI12 showing form of construction.}
\label{fig:ABXSec}
\end{figure}

The invert of the tunnel is formed of a shotcrete base 200~mm deep with a 250~mm deep first phase slab cast {\em in situ} beneath a final cast {\em in situ} concrete infill slab 500~mm deep. Within the first phase invert slab, there is a 200~mm diameter PVC drain surrounded by gravel infill running longitudinally down the centre of the tunnel. Between the outer layer of shotcrete and the cast {\em in situ} tunnel structure, there is a waterproofing membrane,  allowing any groundwater seepage to drain into the longitudinal drain. Towards the bottom of tunnel TI12, there is a transverse drainage channel or caniveau, as shown in \figref{ABPlan}.  The caniveau connects into the longitudinal drainage system.

\begin{figure}[tbp]
\centering
\includegraphics[width=0.95\textwidth]{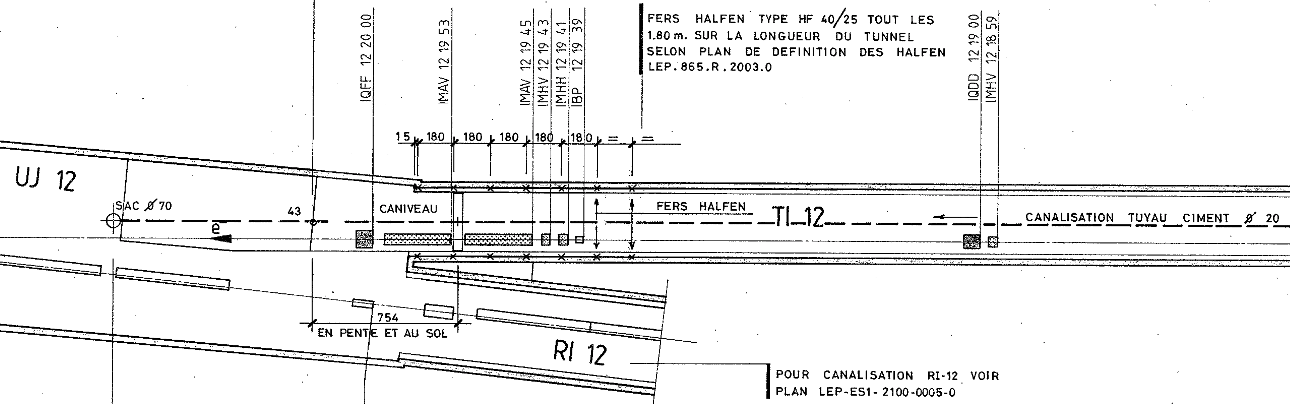} 
\caption{As-built plan view of tunnel showing arrangement of existing drainage infrastructure.}
\label{fig:ABPlan}
\end{figure}

\subsection{Requirements}

The axis of the LOS crosses TI12 at an oblique angle to the tunnel in plan and emerges through the tunnel floor close to TI12's junction with cavern UJ12. A trench will be required to create the space required for the experimental arrangement so it can be positioned centrally along the axis of the LOS. The dimensions and position of the trench in relation to the tunnel floor and the LOS are shown in \figref{trenchsection}.

\begin{figure}[t]
\centering
\includegraphics[width=0.95\textwidth]{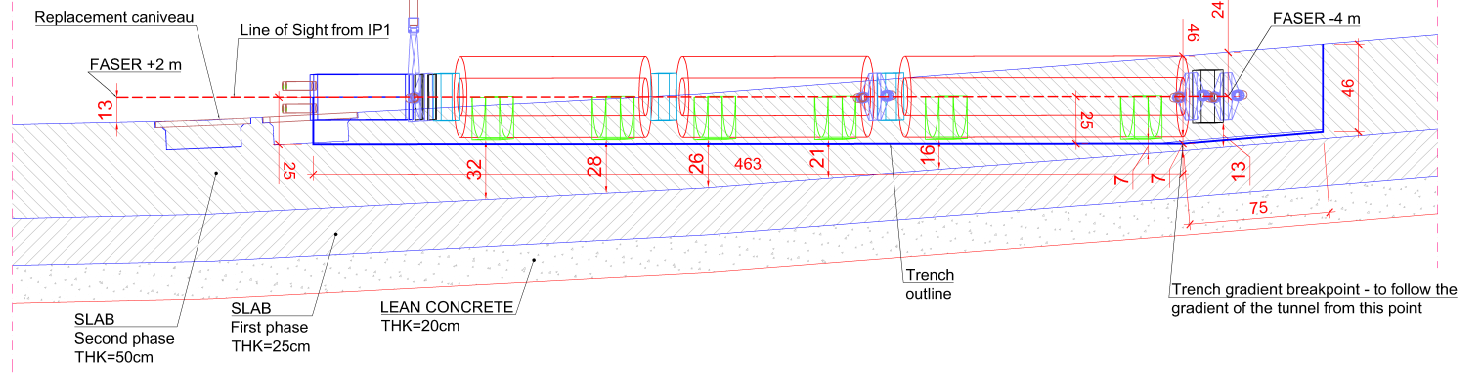} 
\caption{Section showing trench and experiment in relation to TI12 floor and LOS.}
\label{fig:trenchsection}
\end{figure}

The location of the trench and experiment interferes with existing drainage systems and will necessitate the removal of an existing transverse drain. This will need to be replaced with two new transverse drainage channels, above and below the experiment. The new drains will need to be connected into the existing tunnel drainage systems, which in turn outfall into those in UJ12.

In case of any unforeseen seepage through the existing tunnel structure, the trench for the experiment will include an outlet connection to the lower transverse drain to ensure no water can build up within the trench.

As part of the project, CERN's transportation team will also need to be involved, both during the works and for installation and maintenance of the experimental arrangement. To enable their works, lifting points will need to be installed in the tunnel soffit at regular intervals.

\subsection{Design of Modifications}

\begin{figure}[t]
\centering
\includegraphics[width=0.95\textwidth]{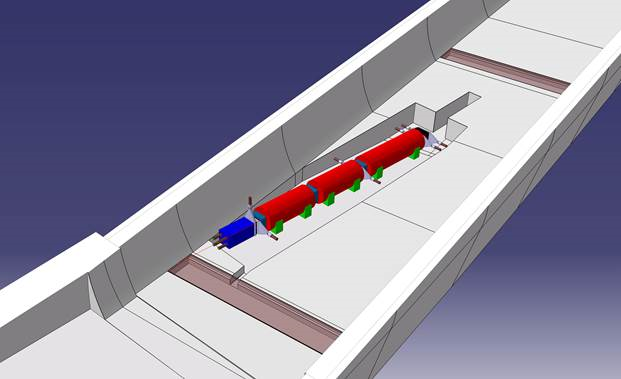}
\caption{View showing extents of trench and proposed drainage in relation to tunnel TI12.}
\label{fig:3Dmodel}
\end{figure}

An outline feasibility design has been carried out to consider what CE enabling works are required for FASER. To enable the feasibility study, a 3D scan of the area has been carried out and matched with the existing 3D model of TI12 to corroborate the accuracy and position of the tunnel. A 3D model of the experimental arrangement has been positioned along the axis of the LOS (provided by CERN's Survey team) in combination with the 3D model of the tunnel shown in \figref{3Dmodel}. This has allowed the interaction between experiment and civil engineering infrastructure to be studied at a preliminary level. Further study will be required: in particular, a detailed structural analysis will be needed to confirm the structural stability of the proposed arrangement. The proposed design will continue to be optimised as all constraints and requirements are clarified. The current preliminary design is shown in \figref{Planview}.

\begin{figure}[t]
\centering
\includegraphics[width=0.95\textwidth]{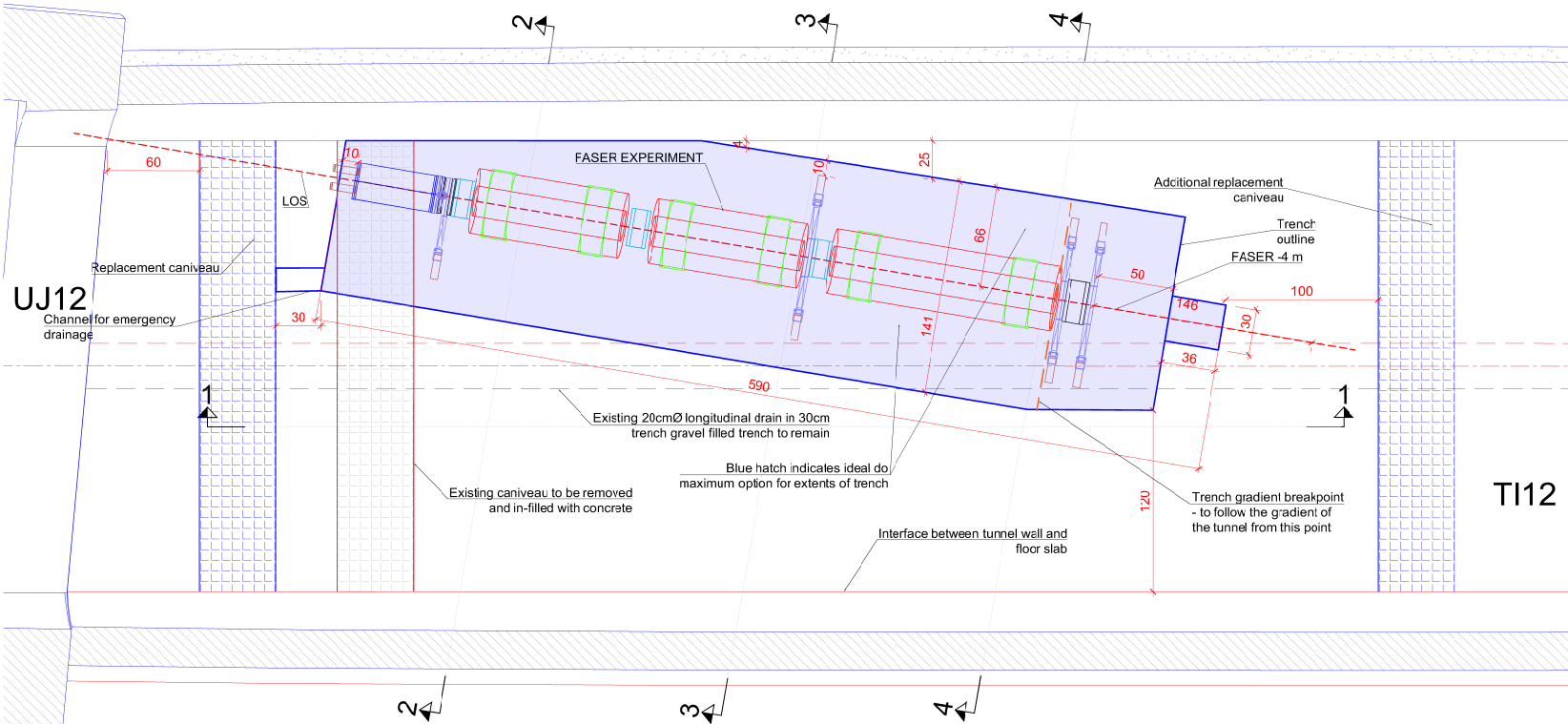} 
\caption{Plan view showing proposed arrangement. }
\label{fig:Planview}
\end{figure}

The proposed trench dimensions have been determined to allow for the installation and maintenance of the experimental arrangement. These dimensions have been reduced as far as possible to minimise the impact on the existing tunnel's structural elements. The trench will still have an effect on the stability of the tunnel, however, which may still require some local strengthening. The location of the experiment in relation to the structure of the tunnel is illustrated in \figref{Sections}.

\begin{figure}[t]
\centering
\includegraphics[width=0.95\textwidth]{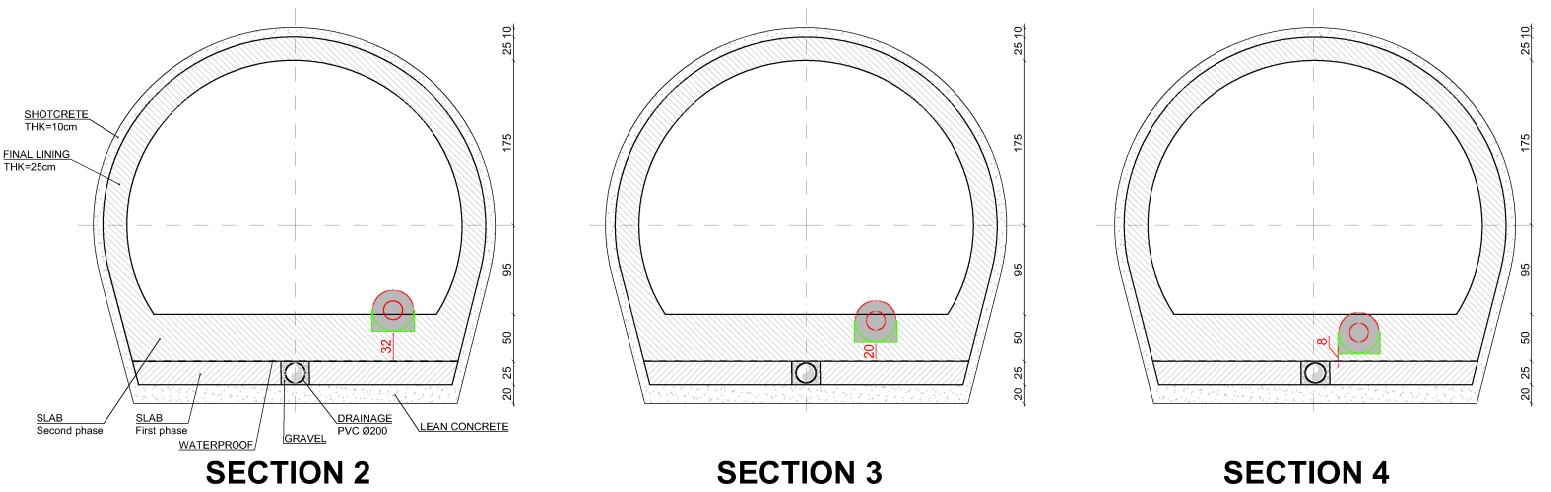} 
\caption{Proposed sections showing depth of experiment along LOS in relation to the structure and drainage of  tunnel TI12 (trench omitted).}
\label{fig:Sections}
\end{figure}

The design has been carried out with a view to the construction works being undertaken during Long Shutdown 2 (LS2). This has meant aiming to minimise the complexity of civil engineering work required so it is feasible to plan, design, and implement the works within a suitable timeframe. To avoid more significant works, for example, the depth of trench has been limited to avoid damage to the longitudinal drain and drainage membrane, 500 mm below the tunnel invert. The position of the experiment along the line of sight has also been optimised with the aim of avoiding the need for openings or cores into the tunnel arch barrel. 

Following finalization of the experimental requirements, a full structural stability analysis will be undertaken. The analysis will be carried out in line with Eurocode design standards. The assessment will consider the removal of concrete as part of the trench and drainage along with the lifting points and associated loading required to transport, install, and maintain FASER's equipment. The analysis will determine whether there is any effect on stability and whether any modification is required to the design to enable the project. At this stage, it is considered likely that some local strengthening will be required. This may consist of additional rock bolts or installation of steel reinforcement in the tunnel invert, but analysis and design will be needed to confirm provisions.

The drainage systems will be designed in line with standard CERN specifications and will be of the same or greater standard as the existing system. 

The CERN Radio-Protection (RP) group has been consulted and will advise on working methods and waste disposal as the project develops. The concrete in this location should be considered as potentially activated. At the beginning of LS2, representative sample(s) will be taken for analysis to measure the level of activation.  

\subsection{Construction Methodology}

This type of construction work is not unusual at CERN, meaning that the work can be carried out by existing framework contractors, which will negate the need for lengthy procurement processes. The works will consist of the following main activities:

\begin{itemize}
\item Isolate works area (and provide temporary services and ventilation);

\item Install transport lifting fixtures in tunnel soffit;

\item Saw cut new trenches;

\item Remove existing transverse drain and repair tunnel infill slab;

\item Install transverse drains and connect to existing longitudinal drains;

\item Undertake strengthening works;

\item Remove concrete to 40~mm below finished level;

\item Cast screed base to trench to level;

\item Clear site removing all arisings and isolation.

\end{itemize}

Concerns have been raised about the works creating dust with the potential to affect existing CERN operations. This will be actively managed throughout the works to avoid any issues. As noted, the first activity will be to isolate the works area. This will be achieved by putting in place a double or triple skin polythene sheet wall system (or a `SAS' as they are commonly known at CERN) to prevent any dust from escaping the works area. \figref{SAS} shows an example system as used in civil engineering work in the AWAKE experimental area at CERN. Filtration systems will be put in place to scrub dust from the air. A negative pressure will be created in the works area as necessary to ensure any small quantity of dust created remains within the confined zone. Suitable respiratory equipment will be utilised where required by construction workers. Framework contracts are in place at CERN with companies experienced in delivery of such confinement works. The methods that will be used are industry standard in the field of asbestos removal and treatment.

The dust created will also be kept to an absolute minimum. Dust suppression techniques will be applied to all activities with the potential to create fine airborne material. Water suppression will be used during diamond saw cutting to prevent mobilisation of dust, with arisings and run off contained and collected. Following saw cutting, typical concrete excavation/demolition techniques using a mechanical breaker will be substituted for rig-mounted diamond coring. \figref{SAS} shows an example of the coring tool to be used. The trench and drainage channels will be cored out to a suitable depth with drill-powered coring rigs in combination with water suppression, waste water collection and filtration systems, producing a minimum of airborne material. This change of technique will have the dual benefit of producing less dust and allowing more accurate control of the depth of excavation to ensure the drainage membrane remains intact. Transportation fixings will be installed by means of rig mounted drills with dust extraction/collection systems, followed by resin anchoring fixings into the tunnel soffit.

\begin{figure}[t]
\centering
\includegraphics[width=0.435\textwidth]{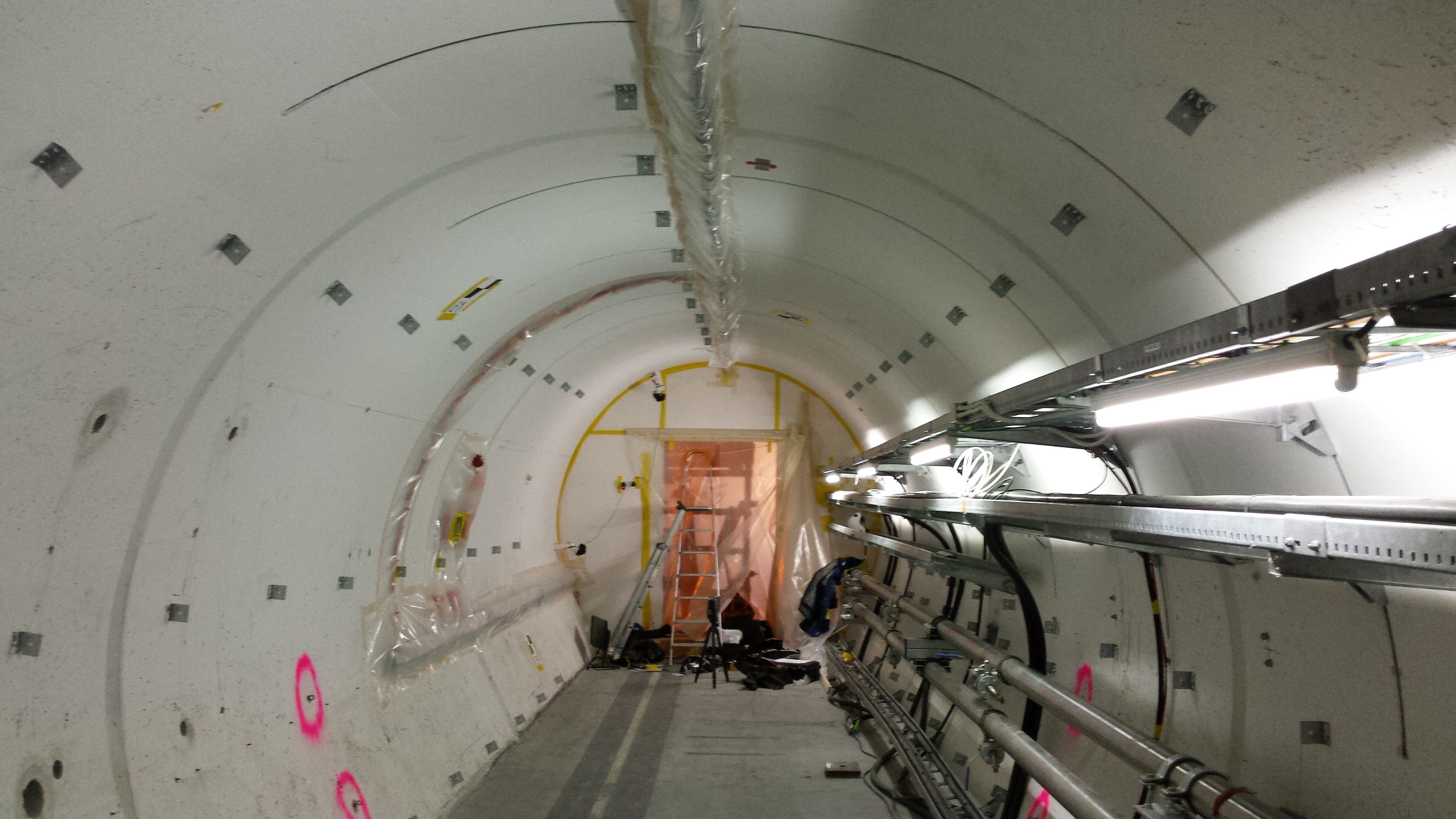} 
\hfill
\includegraphics[width=0.54\textwidth]{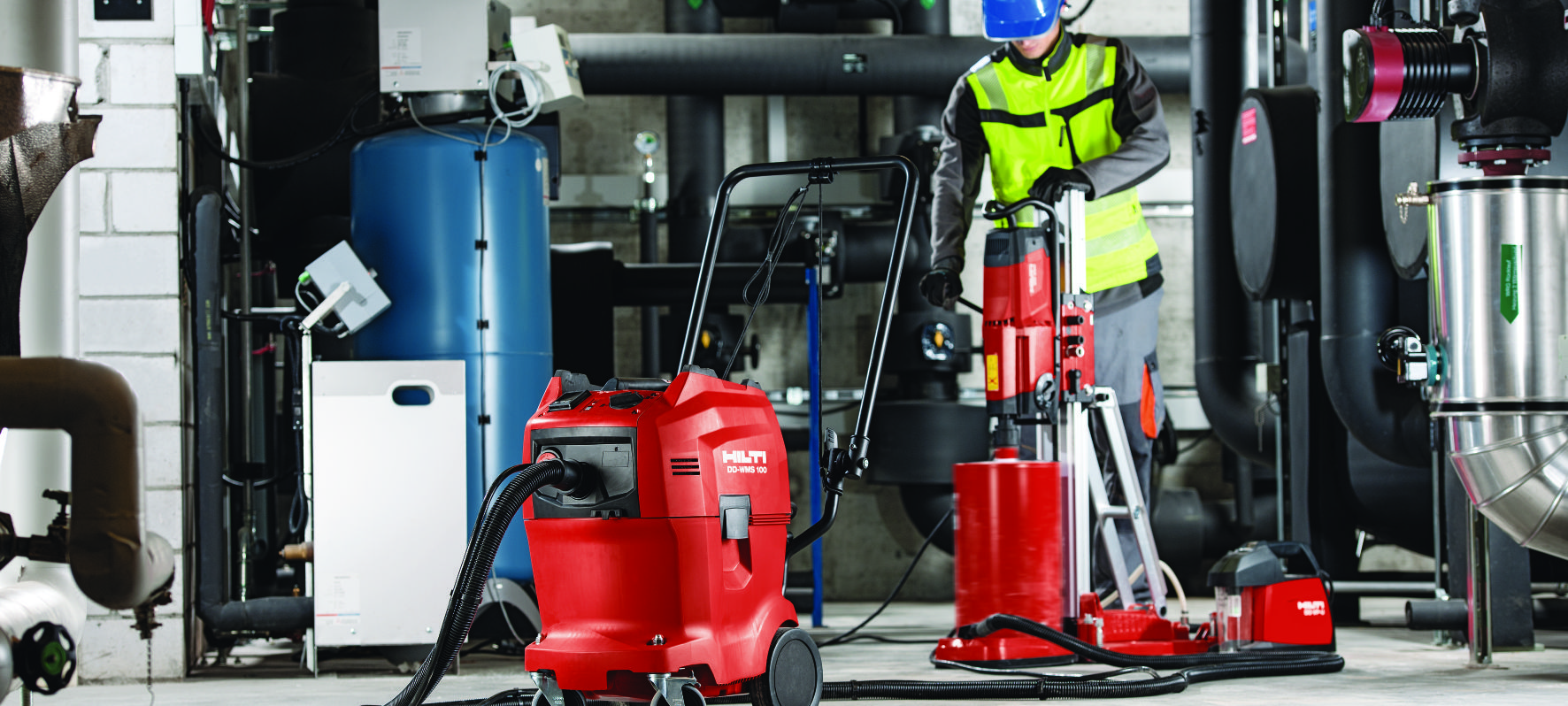} 
\caption{Left: Example of the SAS dust protection system as used in CE works at the AWAKE experimental area at CERN. Right: Example of the coring tool to be used for the CE works.}
\label{fig:SAS}
\end{figure}

CERN's RP department have advised that the use of a sealed SAS will be sufficient to prevent any spreading of potentially activated dust. Further planning will be carried out in conjunction with the RP team.

Prior to leaving the site, the site will be thoroughly cleaned. Air testing to confirm results can be carried out to ensure no dust remains. All systems and procedures will be agreed with the relevant parties in advance of works.

Further study will be necessary to confirm the exact arrangement, particularly in terms of ventilation, but the techniques detailed are feasible, in widespread use within the construction industry, and have been used numerous times in similar situations at CERN.

\subsection{Required Further Studies}

Some additional studies will need to be carried out in advance of construction, including:
\begin{itemize}
\item A suitably detailed structural analysis of the existing tunnel should be carried out to confirm that the works will have no impact on tunnel stability or to allow the design of local strengthening measures necessary.
\item A ground penetrating radar survey of the existing tunnel invert structure should be undertaken to confirm the exact depth of the existing drainage and drainage membrane to facilitate works and confirm the design.
\item A CCTV survey of the existing drainage systems should be carried out to confirm their condition and operation prior to connection.
\item Representative samples will need to be taken to determine the level of concrete activation. 
\end{itemize}

\begin{figure}[t]
\centering
\includegraphics[clip, trim=0.5cm 4.9cm 0.5cm 0.5cm,width=0.95\textwidth]{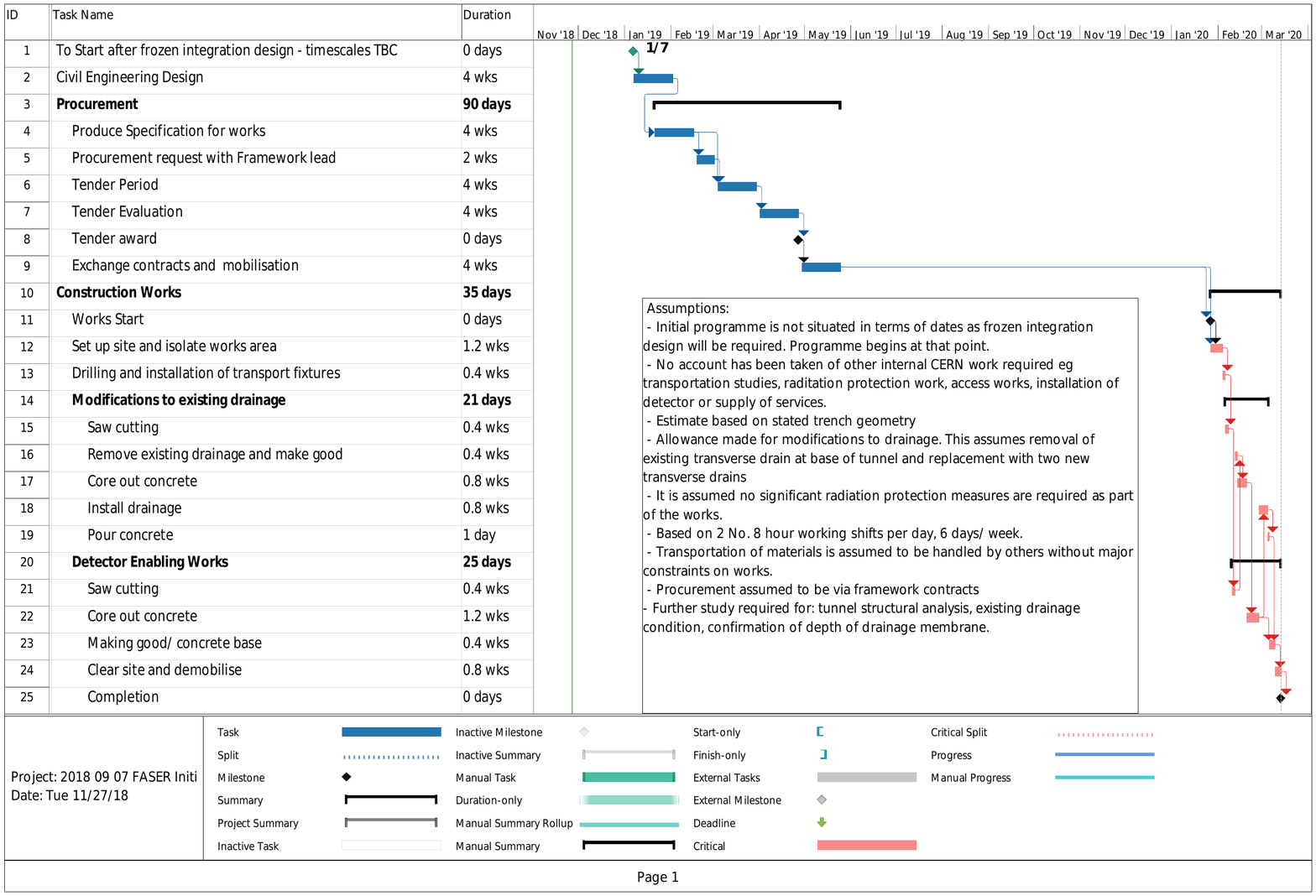}
\caption{Draft schedule for civil engineering work to be done in 2019.}
\label{fig:civilEngineeringTimeLine}
\end{figure}

\subsection{Cost and Schedule}
\label{sec:Estimate}
 
The cost for the civil engineering work is estimated as 83.3~kCHF. This has been based on the layout presented in this chapter and includes costs for detailed design, construction, and construction management, excluding personnel costs for CERN resources. The estimate does not include development costs, materials or personnel costs in advance of detailed design and construction. The estimate and schedule do not include strengthening works, as these cannot be assessed until requirements are confirmed by structural analysis. Rates used have been based on existing framework rates with some tweaking based on comparative analysis of other similar schemes, as well as knowledge of the specific requirements of FASER.

The cost and has been estimated using the following assumptions:
\begin{itemize}
\item Costs and schedule are based on the design as detailed without allowance for significant tunnel strengthening works if required.
\item The basis of the estimates is that the trench is no deeper than 460mm and does not interfere with drainage.
\item The proposed drainage can be connected into existing tunnel drainage without significant repairs or maintenance of the existing.
\item No allowance has been made in tunnel geometry for transport, logistics, fire compartmentalization, access staircases, or other alcoves.
\item Provision of tunnel services, e.g., ventilation, electricity, etc. are not included.
\item No allowance made for network service diversions.
\item Costs are based on dust suppression methods as detailed earlier in this section.
\end{itemize}
 
The accuracy of the cost estimate is Class 4---Study or Feasibility which could be 15-30\% lower or 20-50\% higher (in line with Ref.~\cite{christensen2005cost}, as has been used for previous CERN projects). Until the project requirements are further developed, it is suggested that the maximum band be adopted, i.e., $-15$\% to +50\%.

The above costing assumed that the works could be done with one shift per day, and no work on weekends. Given scheduling constraints in LS2, it is now known 
that the work will need to be done with two shifts per day and work on Saturdays. In addition some work may need to be done with night shifts as when survey work is ongoing in sector 81, CE work will not be possible during the day. Based on this the cost is roughly expected to double to 160~kCHF (with a larger uncertainty than discussed above).
 
The estimated schedule for the work is shown in \figref{civilEngineeringTimeLine}. The Civil engineering work in TI12 is constrained to week 5-10 in 2020 due to other work already scheduled in the LHC, additional CE works can be done in weeks 11-14 with the work taking place at night due to ongoing survey work in the daytime.

\clearpage
\section{Installation and Integration}
\label{sec:installation}

\subsection{Transport}

The biggest and most complicated detector component to transport is the 1.5~m-long, 50~cm-diameter, 1.5~tonne permanent magnet. The lift at point-1 can carry up to 3 tonnes, and the roof bracket at TI12 can support 8 tonnes. It is therefore not expected to be a problem to transport such an object to TI12. Analysis of the laser scan in the TI12 region indicates that there is sufficient room above the LHC machine to carry the magnet over the machine (and the corresponding QRL cryogenic line) using a simple crane attached to the roof bracket of the main tunnel. The magnet would be carried along the LHC tunnel using an electric tractor. For this, care must be taken as the trolley is made of steel, so the magnetic field must be taken into account in the detailed transport work planning. 

The CERN transport group have studied the transport scenario; \figref{transport} shows two pictures from the 3D model of the transport work. This work requires the following tooling:
\begin{itemize}
\item A 2-3t hoist for the rail into UJ12 (the original ones have been removed and reinstalled on UJ62),
\item Two rails into TI12 (design, production installation), and the transverse bar to make a simplified crane,
\item Two 2t hoists to be installed on this rail.
\end{itemize}
This set of tools will be installed and tested at full load in collaboration with the CERN safety group (HSE). 
The whole of the process of installing and testing the tooling will take 2 weeks, and is estimated by the transport group to cost 50~kCHF. 
In addition a protection will be installed over the QRL cryogenic line, and the LHC dipole magnet to act as a final protection in case of a failure during the transport. A preliminary design of this is shown in~\figref{QRLprotection}. This has not been costed, but for now we assume a rough cost estimate of 40~kCHF. 

\begin{figure}[tbp]
\centering
\includegraphics[width=0.95\textwidth]{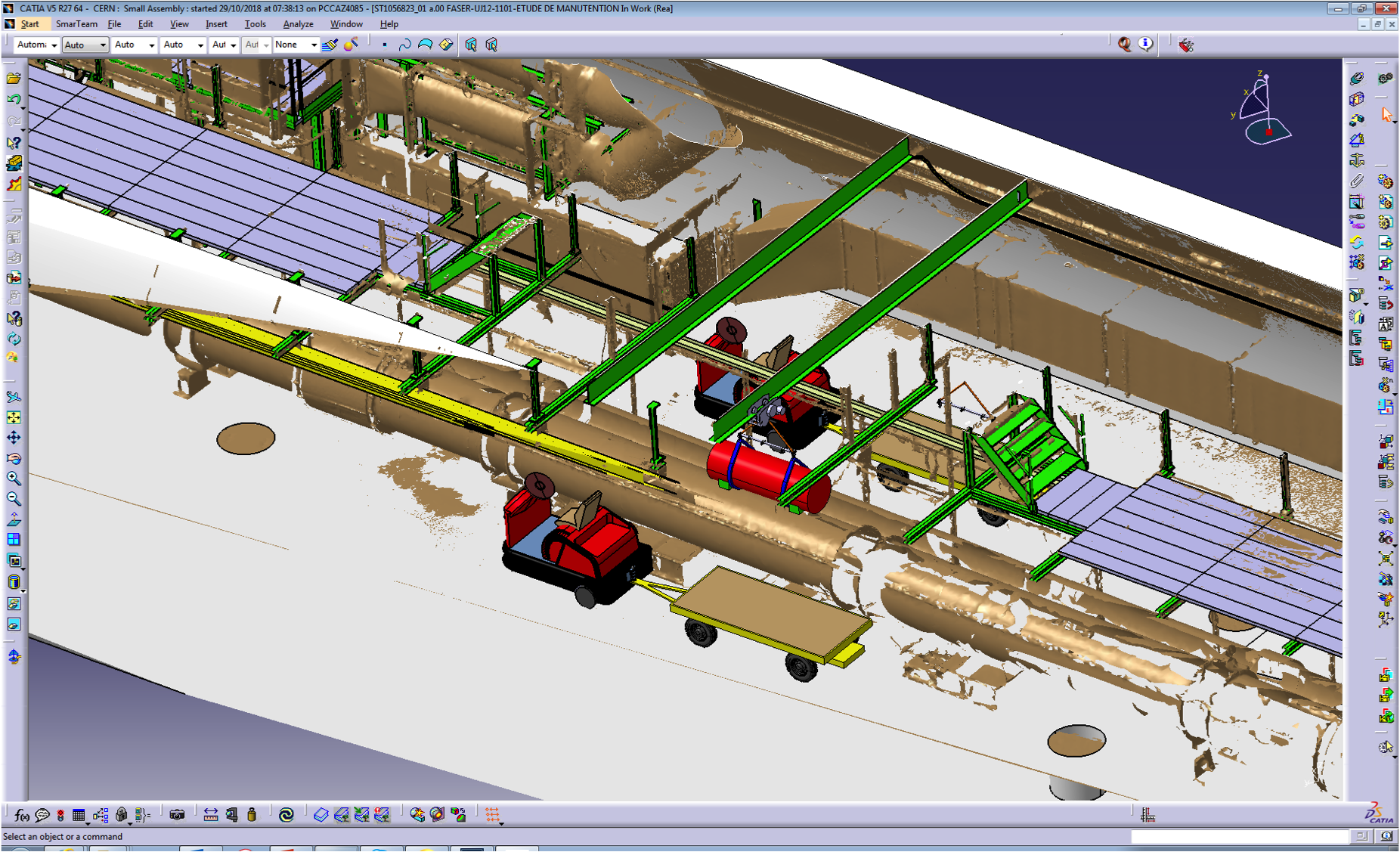}
\includegraphics[width=0.95\textwidth]{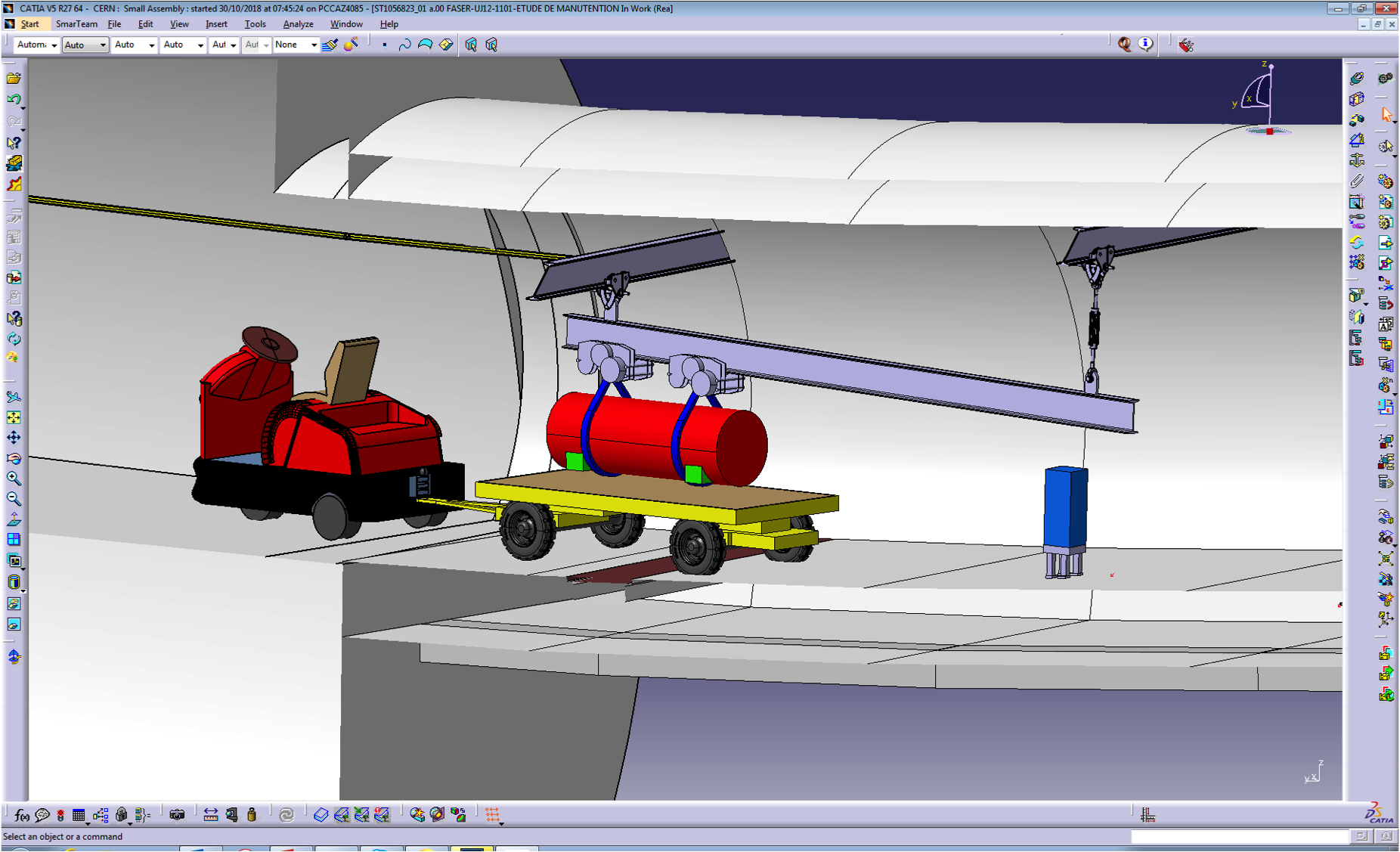}
\caption{Pictures from the study of transporting the detector components of FASER to the detector location. Both pictures show the transport of the largest magnet. The top picture shows how this will be lifted over the LHC machine from one tractor to another. The bottom picture shows how this will be transported into position in the TI12 tunnel.}
\label{fig:transport}
\end{figure}

\begin{figure}[tbp]
\centering
\includegraphics[width=0.95\textwidth]{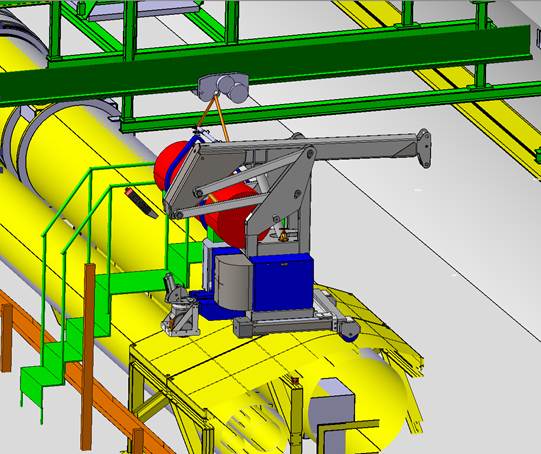}
\caption{A preliminary design of the protection to be installed over the QRL cryogenic line, and the LHC dipole magnets for safety during the transport of FASER components over the machine.}
\label{fig:QRLprotection}
\end{figure}

The transport of the FASER detector components is estimated to take 3 days, and cost 5~kCHF for personnel (this is assuming the transport of 10 pieces).  This work needs to be integrated into the master LS2 schedule to avoid periods when work is ongoing in the LHC tunnel between IP1 and TI12. 

For the transport and installation of the detector, unused ventilation tubes in TI12 need to be removed. This work will be done by the relevant CERN group (EN-CV) and is estimated to take about 1 week, and cost 4~kCHF. Unused cable trays in UJ12 and TI12 will also need to be removed, which can be done by the relevant group (EN-EL).

The access at point-1 is through the PM15 LHC shaft and the bypass tunnel, which joins the LHC beam line just after the triplet magnets, thus bypassing the ATLAS cavern.  This means FASER work will not interfere at all with any LS2 work of the ATLAS experiment, which has its own access shaft. 
Transporting the digging tools for the civil engineering work, and removing the spoil from the excavation (estimated to be $\approx 3 $m$^3$) is 
not included in the above cost estimates yet.

\subsection{Services}

The following services are needed for FASER:
\begin{itemize}
\item Cooling: As discussed in \secref{cooling}, cooling will be provided by a standalone water chiller situated in the TI12 tunnel close to the detector;
\item Dry air: As discussed in \secref{cooling}, compressed air with a dew point below $-40^\circ$C will be provided by EN-CV. The estimated cost of connecting the compressed air system into FASER at TI12 is 6~kCHF and it is one day's work, although this needs to be planned in the LS2 schedule;
\item Power: The estimated power consumption for FASER is 3.4~kW (a breakdown is shown in Table~\ref{tab:power}). After discussion with the relevant CERN group (EN-EL), the following scheme for powering the experiment is assumed, although its feasibility remains to be confirmed. The powering would use the 400~V at 16~A outlet available in UJ12 about 50~m from FASER. EN-EL would lay 50~m of cable and install a switchboard with a master breaker and 3 breaker circuits separated for the cooling, readout, and powering. Two emergency cutoff switches (AUG) would also be installed at either end of the detector. This solution is estimated to cost about 10~kCHF, including installing lighting in TI12. The above solution would not provide UPS power for FASER, which would be significantly more expensive. The need for UPS power is still under evaluation. 
\item Signals, readout and networking: Data will come in/out of the experiment via 3 dedicated optical fibers linking TI12 to the surface building in point-1 (SR1). These fibers will be installed during LS2 by EN-EL along with several spare fibers (TBC). Two fibers will connect Ethernet switches for data and commands (this gives sufficient bandwidth for the experiment readout as discussed in \secref{tdaq}), while the third fiber provides the BST signal from the BE-BI group to the BOBR module in TI12. On the surface, rack space will be needed for an Ethernet switch and the four readout/controller PC's. The Ethernet switches will be provided and managed by CERN-IT (the cost for this is included in the TDAQ budget) and provides a private network for FASER which is not directly accessible from outside the CERN network. The surface switch will be connected to the CERN general purpose network (GPN) in SR1 and will be configured to only provide access to and from the GPN for the three TDAQ PCs. The exact surface location and connection points are still under discussion.
\end{itemize}

\begin{table}[tbp]
\centering
\begin{tabular}{|l|c|c|c|}
\hline
\  {\bf Component} \ &  \ {\bf Power [W]} \  \\ \hline
SCT module	&	540 \\ \hline
SCT readout	&	450 \\ \hline
SCT power supply	&	200 \\ \hline
Chiller	&	1620 \\ \hline
PMT HV	&	180 \\ \hline
PMT Readout	&	60 \\ \hline
Trigger logic	&	50 \\ \hline
VME Crate	&	120 \\ \hline
Network switch	&	180 \\ \hline
{\bf Total}	&	{\bf 3400} \\ \hline
\end{tabular}
\caption{Expected power consumption of FASER experiment in TI12.}
\label{tab:power}
\end{table}

\subsection{Integration}

The CERN integration team have started to develop a model for the TI12 area, including the FASER detector and its associated infrastructure (chiller, electronics, power supplies, etc.). \Figref{integrationModel} shows a screen-shot from the integration model of the TI12 area.

\begin{figure}[tbp]
\centering
\includegraphics[width=0.95\textwidth]{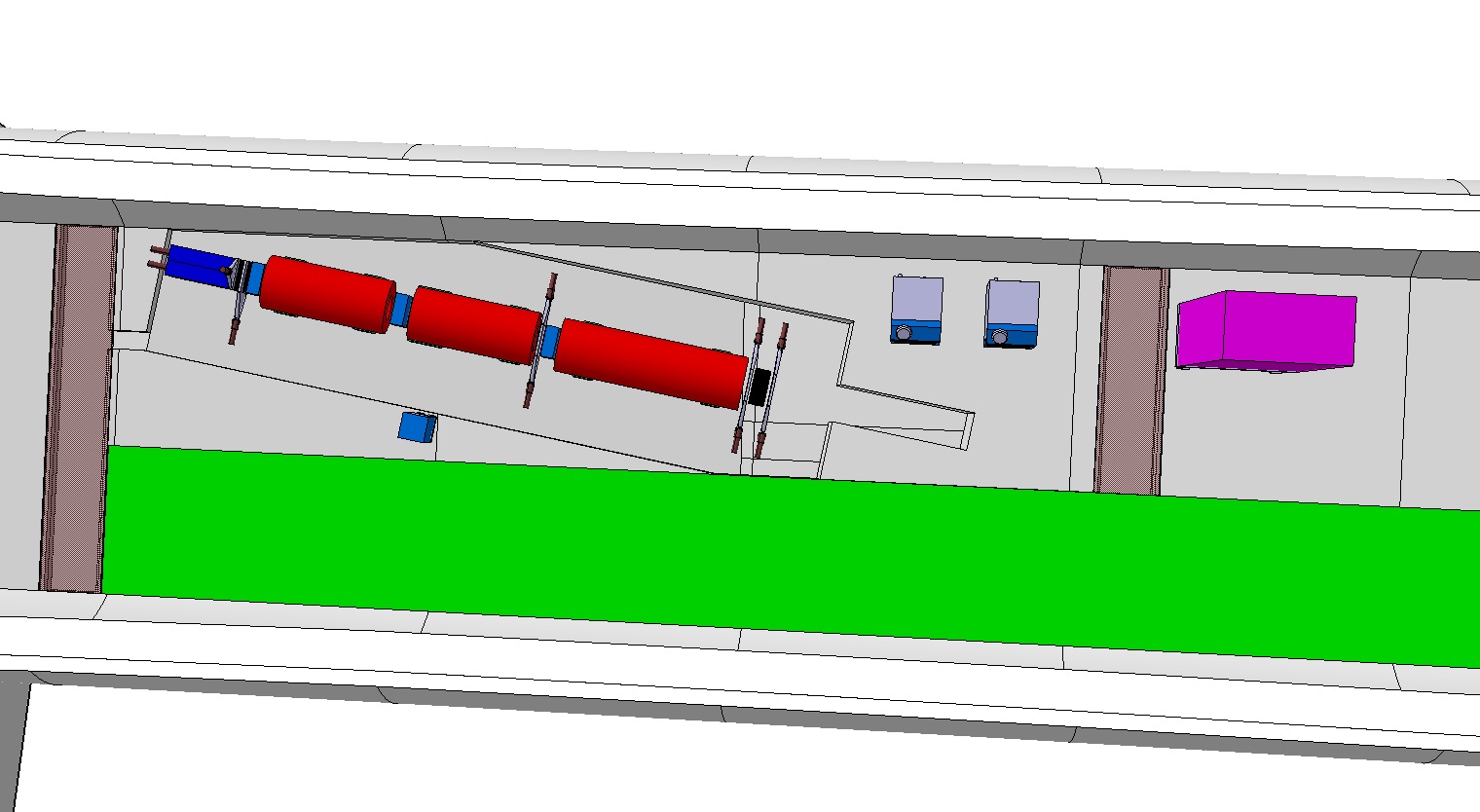}
\includegraphics[width=0.95\textwidth]{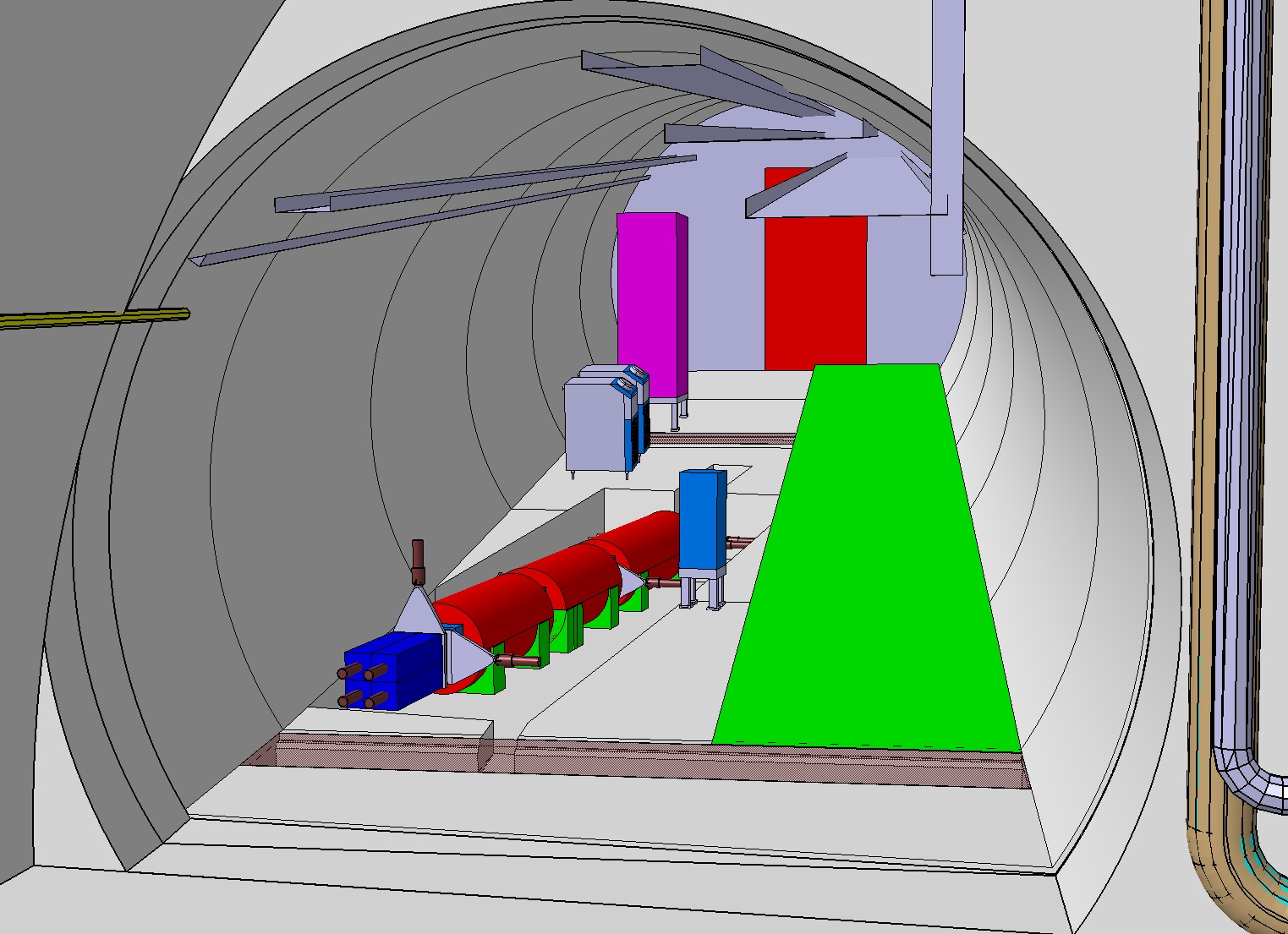}
\caption{Pictures from 3D model of integration of the experiment in the TI12 tunnel.  The model includes: the detector, the mini-crate for the tracker readout electronics, a VME crate for the TDAQ and power supplies, and the chillers (one is the spare). The green space is 1.2~m reserved for access.} 
\label{fig:integrationModel}
\end{figure}

\subsection{Schedule}

The exact schedule and precise time estimates for the various preparatory and infrastructure work still needs to be worked out and fitted into the availability of the various groups. An indicative schedule is shown in \figref{Integration}. The installation window for major equipment such as the magnets is constrained to week 20-22 in 2020, but is only expected to about three days as noted above.

\begin{figure}[t]
\centering
\includegraphics[clip, trim=0.5cm 5.2cm 0.5cm 0.5cm,width=0.95\textwidth]{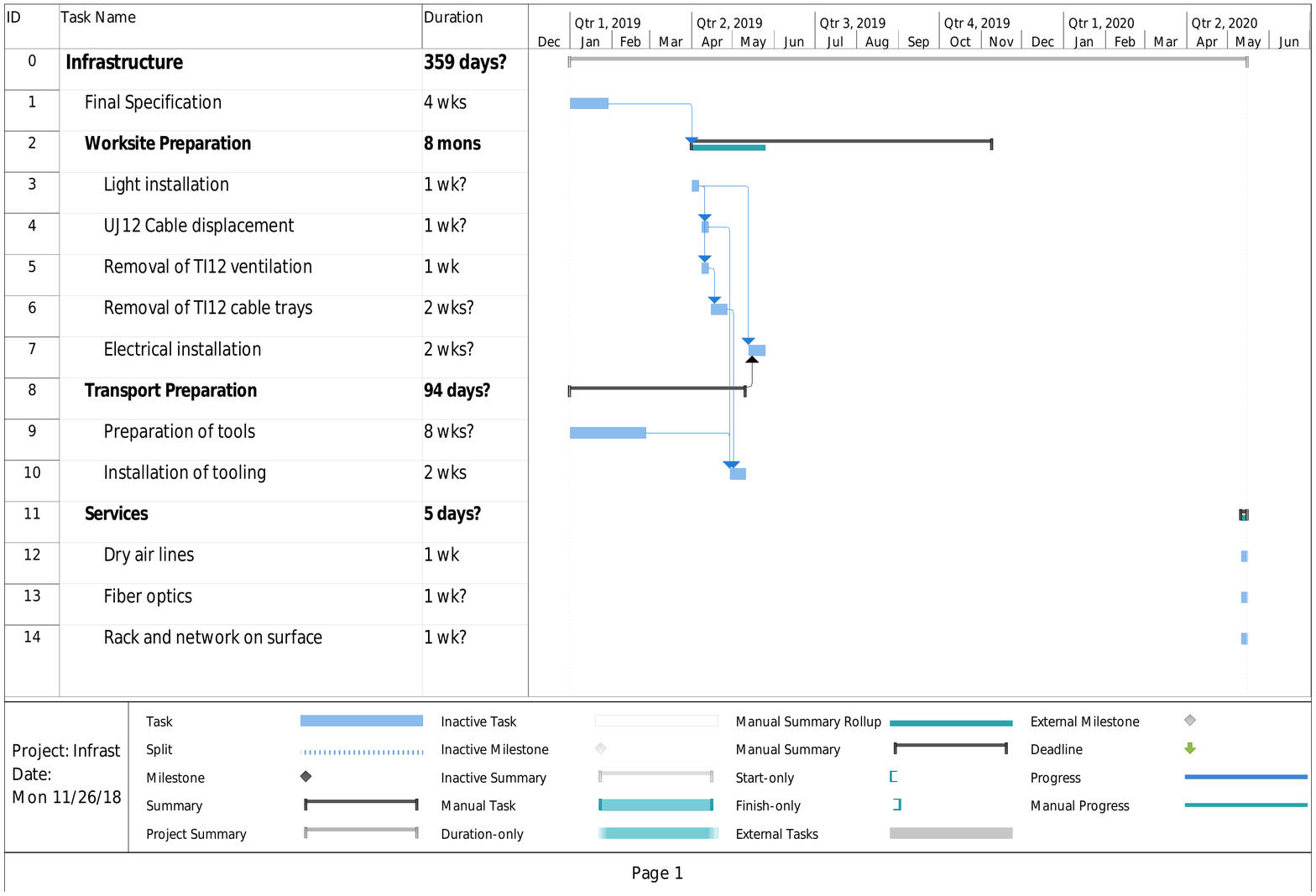}
\caption{Draft schedule for the infrastructure work. The exact scheduling of the various preparatory work still has to be integrated in the LS2 schedule, but is expected to be able to take place in the indicated time intervals in 2019 while the services will be installed in the first half of 2020.}
\label{fig:Integration}
\end{figure}

\clearpage
\section{Commissioning}
\label{sec:commissioning}

Five commissioning phases are planned for the FASER experiment:
\begin{enumerate}
\item {\bf Individual component testing.} Every individual component of the FASER experiment will be separately tested in the place of construction or assembly. This phase includes quality assurance tests for existing components, such as the SCT modules, and testing of cables and boards with pulses. Individual components will need to be tested under magnetic field to ensure proper functioning in realistic conditions. 
\item {\bf Integration on surface.} All the components, except for the magnets, will be assembled for integration on the surface. The commissioning will proceed with cosmic runs in a setup where the assembled components will be placed at an angle with respect to the horizontal line. It is expected that the full detector and readout pieces will participate to such cosmic runs, with the exception of the magnets. Stability runs for extended periods of time are expected to take place.
\item {\bf Dry installation on the surface.} The magnets will be installed in the final support structure on the surface together with all of the detector components to ensure everything fits well. It is foreseen that the full system will be exercised for some period to ensure the readout system is not affected by the magnetic field.
\item {\bf Commissioning in the tunnel, without beam.} Following the installation of individual components in the tunnel, basic testing similar to what was done for `Individual component testing' will be performed to ensure that there is no damage from manipulation during installation. Once all components (including the magnets) are installed in their final position, combined runs will take place with test pulses.  Integration with the external, BST-based, clock will also be done during this period.
\item {\bf Commissioning with beam.} The final commissioning of the FASER experiment will happen with beams. Initial beam commissioning will be attempted with beam-gas interactions, which are expected to be higher after a long shutdown due to worse vacuum conditions, while the final commissioning will be done with the first collision data. Extended calibration runs for the detector are expected to take place.
\end{enumerate}

A preliminary schedule for the commissioning phase is shown in \figref{CommissioningTimeLine}.

\begin{figure}[b]
\centering
\includegraphics[clip, trim=0.5cm 12.5cm 0.5cm 0.5cm,width=0.95\textwidth]{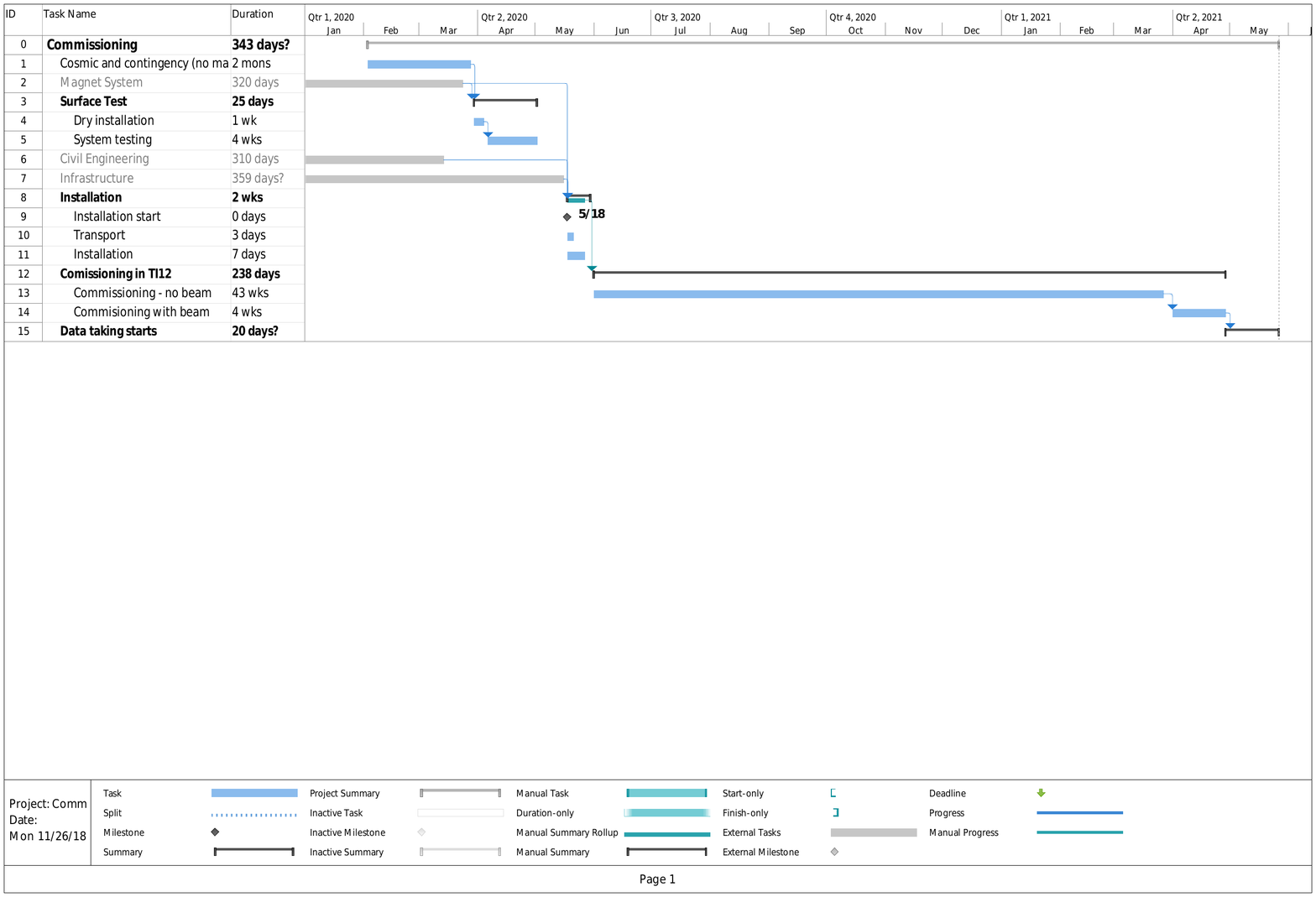}
\caption{Schedule for FASER commissioning on surface and in TI12 as well as for installation in TI12.}
\label{fig:CommissioningTimeLine}
\end{figure}

\clearpage
\section{Safety}
\label{sec:safety}

The FASER project is followed by the Project and Experiment Safety Support (PESS) of the HSE-OHS group. An HSE correspondent has been named, and the following domains are concerned: civil/structural engineering including worksite aspects, asbestos detection and eventual removal, environmental protection,  mechanical safety, HVAC, fire and chemical safety, non-ionizing radiation, and magnetic and electrical safety.  In particular, it has been individuated that the civil and structural engineering domain presents major safety implication aspects and, therefore, a safety clearance shall be released by the HSE unit head before starting the civil works (see Ref.~\cite{CERN-SR-SO}).
A team of HSE specialists is currently drawing the Launch Safety Agreement, listing the applicable rules and standards in the mentioned safety domains. The HSE correspondent, along with the EP DSO office, will support the project leader in gathering the required documentation to complete the safety file.

\clearpage
\section{Offline Software and Computing}
\label{sec:software}

\subsection{Detector Simulation}

A GEANT4-based~\cite{Agostinelli:2002hh} model of the detector generates simulated data for detector optimization and preliminary performance studies. The model includes the nominal geometry of the ATLAS SCT modules and LHCb ECAL modules, along with a simplified approximation (0.6~T uniform dipole) of the magnetic field. No fringe fields, scintillator planes or mechanical support structures are included at present. For the purpose of simulating multiple scattering, the amount of material along the paths of particles through the tracker is accurate to within $\pm 15\%$.

In addition to the generic ``particle gun'' functionality provided by the toolkit, an internal generator can simulate decays of dark photons with user-specified mass, momentum, and daughter particles. Simulated dark photons arrive with a uniform solid angle distribution, assuming a source located at the ATLAS interaction point, and decay uniformly along the length of the detector. \Figref{TrackEventDisplay} shows two views of a simulated dark photon decay recorded in FASER.

The tracker simulation produces digitized strip firing data using a simplified electronics model with Gaussian charge smearing and a uniform threshold, along with information to determine which Monte Carlo truth particles contributed energy to them.  Each simulated ECAL module produces a single value corresponding to the total energy deposited in its scintillating layers. The time structure of the signal, which will be available for the real ECAL modules, is not yet simulated.

\begin{figure}[tbp]
\centering
\includegraphics[width=0.9\textwidth]{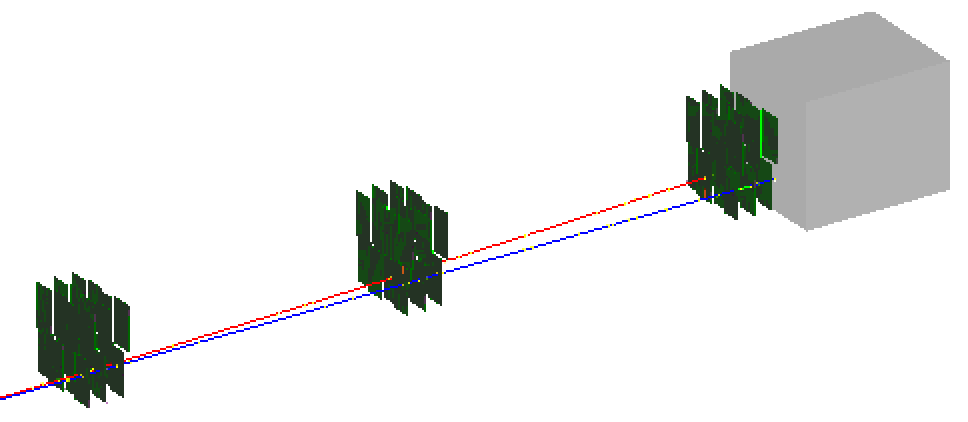}
\includegraphics[width=0.5\textwidth]{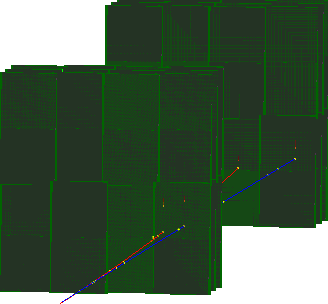}
\caption{Two views of a simulated dark photon decay to $e^+ e^-$ recorded in FASER. Above: Top view showing the two charged tracks separating in the magnetic field as they pass through the tracker and enter the calorimeter at top right. Below: Close-up showing the silicon strips fired in the first planes of the first two tracking stations.}

\label{fig:TrackEventDisplay}
\end{figure}

\subsection{Data Reconstruction}
\subsubsection{Strategy}

The strategy for reconstruction of tracker data will be similar to that used in ATLAS, although the much lower occupancy in FASER allows considerable simplification. Neighboring hit strips on a single side of a tracker plane are grouped into clusters. Then clusters on opposite sides of a plane are combined to form space-point candidates. Combinatorial pattern recognition will be used to identify track candidates passing through space-point candidates in different tracker planes. Track candidates could pick up additional compatible space-points in a road around their trajectory. The final track fit will use all available information, with extrapolation through a realistic model of the detector material (including energy loss and multiple scattering) and a detailed map of the magnetic field.

\subsubsection{Performance studies}

FASER has developed ``proof-of-concept" clustering, space-point and track reconstruction algorithms to optimize and validate the detector design. Track identification and hit association (pattern recognition) is simulated, assuming perfect efficiency, by using Monte Carlo truth information. Misalignment effects are not included.

The tracker must resolve closely-separated, oppositely-charged tracks to identify the dark photon signal. \Figref{TwoTrackSeparation} shows the efficiency to reconstruct two isolated space-points in the first tracker plane for several dark photon masses and momenta.

\begin{figure}[tbp]
\centering
\includegraphics[width=0.9\textwidth]{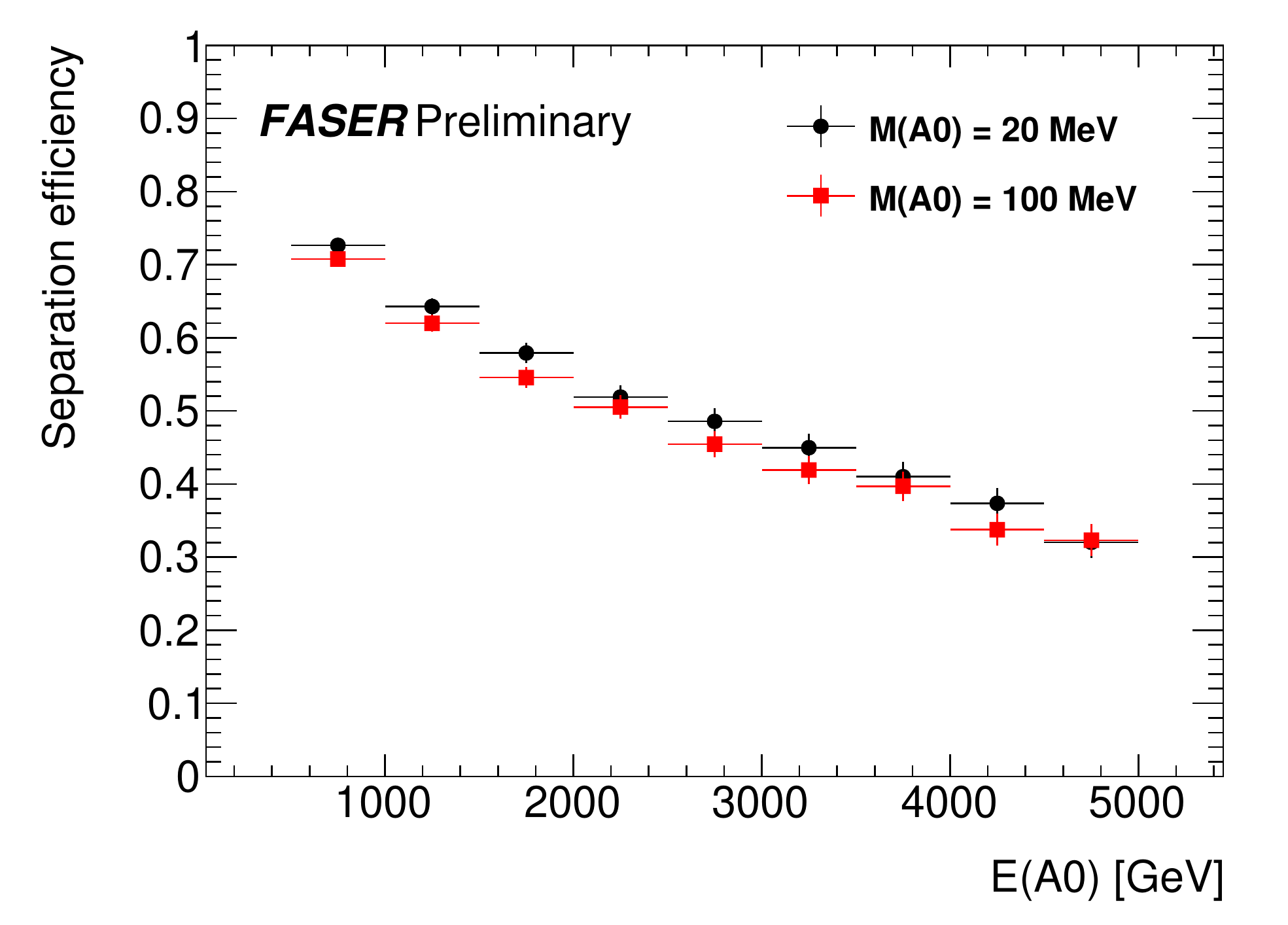}
\caption{Two-track separation efficiency based on isolated space-points in the first tracker plane for the indicated dark photon masses and momenta. Dark photon decays are uniformly distributed over the length of the 1.5~m decay volume. Because most of the separation comes from magnetic bending, rather than the transverse momenta of the decay products, the separation efficiency does not depend on the dark photon mass. Higher dark-photon energies, on the other hand, significantly degrade efficiency.}
\label{fig:TwoTrackSeparation}
\end{figure}

\Figref{SpacePointResolution} shows the reconstructed space-point resolution in the first tracker plane for simulated dark photon decays. The resolution achieved is comparable to that expected from the ATLAS SCT~\cite{Aad:2014mta}.

\begin{figure}[tbp]
\centering
\includegraphics[width=0.45\textwidth]{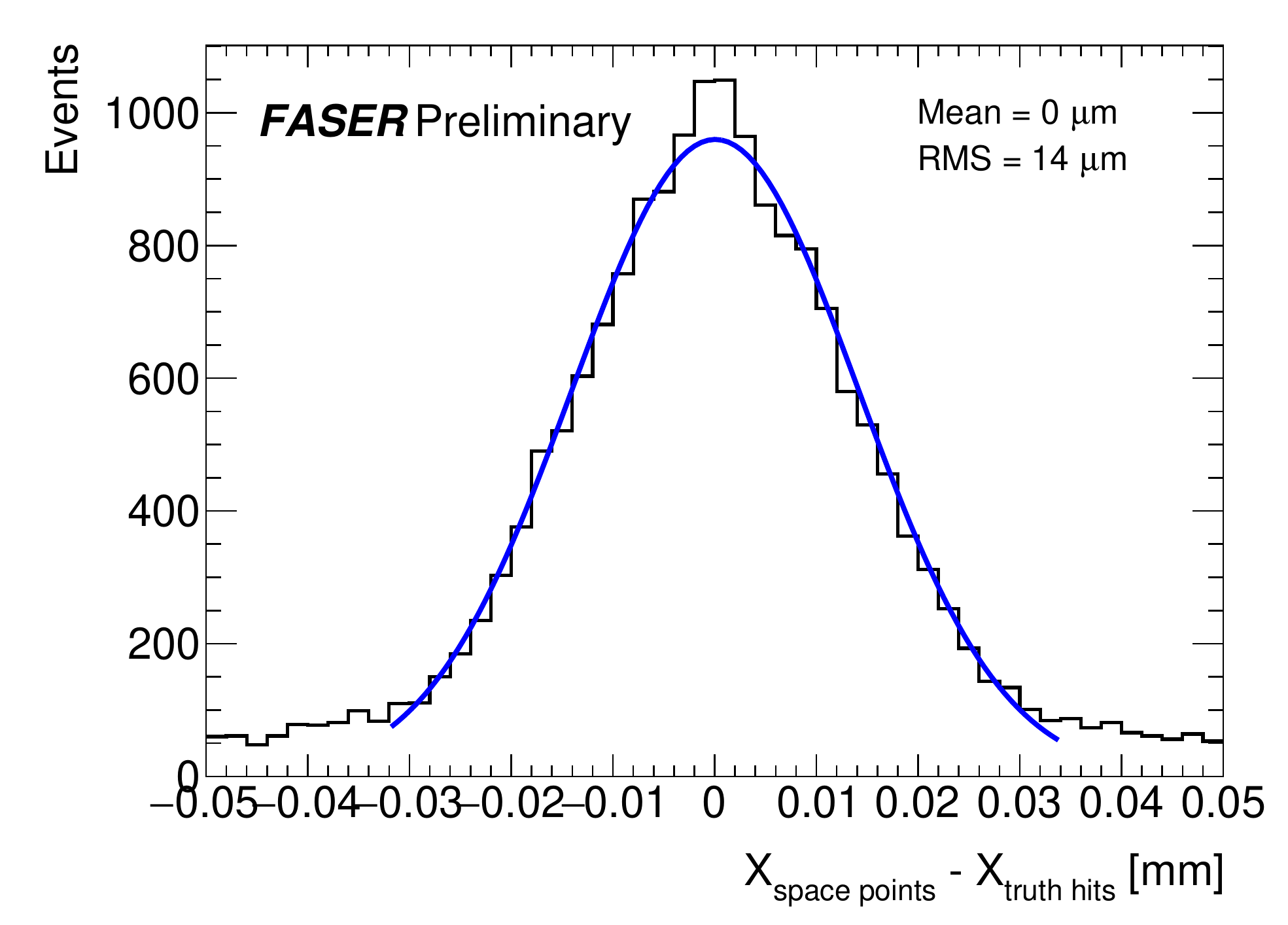}
\includegraphics[width=0.45\textwidth]{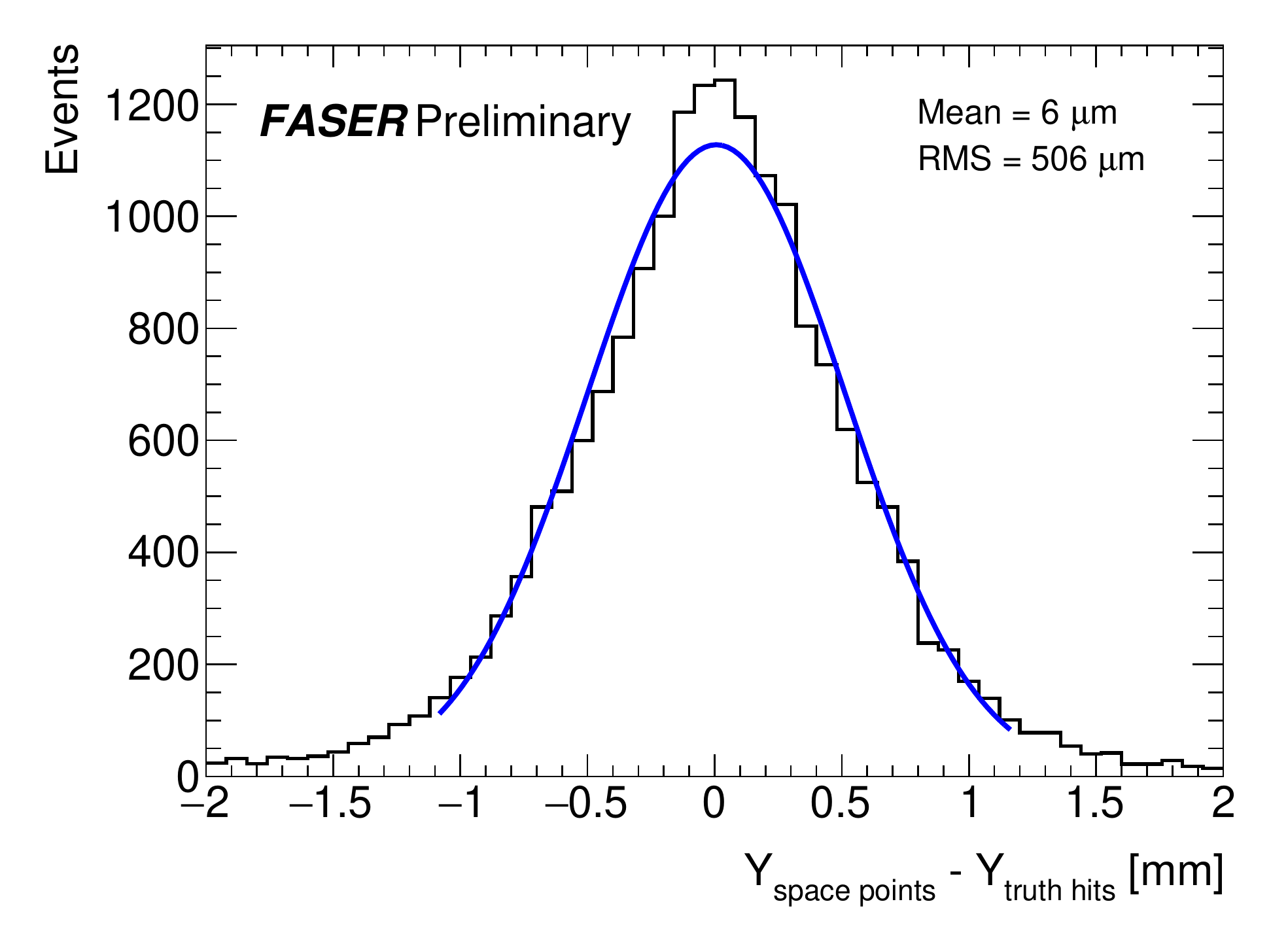}
\caption{Spatial resolution of reconstructed space-points in the magnetic bending direction (left) and along the strip direction (right), with respect to Monte Carlo truth, for the first detector plane.}
\label{fig:SpacePointResolution}
\end{figure}

Karimaki~\cite{Karimaki:1997ff} calculated the expected momentum and other resolutions of a magnetic spectrometer from first principles, under assumptions valid for FASER. \Figref{MomentumResolution} compares this theoretical performance, including measured space-point resolution and expected multiple scattering, with the results of a global $\chi^2$ fit to reconstructed space-points. The preliminary track fit does somewhat worse than the theoretical prediction at higher momentum, but still gives acceptable resolution above 1~TeV.

\begin{figure}[tbp]
\centering
\includegraphics[width=0.9\textwidth]{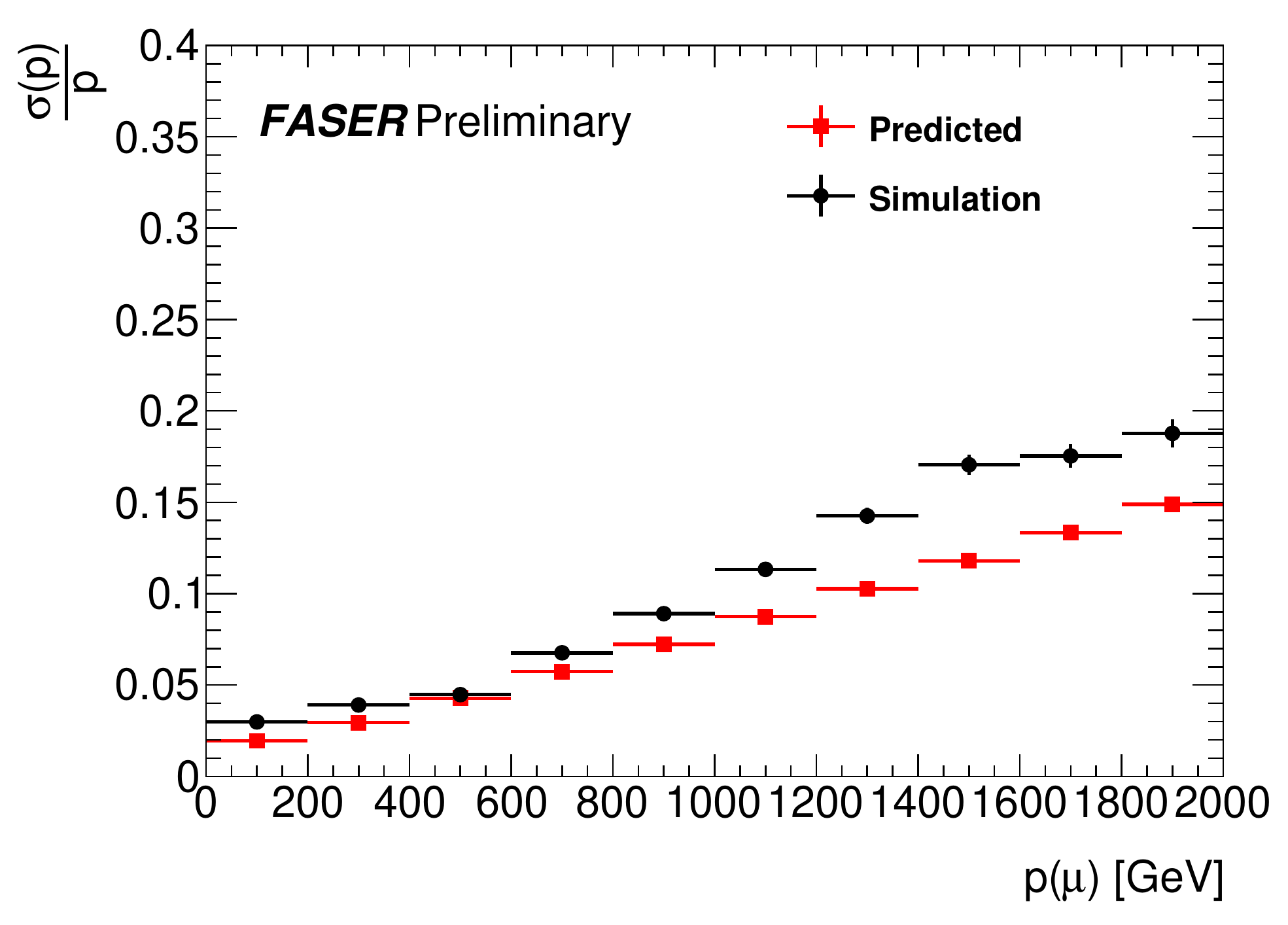}
\caption{Fractional momentum resolution $(\sigma_p/p)$ for reconstructed muon tracks as a function of momentum, compared to the predicted resolution from Karimaki~\cite{Karimaki:1997ff}.}
\label{fig:MomentumResolution}
\end{figure}

To summarize, in these and other studies, using relatively unrefined reconstruction software, the simulated performance of the detector closely tracks our design expectations.

\subsubsection{ACTS for FASER}

The prototype reconstruction software used for validation and optimization of the detector design is inadequate for physics use because it lacks the ability to accommodate a realistic magnetic field or a detailed description of the detector's material. More sophisticated pattern recognition, extrapolation, and track-fitting will be required.

Moving forward, the ACTS (``A Common Tracking System'') project~\cite{Gumpert:2017wrm} offers FASER the opportunity to leverage a fully-featured, open-source reconstruction framework based on the ATLAS tracking software that several FASER members are already familiar with.  ATLAS plans to replace its original tracking software with this modernized version after Run 3, and so it will be commissioned and then supported during the lifetime of FASER.

FASER members have worked with the ACTS team over the last six months, and a proof-of-concept implementation of the FASER geometry in the ACTS detector description framework has already been achieved. The ACTS developers have embraced FASER as a useful milestone on their road to the more demanding requirements of ATLAS itself. Thus, we are confident FASER will have access to world-class, collaboratively developed and maintained track reconstruction software, with a far more realistic commitment of effort than would be required to deliver the same functionality ourselves.

\subsection{Offline Software Infrastructure and Schedule}

In keeping with its economical design and compact scale, FASER's offline software will be much simpler than the larger and more complex flagship LHC experiments. ROOT~\cite{Brun:1997pa} and related technology will be sufficient for many of FASER's geometry, conditions, data quality and other database needs. The challenge will not be to find software capable of meeting our needs, but avoiding (as much as possible) the overhead of tools that far exceed them.

\subsection{Computing}

FASER's computing and storage requirements will be orders of magnitude smaller than the flagship LHC experiments. The detector's 72 ATLAS SCT modules represent less than 2\% of the corresponding ATLAS sub-detector, and the occupancy is lower by a similar factor.  As discussed in \secref{tdaq}, the bulk of FASER's raw data will consist of waveform information from the trigger/timing scintillator PMTs and LHCb ECAL modules, with an expected average event size of 25 kB. Pessimistically, at expected Run 3 luminosity, and with no trigger pre-scaling or data reduction, the expected raw data rate from the experiment is $16\,\hbox{MB}/\hbox{s}$ or roughly $1\, \hbox{TB}/\hbox{fb}^{-1}$. If raw waveform data from events firing the veto can be discarded or summarized, the event size shrinks to about 5 kB and the data rate falls to $\sim 200 \, \hbox{GB}/\hbox{fb}^{-1}$ even without further data reduction. This rate is unlikely to tax the capacity of the CASTOR data storage system.

Reconstruction of FASER data will consume negligible CPU time. Simulation time (for signal events) is completely dominated by the calorimeter in the present, unoptimized implementation. Because the calorimeter serves primarily as a tag for energetic showers, fast simulation or parameterization of its response can be used to avoid the much more costly full simulation. 

TeV muons in FASER can be simulated on a single desktop PC at rates exceeding the realtime trigger rate at peak luminosity. Neutrino background simulations will likely be the most CPU-intensive: for an integrated luminosity of $150\, \hbox{fb}^{-1}$, roughly 100,000 charged current neutrino interactions above 100 GeV are expected in FASER's magnets; scaling the single-muon simulation time by the expected multiplicity (30), this corresponds to approximately 250 CPU hours on a single-core PC to simulate twenty times the expected background sample. FASER should require only a token allocation of shared capacity on the CERN batch system to meet its computing needs.

\subsection{Cost and Schedule}

Planning for offline software development and commissioning is still in its earliest stages, but one of the first steps will be to implement a prototype geometry database in the first half of 2019, to provide a common reference for existing simulation and evolving ACTS reconstruction software.

As graduate students join the experiment, they will help integrate simulation and reconstruction tools with the geometry database.  Demonstration of ``full-chain'' data simulation and reconstruction with early versions of the production software by the end of 2019 is a realistic milestone.

Conditions data and associated tools for calibration, alignment, data quality, and luminosity, along with physics analysis code, would be built on top of this foundation as it continues to mature in 2020.

CERN's IT Department estimates the cost of FASER's computing and data storage needs at 160~kCHF total over the lifetime of the experiment (2020 -- 2024) based on the following assumptions:

\begin{itemize}
\item Collecting 1 petabyte (PB) of real and simulated data per year onto tape media;
\item 1 PB of disk space for raw data, Monte Carlo, derived datasets and tape  staging;
\item 1000 CPU cores available for processing, simulation and analysis (equivalent to 1000 concurrent running jobs);
\item Home directories for users;
\item Project space for central datasets (ntuples);
\item On-demand databases for bookkeeping and metadata;
\item 20\% available in 2020, full capacity in 2021 -- 2024 (three years of running and one year post-analysis).
\end{itemize}

In addition, we expect at least one FASER institution (to be determined) will provide off-site computing infrastructure, for example by sharing resources with an existing ATLAS Tier-3 grid facility.

\clearpage
\section{Overall Cost and Schedule}
\label{sec:costing}

\subsection{Cost}
The overall cost of the experiment hardware, to be borne by the FASER collaboration, is summarized in Table~\ref{tab:overAllCost}. The costs for each major subsystem are itemized in the corresponding sections above. Spares are shown separately in the table, as in some cases these can be shared between different systems. To ensure adequate funding a contingency of 20\% will be included on top of the costs shown. 
The cost estimates for the main infrastructure work that is assumed to be borne by CERN is detailed in Table~\ref{tab:CERNcosts}. 

\begin{table}[bp]
\centering
\begin{tabular}{|l|c|c|}
\hline
\  {\bf Detector component} \ &  \ {\bf Cost [kCHF]} \  &  \ {\bf Detailed Table} \  \\ \hline
\ Magnet \ &  420 \quad & Table~\ref{table:magnet-budget}  \\ \hline
\ Tracker Mechanics \ &  66 \quad  & Table~\ref{table:trackerMechanicsBudget} \\ \hline
\ Tracker Services  \ &  105 \quad  & Table~\ref{table:trackerBudget} \\ \hline
\ Scintillator Trigger \& Veto  \ &  52 \quad  & Table~\ref{table:scintillatorBudget} \\ \hline
\ Calorimeter \ &  13 \quad  & Table~\ref{table:calorimeterBudget} \\ \hline
\ Support structure  \ &  60 \quad  & Table~\ref{table:SupportStructure} \\ \hline
\ Trigger \& Data Acquisition \ &  52 \quad  & Table~\ref{table:tdaqBudget} \\ \hline
\  {\bf Total} \ & {\bf 768}  \quad & - \\ \hline 
\ Spares \ &  56 \quad  & - \\ \hline
\end{tabular}
\caption{Overall budget for FASER experiment hardware. The TDAQ system includes the readout for all detectors (including the Tracker).} 
\label{tab:overAllCost}
\end{table}

\begin{table}[bp]
\centering
\begin{tabular}{|l|c|}
\hline
\  {\bf Work} \ &  \ {\bf Cost [kCHF]} \  \\ \hline
\ Civil Engineering \ &  160** \quad   \\ \hline 
\ Transport \ &  95** \quad   \\ \hline
\ Optical Fiber \& Network Connection \ &  10* \quad \\ \hline
\ Power Connection \ &  10 \quad \\ \hline
\ Compressed Air Connection \ &  6 \quad \\ \hline
\ Preparation of TI12 \ &  10* \quad \\ \hline 
\  {\bf Total} \ & {\bf $291$}  \quad \\ \hline 
\end{tabular}
\caption{Budget for infrastructure work and computing support whose cost is assumed to be borne by CERN. Items marked by a * are not full quotes but are rough cost estimates, whereas items marked with ** have parts that have been fully costed, but an additional part that is a more rough cost estimate.
}
\label{tab:CERNcosts}
\end{table}

\subsection{Schedule}

Preparation of the TI12 tunnel (removing ventilation pipes, installing lights) must happen in 2019 to allow the civil engineering work to proceed in week 5-10 of 2020. Magnet construction (the other main schedule driver) will progress in parallel during 2019, along with the detector construction and commissioning tasks on the surface. The target for installation and testing of the transport tooling, followed by detector installation, is week 20-22, before the LHC is re-filled with Helium. After installation, commissioning the detector {\em in situ} (first with access, and then without) will become the main activity.

\begin{figure}[t]
\centering
\includegraphics[clip, trim=0.5cm 2.45cm 0.5cm 0.5cm,width=0.92\textwidth]{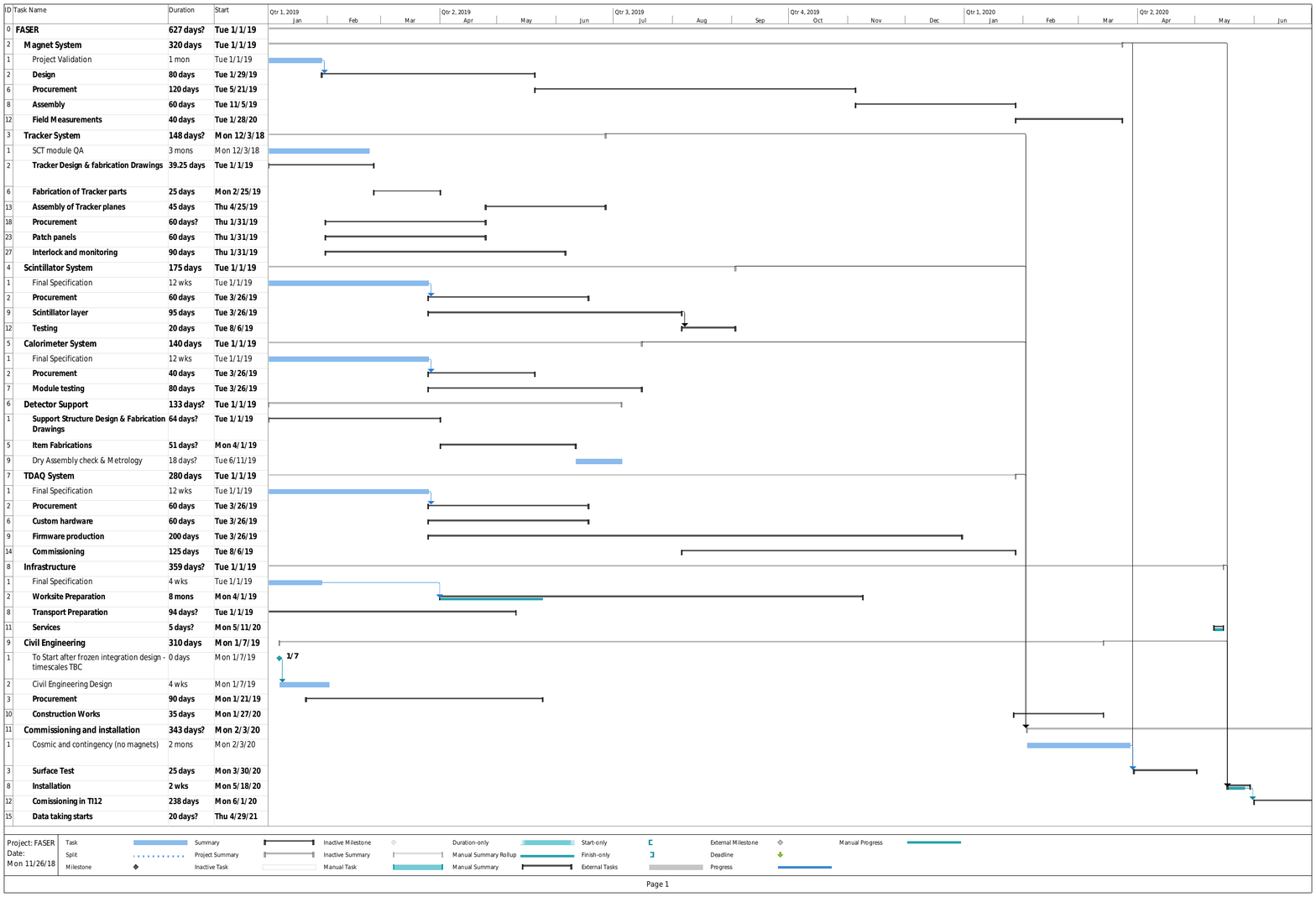}
\caption{Overall schedule for FASER construction, installation, and commissioning in TI12.}
\label{fig:MasterTimeLine}
\end{figure}

\acknowledgments

We are grateful to the ATLAS SCT project and the LHCb Calorimeter project for letting us use spare modules as part of the FASER experiment. 
In addition, FASER gratefully acknowledges invaluable assistance from many people, including 
the CERN Physics Beyond Colliders study group; the LHC Tunnel Region Experiment (TREX) working group; 
Rhodri Jones, James Storey, Swann Levasseur, Christos Zamantzas, Tom Levens, Enrico Bravin (beam instrumentation); 
Dominique Missiaen, Pierre Valentin, Tobias Dobers (survey); 
Caterina Bertone, Serge Pelletier, Frederic Delsaux (transport); 
Andrea Tsinganis (FLUKA simulation and background characterization); 
Attilio Milanese, Davide Tommasini, Luca Bottura (magnets); 
Burkhard Schmitt, Christian Joram, Raphael Dumps, Sune Jacobsen (scintillators); 
Dave Robinson, Steve McMahon (ATLAS SCT); 
Yuri Guz (LHCb calorimeters); 
Stephane Fartoukh, Jorg Wenninger (LHC optics), Michaela Schaumann (LHC vibrations);
Marzia Bernardini, Anne-Laure Perrot, Katy Foraz, Thomas Otto, Markus Brugger (LHC access and schedule);
Simon Marsh, Marco Andreini, Olga Beltramello (safety); 
Stephen Wotton, Floris Keizer (SCT QA system and SCT readout); 
Yannic Body, Olivier Crespo-Lopez (cooling/ventilation);
Yann Maurer (power);
Marc Collignon, Mohssen Souayah (networking); 
Gianluca Canale, Jeremy Blanc, Maria Papamichali (readout signals);
Bernd Panzer-Steindel (computing infrastructure);
and Fido Dittus, Andreas Hoecker, Andy Lankford, Giovanna Lehmann, Ludovico Pontecorvo, Michel Raymond, Christoph Rembser, Stefan Schlenker (useful discussions). 

J.L.F., F.K., and J.S.~are supported in part by U.S.~National Science Foundation Grant No.~PHY-1620638.  
J.L.F.~is supported in part by Simons Investigator Award \#376204. 
I.G.~is supported in part by U.S.~Department of Energy Grant DOE-SC0010008.  
S.T.~is supported in part by the Lancaster-Manchester-Sheffield Consortium for Fundamental Physics under STFC grant ST/L000520/1, by the Polish Ministry of Science and Higher Education under research grant 1309/MOB/IV/2015/0, and by the National Science Centre (NCN) research Grant No.~2015-18-A-ST2-00748.
This work is supported in part by the Swiss National Science Foundation.

\bibliography{FASER_TP_arxiv}
\end{document}